\documentclass{aa}
\PassOptionsToPackage{desactivate}{linenoaa}
\usepackage{graphicx}
\usepackage{txfonts}
\usepackage{xfrac}
\usepackage[colorlinks=true,linkcolor=blue,citecolor=blue,urlcolor=blue]{hyperref}
\usepackage[flushleft]{threeparttable}
\usepackage{enumitem}
\usepackage{linenoaa}
\usepackage{orcidlink}
\let\orcid\orcidlink

\newcommand{\ie}{{i.e.}\xspace}
\newcommand{\eg}{{e.g.}\xspace}

\newcommand{\redshift}{\ensuremath{{\rm z}}\xspace}
\newcommand{\Rvir}{\ensuremath{R_{\rm vir}}\xspace}
\newcommand{\Mvir}{\ensuremath{M_{\rm vir}}\xspace}

\newcommand{\Gyr}{\ensuremath{\,{\rm Gyr}}\xspace}
\newcommand{\Mpc}{\ensuremath{\,{\rm Mpc}}\xspace}
\newcommand{\cMpc}{\ensuremath{\,{\rm cMpc}}\xspace}
\newcommand{\kpc}{\ensuremath{\,{\rm kpc}}\xspace}
\newcommand{\pc}{\ensuremath{\,{\rm pc}}\xspace}
\newcommand{\sm}{\ensuremath{\,{\rm M_{\odot}}}\xspace}
\newcommand{\slu}{\ensuremath{\,{\rm L_{\odot}}}\xspace}

\newcommand{\degree}{\ensuremath{{\rm ^o}}\xspace}
\newcommand{\as}{\ensuremath{\,{\rm arcsec}}\xspace}

\newcommand{\mags}{\ensuremath{\,{\rm mag}}\xspace}

\newcommand{\Rr}{\ensuremath{\Re_{\star}}\xspace}

\newcommand{\rhoavf}{\ensuremath{\langle\bar{\rho}^{\star}_{2h\star}\rangle}\xspace}

\newif\ifaddbold
\addboldfalse  % flag to remove bold changes

\newif\ifaddcom
\addcomfalse  % flag to remove bold comments

\ifaddbold
    \newcommand\cMBII[1]{\textcolor{black}{\textbf{#1}}}
    \newcommand\cTPII[1]{\textcolor{black}{\textbf{#1}}}
    \newcommand\cYOBII[1]{\textcolor{black}{\textbf{#1}}}
\else
    \newcommand{\cMBII}{}
    \newcommand{\cTPII}{}
    \newcommand{\cYOBII}{}
\fi

\ifaddcom

\else

\fi

\newcommand{\cMB}{}

\begin{document} 

\title{Diving into dangerous tides: The impact of galaxy cluster tidal environments on satellite galaxy mass densities}

\titlerunning{Diving into dangerous tides}

\author{Mat\'ias~Bla\~na\inst{1}\orcid{0000-0003-2139-0944},
Thomas~H.~Puzia\inst{2}\orcid{0000-0003-0350-7061},
Yasna~Ordenes-Brice\~no\inst{3}\orcid{0000-0001-7966-7606},
Patricia~B.~Tissera \inst{2,4}\orcid{0000-0001-5242-2844},
Marcelo~D.~Mora\inst{5}\orcid{0000-0003-4385-0411},\,\,\,\\ 
\,\,\,\,\,
Diego~Pallero\inst{6}\orcid{0000-0002-1577-7475}, 
Evelyn~Johnston\inst{3}\orcid{0000-0002-2368-6469}, 
Bryan~Miller\inst{7}\orcid{0000-0002-5665-376X},
Tuila~Ziliotto\inst{8}\orcid{0000-0001-8538-2068},
Paul~Eigenthaler\inst{2}\orcid{0000-0001-8654-0101},
\and Gaspar~Galaz\inst{2}\orcid{0000-0002-8835-0739}
}
\institute{
Vicerrector\'ia de Investigaci\'on y Postgrado, Universidad de La Serena, 
La Serena 1700000, Chile
\\\email{matias.blana.astronomy@gmail.com}
\and Instituto de Astrof\'isica, Pontificia Universidad Cat\'olica de Chile,
Av. Vicu\~na Mackenna 4860, Santiago, 7820436, Chile
\and Instituto de Estudios Astrof\'isicos, Facultad de Ingenier\'ia y Ciencias, Universidad Diego Portales, 
Av. Ej\'ercito Libertador 441, Santiago, Chile
\and Centro de Astro-Ingenier\'ia, Pontificia Universidad Cat\'olica de Chile
Av. Vicu\~na Mackenna 4860, Santiago, 7820436, Chile
\and  Las Campanas Observatory, Carnegie Observatories, 
Casilla 601, La Serena, 7820436, Chile
\and Departamento de F\'isica, Universidad T\'ecnica Federico Santa Mar\'ia, 
Avenida España 1680, Valparaíso, Chile
\and Gemini Observatory, South Operations Center, 
Casilla 603, La Serena, Chile
\and Department of Physics and Astronomy Galileo Galilei, University of Padua, 
Via Marzolo, 8-35131, Padua, Italy\\
}

\date{Received 7 February 2024 / Accepted 13 May 2025}
\abstract 
{
Satellite galaxies endure powerful environmental tidal forces that drive mass stripping of their outer regions.
Consequently, satellites located in central regions of galaxy clusters or groups, where the tidal field is strongest, are expected to retain their central dense regions while losing their outskirts. This process produces a spatial segregation in the mean mass density with the cluster-centric distance (the $\bar{\rho}-r$ relation).
To test this hypothesis, we combined semi-analytical satellite orbital models with cosmological galaxy simulations. We find that not only the mean total mass densities ($\bar{\rho}$), but also the mean stellar mass densities ($\bar{\rho}^{\star}$) of satellites exhibit this distance-dependent segregation ($\bar{\rho}^{\star}\!\!-\!r$). The correlation traces the host's tidal field out to a characteristic transition radius at $\Rr\!\approx\!0.5\,\Rvir$, beyond which the satellite population's density profile can have a slight increase or remain flat, reflecting the weakened tidal influence in the outskirts of galaxy clusters and beyond.
We compare these predictions with observational data from satellites in the Virgo and Fornax galaxy clusters, as well as the Andromeda and Milky Way systems. Consistent trends in the satellite mean stellar mass densities are observed across these environments.
Furthermore, the transition radius serves as a photometric diagnostic tool: it identifies regions where the stellar components of satellites underwent significant tidal processing and probes the gravitational field strength of the host halo.}


\authorrunning{Bla\~na et al.}
\maketitle

\section{Introduction}
\label{sec:intro}
Satellite galaxies experience significant changes as they plunge into denser galaxy ecosystems and interact\footnote{\scriptsize Everything is interaction (in German, Alles ist Wechselwirkung), from Alexander von Humboldt diaries, 1803/04: \url{http://resolver.staatsbibliothek-berlin.de/SBB0001527C00000000}}\,with more massive structures.
Many studies have extensively discussed important related subjects, such as the hierarchical assembly and clustering of galaxies \citep[][and references therein]{White1978,Merritt1985,Navarro1996,MoBoWh2010}, 
and the environmental processes that shape satellite galaxies, such as
the interactions with the harsh gas environment in galaxy clusters and groups through ram pressure stripping, instabilities, and thermal heating \citep[\eg][]{Gunn1972,Mori2000,Boselli2006,Peng2010,Armillotta2017,Boselli2022}.

\cMBII{Satellite evolution can also be driven by the gravitational field of their host systems and/or by interactions with other satellites \citep[harassment;][]{Toomre1972,Merritt1983,Merritt1984,Gnedin2003,Smith2016}.
The host's gravitational tidal field can change the satellite mass distribution through tidal (de-)compressions, which can change the mass densities in satellites \citep{Valluri1993,Dekel2003,Masi2007,Renaud2009}, 
and through tidal stripping and total satellite disruption \citep[cannibalism;][]{Merritt1985,Read2006a,Read2006b,Fellhauer2006,Fellhauer2007b,Fellhauer2008,Penarrubia2008,Blana2015,Ogiya2022a,Errani2021a,Errani2022,Errani2023,Stucker2023,Montero-Dorta2024}.
As a consequence, the long-time accreted satellites that continue to survive will only have their central dense regions gravitationally bound, whereas their external, less dense regions, as well as most satellites, are cannibalised by the hosting cluster, especially in the central regions of clusters. 
First-infall satellites, on the other hand, will have their outer regions stripped only where the host tidal field becomes stronger than the local gravity of the satellite.
This establishes a relation between the mass densities of satellite galaxies and their distances from the host (the $\bar{\rho}\!-\!r$ relation). 
Previous studies of the Milky Way (MW) satellite galaxies located within $\sim\!80\kpc$ found such a relation \citep{Kaplinghat2019,Robles2021,Pace2022,Cardona-Barrero2023,Hammer2024}.}

In this article, we investigate this relation further
\cMBII{by analysing more massive environments, such as galaxy clusters, while also analysing the mean mass densities of different components (stellar, gas, and dark matter).}
\cMBII{We also adopt three approaches to study this relation: (1) semi-analytical satellite orbital models with a simplified stripping model,} (2)cosmological galaxy simulations, and (3) observations of satellites in galaxy clusters and in the Local Group.
The sections of the article are ordered as follows:
In Sect.~\ref{sec:theo} we explain the theoretical framework. 
In Sect.~\ref{sec:met} we describe the methodology used for modelling, analysis, and the data from simulation and observations.
In Sect.~\ref{sec:res} we present the results, followed by a discussion in Sect.~\ref{sec:dis}.
We end in Sect.~\ref{sec:con} with our conclusions.

\section{\cMB{Theoretical background}}
\label{sec:theo}

\subsection{\cMB{The relation between the satellite mean mass densities, pericentres, and the host tidal field}}
\label{sec:theo:tdenrel}

When satellite galaxies enter their hosting cluster or group, they are affected by the gravitational tidal field of the main host.  
The closer the satellite is to the centre of the host, the stronger the tidal field becomes, 
stripping and accreting the material in the outskirts of the satellite.
In consequence, satellites that have long been accreted and still survive the tidal forces have only their dense central regions remaining and gravitationally bound.
This is expected to produce the ($\bar{\rho}-r$) relation, which is an (anti-)cor\-relation between satellite mass densities and their orbital pericentres ($r_{\rm peri}$).
For example, \citet[][see their Fig.2]{Kaplinghat2019} and \citet[][their Fig.1]{Cardona-Barrero2023} quantified the correlation between the central dynamical mass densities of MW satellites measured at a radius of 150\pc from the satellite centre ($\rho_{150}$), and their orbital pericentres for a few MW satellites within 90\kpc.
Similarly, \citet[][their Fig.8]{Hayashi2020} compared the values of the MW satellite $\rho_{150}$ with data from cosmological galaxy simulations.
Although these studies find a measurable correlation, they used the dynamical mass densities of the satellites measured at a fixed radius of 150\pc, which does not consistently trace environmental tidal effects.
In fact, the quantity $\rho_{150}$ was instead proposed to probe the impact of stellar feedback on the central dark-matter densities of dwarf galaxies \citep{Read2019a}. 
Furthermore, \citet[][their Fig.5]{Pace2022} do explore the density-pericentre relation ($\bar{\rho}-r$) comparing the satellite mean dynamical mass densities measured within their stellar half-mass radii ($\bar{\rho}_{h\star}$) with the orbital pericentres and with the MW tidal field, finding that most satellites have densities above the tidal field strength equivalent.
\cMBII{Moreover, the analysis of the stellar mass densities of MW star clusters also shows a correlation with their pericentres and the MW tidal field strength, as these systems experience the same environmental tidal forces as satellite galaxies.} \citep[][and references therein]{Gieles2011,Carballo-Bello2012,Webb2013}. \cTPII{This has also been demonstrated in the globular cluster system of NGC\,1399, the central giant ellitpical galaxy in the Fornax galaxy cluster, where the globular clusters in the circum-nuclear region were found to be systematically smaller and denser as a result of the strong tidal field \citep{Puzia2014}.}
However, while \textit{Gaia} has made it possible to estimate the orbital pericentres of the MW satellites in the inner region and in its outskirts, albeit with large observational uncertainties \citep{McConnachie2020,McConnachie2021} \citep[see also][]{Blana2020}, this becomes technically impossible for distant extra-galactic satellite systems.

\cMBII{\cYOBII{In this work, we develop a model to investigate how the gravitational tidal field of a host galaxy cluster or group influences the mean mass densities of its satellite galaxy population, and to identify potentially observable signatures of these effects.} }
To this end, we employ a simplified tidal stripping model presented in \citet[][see Eq.~8.91, 8.92, 8.107, 8.108]{Binney1998} \citep[for a more detailed model see][]{Read2006a}.
\cYOBII{A satellite of total mass $m$, subject to the tidal field $|\tau|$ of its host cluster or group, defines a region around the satellite with a maximum volume of $\sfrac{4}{3}\pi r_{\rm J}^3$ with a (Jacobi) radius $r_{\rm J}$, in which the material remains bound to the satellite. The combination of this volume and the satellite mass $m$ yields the Jacobi mean mass density ($\bar{\rho}_{\rm J}$), which depends on the tidal field as follows:}
\begin{align}
& \bar{\rho}_{\rm J} = |\tau|
\label{eq:tdenj}.
\end{align}
The tidal field (magnitude) of the host system ($\tau$) depends on the gradient of the acceleration field of the host's mass distribution. For simplicity, we adopt the approximation for a spherically symmetric system, which depends on the mean mass density profile \cMBII{of the host} ($\bar{\rho}^{\rm H}$) as a function of the \cMBII{distance from the host centre} $r$ given as
\begin{align}
& |\tau| := 3f^{-3}\bar{\rho}^{\rm H}\left(r\right) = 3f^{-3} \frac{M^{\rm H}(r)}{\sfrac{4}{3}\,\pi r^3} \,\,{;\rm where} \,\, f=\left(1- \frac{1}{3}\frac{{\mathrm d}\ln M^{\rm H}}{{\mathrm d} \ln r} \right)^{-1/3}, \label{eq:tfield}
\end{align}
where $M^{\rm H}(r)$ is the cumulative mass profile of the host and the function $f$ improves the modelling for extended host mass distributions, which simplifies to $f=1$ for a point such as mass or the far-field approximation. Considering typical mass profiles for cluster haloes, such as the NFW density profile \citep{Navarro1996}, results in a tidal field profile that decreases with distance ($|\tau|\propto1/r$).

The Jacobi density of the satellite ($\bar{\rho}_{\rm J}$) is the mean mass density calculated within $r_{\rm J}$, \ie:
\begin{align}
& \bar{\rho}_{\rm J} :=\frac{m\left(r_{\rm J}\right)}{\sfrac{4}{3}\pi r_{\rm J}^3}
\,\,{,\rm where} \,\,r_{\rm J} = \left(\frac{m}{\sfrac{4}{3}\pi\, |\tau(r)|}\right)^{1/3}
\label{eq:denj}
\end{align}
is the instantaneous Jacobi radius, or also called Hills radius or tidal radius.

\cYOBII{As a satellite orbits within a galaxy cluster and approaches its centre, the tidal field strength $|\tau|$ increases, causing the Jacobi radius to shrink. If $r_{\rm J}$ becomes smaller than the satellite's mass truncation radius ($r_{\rm tr}$), the satellite begins to lose mass through tidal mass stripping, further reducing $r_{\rm J}$.}
\cMBII{Using Eq.~\ref{eq:tdenj} and assuming a mass $m(r')$ and mean mass density $\bar{\rho}(r')$ profiles for the satellite, where $\bar{\rho}(r')\!=\!m(r')/(\sfrac{4}{3}\pi r'^3)$ depends on the satellite-centric coordinate $r'$, we can measure the value of $r'$ at which the satellite's mean mass density profile exceeds the tidal field and, thus, remains gravitationally bound to the satellite; namely, where}
\cMBII{
\begin{align}
 \bar{\rho}_{\rm tr} &\geq \bar{\rho}_{\rm J} = |\tau|  \label{eq:tdentr},
\end{align}
with $\bar{\rho}_{\rm tr}\!=\!\bar{\rho}(r'\!=\!r_{\rm tr})$. With this density we can determine the mass that is still bound and within the truncation radius $m_{\rm tr}\!=\!m(r'\!=\!r_{\rm tr})$, and the truncation radius constrained by $r_{\rm J}$ according to
\begin{align}
r_{\rm tr}\!&\leq r_{\rm J}= \left(\frac{m_{\rm tr}}{\sfrac{4}{3}\pi|\tau(r)|}\right)^{1/3}.
\label{eq:rtrmin}
\end{align}
}

\cMBII{Using Eq.~\ref{eq:tdentr} we can predict how the mean mass density of a satellite will evolve as it falls into a cluster. For example, if a satellite has an extended mass distribution with a large initial truncation radius $r_{\rm tr, init}$ and a low initial mean mass density ($\bar{\rho}_{\rm tr,init}$), we can predict that, as soon as the satellite approaches the cluster centre, the tidal field becomes stronger and the outer mass layers of the satellite becomes stripped. This shrinks the truncation radius ($r_{\rm tr}<r_{\rm tr,init}$) and the mean mass density within the new truncation radius increases. This shows that for satellite tidal stripping to occur, neither mass nor size individually matters, but the resulting mean mass density.}

\cMBII{Furthermore, as a satellite orbits its host, it can experience different strengths of the tidal field for non-circular orbits, where typically the host tidal field anti-correlates with distance ($|\tau|\propto1/r$).
In particular, a satellite that approaches its host for the first time with a radial orbit experiences an increasingly stronger tidal field strength as it approaches the host, reaching its maximum at the orbital pericentre ($r_{\rm peri}$):
\begin{align}
|\tau|_{\rm max}&=|\tau(r_{\rm peri})| \label{eq:tmax}.
\end{align}
This implies that the maximum tidal field experienced by a satellite is inversely proportional to its pericentre distance ($|\tau|_{\rm max}\propto 1/r_{\rm peri}$).
Therefore, according to Eq.~\ref{eq:denj}, the maximum mean mass density due to tidal truncation is reached at the pericentre, where the tidal field is maximised, \ie
\begin{align}
\bar{\rho}_{\rm tr}^{\rm max} &= \bar{\rho}_{\rm J}^{\rm max} = |\tau|_{\rm max},
\label{eq:dentrmax}
\end{align}
which is expected for most mass density profiles that we can adopt for the hosting cluster or group \cYOBII{\citep[\eg NFW,][]{Burkert2000}}.~Consequently, the smallest mass and truncation radius are reached at the pericentre, $m_{\rm tr}^{\rm min}=m(r'\!=\!r^{\rm min}_{\rm tr})$ and $r_{\rm tr}^{\rm min}$, and the maximum density anti-correlates with the pericentres, $\bar{\rho}_{\rm tr}^{\rm max}\propto 1/r_{\rm peri}$, allowing us to obtain the anti-correlated mass-density-distance relation ($\bar{\rho}-r$).}\\

\cMBII{\cYOBII{However, since measuring the truncation radius to determine the total mean mass densities can be impractical or difficult in observations or simulations, one alternative is to use} the half-mass radius of the remaining bound mass of the satellite within its truncation radius. This is defined as
\begin{align}
m(h:=r')\!=\!\sfrac{1}{2}\,m_{\rm tr},
\label{eq:rh}
\end{align}
which depends not only on the shape of the satellite's mass profile, but also on the truncation radius $h\!=\!h(m,r_{\rm tr})$. 
This allows us to define the mean mass densities within one and twice the half-mass radius, as follows:
\begin{align}
\bar{\rho}_{h}&=\frac{\sfrac{1}{2}\,m_{\rm tr}}{\sfrac{4}{3}\pi h^3},
\label{eq:denh}\\
\bar{\rho}_{2h}&=\frac{m_{2h}}{\sfrac{4}{3}\pi (2h)^3},
\label{eq:den2h}
\end{align}
\cMBII{where $m_{2h}\!:=\!m(r'\!\!=\!\!{\rm 2h})$.} Moreover, it can be shown that for standard mass profiles (\eg NFW, Plummer, Exponential), the mean mass densities are higher in the internal regions of a satellite than in its outskirts, \ie $\bar{\rho}_{\rm tr}\leq\bar{\rho}_{2h}\leq\bar{\rho}_{h}$.
This implies that Eq.~\ref{eq:tdentr} also imposes
\begin{align}
& |\tau| \leq \bar{\rho}_{\rm tr} < \bar{\rho}_{2h}<\bar{\rho}_{h}. \label{eq:tdenh}
\end{align}
Furthermore, when the satellite reaches its orbital pericentre, the half-mass radius and mean mass densities can also reach their respective minimum ($r^{\rm min}_{h}$) and maximum values ($\bar{\rho}^{\rm max}_{h}$ and $\bar{\rho}^{\rm max}_{2h}$).}

\cMBII{In the following, we explore whether we can find similar relations for the stellar mass component of satellite galaxies. Eqs.~\ref{eq:tdenj} and \ref{eq:tdentr} relate the host's tidal field to the mean mass density of a satellite within its truncation radius ($\bar{\rho}_{\rm tr}$). We can reformulate $\bar{\rho}_{\rm tr}$ as the sum of the mean mass densities of dark matter ($\chi$) and stars ($\star$), \ie $\bar{\rho}_{\rm tr} = \bar{\rho}^{\star}_{\rm tr}\! +\! \bar{\rho}^{\rm \chi}_{\rm tr}$. For simplicity, we consider a gas-free satellite galaxy, which allow us to re-write Eq.~\ref{eq:tdentr} as
\begin{align}
&\bar{\rho}^{\star}_{\rm tr} + \bar{\rho}^{\rm \chi}_{\rm tr} \geq |\tau|.
\label{eq:tdentrdm*}
\end{align}
}
\cYOBII{Now, in an hypothetical case, we can consider a pristine satellite galaxy that begins its first infall. Using Eq.~\ref{eq:tdentrdm*} we can predict that as the satellite approaches the host halo and $|\tau|$ increases, the satellite begins to lose its outer layers of material (initially mostly dark matter) and, as a consequence, it shrinks its initial truncation radius $r_{\rm tr,init}$ according to Eq.~\ref{eq:rtrmin}, resulting in a mean mass density larger than before its infall (\ie $\bar{\rho}^{\chi}_{\rm tr}>\bar{\rho}^{\chi}_{\rm tr,init}$). Later, as the satellite sinks deeper into the host's tidal field, the outer layers of the stellar material are increasingly more strongly stripped as well, and both mass components reach the same truncation radius $r_{\rm tr}^{\star}\!=\!r_{\rm tr}$. 
As a result, given the typical mass density profiles of satellite galaxies (\eg exponential or Hernquist profiles), a stripped satellite is expected to have a mean stellar mass density higher than its pre-infall value (\ie $\bar{\rho}^{\star}_{\rm tr}>\bar{\rho}^{\star}_{\rm tr,init}$), which is due to the shrinking of the truncation radius ($r_{\rm tr}<r_{\rm tr,init}$).}

\cMBII{\cYOBII{In addition, given that the direct} measurement of the stellar mass truncation radius ($r_{\rm tr}^{\star}$) is also difficult to determine in observations of galaxies with faint substructures, we use instead the stellar half-mass radius ($h\star$), and/or multiples of this quantity, as this depends on the mass profile, but also on $r_{\rm tr}^{\star}$, as follows:}
\begin{align}
m^{\star}(h\star\!:=\!r')\!=\!\sfrac{1}{2}\,m^{\star}_{\rm tr}.
\label{eq:rh*}
\end{align}

Similar to the case of the mean dynamical mass densities, $h_{\star}$ allows us to define the mean stellar mass densities within one and two $h_{\star}$ as 
\begin{align}
\bar{\rho}^{\star}_{h\star}&=\frac{\sfrac{1}{2}\,m_{\rm tr}^{\star}}{\sfrac{4}{3}\pi h_{\star}^3},
\label{eq:denh*}\\
\bar{\rho}^{\star}_{2h\star}&=\frac{m_{2h{\star}}^{\star}}{\sfrac{4}{3}\pi (2h_{\star})^3},
\label{eq:den2h*}
\end{align}
where $m_{2h{\star}}^{\star}\!:=\!m^{\star}(r'\!=\!2h{\star})$, with a dependency on $r_{\rm tr}^{\star}$ through $m^{\star}_{\rm tr}$. 
Moreover, following the arguments of Eq.~\ref{eq:tdenh} between inner and outer densities for typical stellar mass densities, we can also state that for each satellite galaxy:
\begin{align}
\bar{\rho}^{\star}_{\rm tr} <\bar{\rho}^{\star}_{2h\star} < \bar{\rho}^{\star}_{h\star}.
\label{eq:denh*}
\end{align}

\cYOBII{Satellite's stellar masses can be estimated more easily than their dynamical masses, as the former requires adopting stellar-mass-to-light ratios from stellar population models, while the latter requires resolved kinematic observations. Stellar masses can be a useful direct tracer of the radial variation of the mean mass densities of the satellite population.}

\cYOBII{Modelling the tidal shape of the mass densities of a satellite population, with the equations described above, implies an assumption that satellites behave as rigid profiles and mass stripping} removes the outer shells, layer by layer, shrinking $r_{\rm tr}$ and, therefore, also $h$ and $h\star$ to their minimum values when the satellite passes through the pericentre, $r_{\rm peri}$. 
\cYOBII{More complex models in the} literature also find that the half-mass radius \cYOBII{reduces} during tidal stripping. This is expected, as the tidal field and stripping can also induce an internal restructuring of the bound material in a satellite galaxy, as its density reacts to its own potential ($\phi_{\rm sat}$) and the \cMBII{host's} potential ($\phi_{host}$) as $\phi_{\rm sat,tot}=\phi_{\rm sat}+\phi_{host}$. 
Studies of the tidal tracks of \citet[][see their Fig.~2 and 13]{Errani2022} and \citet{Stucker2023} reveal that for typical satellite mass distributions (\eg NFW, Plummer, Exponential), the density profiles are reshaped by the tidal field, preferentially \cYOBII{reducing} the size of the half-mass radii of the total and stellar mass distributions during the stripping process. Moreover, similar results were obtained for MW star clusters, as expected, because they experience the same environmental tidal forces \citep{Webb2013}.
\cMBII{The studies mentioned above suggest that the definitions of $h$ and $h_{\star}$ in Eqs.~\ref{eq:rh} and \ref{eq:rh*} capture the overall shrinking of these half-mass radii produced by \cYOBII{tidal} stripping. Therefore, we use the previous equations to later build in Sect.~\ref{sec:met:toy} a simple model to understand the effects of tidal stripping on a satellite population in galaxy clusters and groups.}
Nevertheless, more extreme tidal mechanisms can generate more complex dynamical configurations, such as tidal shocks \citep[][]{Penarrubia2008,Hammer2023,Hammer2024}, or stages before and after complete tidal disruption that can produce ultradiffuse galaxies \citep{Ogiya2018a} and streams \citep{Smith2013,Blana2015}.

\subsection{\cMB{Tracing the distribution of mass densities and additional properties of a satellite galaxy population}}
\label{sec:theo:distrib}
\cMBII{Most of the studies in the literature that explore the effects of the host tidal field on the mass densities of its satellite galaxies use their pericentre distances 
\citep[$r_{\rm peri}$;][]{Kaplinghat2019,Hayashi2020,Pace2022,Cardona-Barrero2023}, following Eqs.~\ref{eq:tmax} and \ref{eq:dentrmax}.}
However, while the \textit{Gaia} observatory has allowed us to estimate orbital pericentres of satellite galaxies within the MW, and in its outskirts with larger uncertainties, this becomes impossible with the current technology for extragalactic systems beyond the Local Group.
Therefore, the use of observed cluster-centric distances of satellite galaxies ($r$) is the only alternative. 

First infalling satellites have not yet reached their pericentres and, therefore, they are expected to exhibit mass densities that correlate with the local tidal field at their current positions ($r$) according to Eq.~\ref{eq:tdentr}.
\cMBII{Moreover, satellite orbital distributions in equilibrium can show a correlation between radial satellite distances ($r$) and their pericentres ($r\propto r_{\rm peri}$), which can have a spread with a size that depends on the shape of the distribution.
Furthermore, since we are interested in studying the effects of the tidal field of a galaxy cluster or group, we require selecting systems that have a dominant central potential where $|\tau|\propto 1/r$. In addition, we require that the host potential evolves slowly compared to the satellite orbital evolution and, therefore, preferentially avoid clusters that are currently undergoing subcluster mergers where complex tidal forces vary on dynamical timescales similar to those of the satellite orbits.
Studies find that most low-redshift clusters and groups have their central regions already formed and in equilibrium \citep{Merritt1985}, with the accretion of low-mass satellites and only a few cases of ongoing galaxy cluster major mergers having little effect on the large-scale halo potential \citep{Tempel2017,Lokas2023}.
Therefore, in the most common cases, it is possible \cYOBII{to derive a relation between the distributions of the satellites' cluster-centric distances $r$ and their pericentres, $r_{\rm peri}$, showing that, on average, there is a direct correlation between these quantities, such that  $r\propto r_{\rm peri}$.}}

\cMBII{Moreover, given that we are interested in inspecting the collective effect of the host tidal field on its entire satellite population, we use the moving average of the satellite mass densities to characterise their radial transformation.
With this we can probe the (anti)correlation between the distribution of satellite mass densities and cluster-centric distances $\langle \bar{\rho} \rangle \propto \langle 1/r \rangle$.
For this, we define the moving average function as follows:}
\begin{align}
\langle Y\rangle(r_i) &=  K\Big(r_i,Y_i, q\Big) = 10^{\frac{1}{q}\sum_{j=-\,q/2}^{j=+q/2}\log Y_j},
\label{eq:kern}
\end{align}
\cMBII{which depends on the distance to the host's centre, $r$, the number of neighbours, $q$, and the variable, $Y_{i}$. This function is used to calculate a logarithmic moving averaged filter of the same variable $\langle Y_{i}\rangle$.
We applied this to different satellite variables, $Y$: dynamical mass densities ($\bar{\rho}_{\rm tr}$, $\bar{\rho}_{2h}$), stellar mass densities ($\bar{\rho}^{\star}_{2h\star}$), stellar mass ($M^{\star}_{2h\star}$), pericentres ($r_{\rm peri}$), among others.
Here, we adopted a logarithmic weighted average filter because:
i) the typical mass densities in the satellite populations considered in this study are found to show a log-normal behaviour (see Sect.~\ref{sec:res:sim}) and span two orders of magnitude, and ii) the presence of mass density outliers that could introduce strong variations if the kernel were linear. Therefore, we aim at measuring the bulk of the distribution to determine the distance behaviour of the mass density of the whole satellite population. 
In addition, we also tested a median filter that yields similar results, leaving our main conclusions unchanged.}

\section{Methodology}
\label{sec:met}
\cMB{To explore and probe the relation of the mass densities of satellite galaxies and their distances to the host system and the tidal field we used three approaches: (1) a semi-analytical satellite orbital toy model, (2) cosmological galaxy simulations, and (3) observational data of four systems. 
Each approach allowed us to investigate different aspects. 
For example, the toy model allowed us to obtain a simple interpretation of the equations in Sect.~\ref{sec:theo} on the effects of tidal truncation in an environment with controlled variables without including other physical processes that are involved in the evolution of satellites, such as ram-pressure stripping, starvation, etc. 
Moreover, access to the full orbital history of a satellite allowed us to calculate peri-, apo-centres, and timescales. We could also repeat the calculations while considering different setups and potentials, capabilities that are not possible in cosmological galaxy simulations.
However, in this simplified toy model, the satellites were modelled as rigid systems, where the tidal forces strip only outer layers of its dark and baryonic material in an 'onion'-like interpretation of Eqs.~\ref{eq:tdentr} and \ref{eq:dentrmax}, leaving their central mass distributions unperturbed.
Therefore, we also used cosmological galaxy simulations to probe the mass-density distance relation. These simulations included multiple physical processes, such as star formation, dynamical friction, and other processes mentioned above.
Lastly, we included observed satellite systems to compare with the models. However, we cannot here access all the information, such as their full dark-matter distributions, 3D spatial position or velocity, or the full orbital histories of a limited sample.
We explain the setup of these approaches in the following sections.
}

\subsection{A simplified satellite tidal stripping toy model} 
\label{sec:met:toy}
\cMB{This model is an orbital construction of satellite galaxies in a galaxy cluster. 
\cMBII{We chose a system with the mass and size of the Fornax cluster, as this system is larger in number of satellites and in total mass than the Local Group, with a well-studied and resolved satellite population.}
The satellites were modelled as test particles that have analytical mass density profiles for the stellar and dark matter components evolving under the influence of the tidal field of the cluster according to the equations of Sect.~\ref{sec:theo:tdenrel}.
For simplicity, we considered gas-less satellites only.
For the toy model, we began by defining the gravitational force of the host, where we set up a Fornax-type cluster with an NFW profile with a virial mass and radius of $M_{\rm vir}\!=\!7\times10^{13}\sm$ and $R_{\rm vir}\!=\!1\Mpc$ \citep{Drinkwater2001,Schuberth2010}, with a concentration of $c=8.5$, and where we used Eq.~\ref{eq:tfield} to calculate its tidal field profile $|\tau(r)|$.}

\cMBII{To have an observationally motivated satellite population we used the Schechter Luminosity Function to draw their luminosities for a total of $2\times10^3$ satellites, ranging between $M_r\!=\!-22$ and $-9\mags$.
For simplicity, we adopted a constant stellar mass-to-light ratio of $\Upsilon_{r}=2\sm\slu^{-1}$ to obtain the stellar masses, and tested variations between 1 and $4\sm\slu^{-1}$.}
To model the stellar mass density profile $\rho^{\star}(r')$, mean stellar mass density $\bar{\rho}^{\star}(r')$ and stellar mass $m^{\star}(r')$ profile of each satellite and the impact on the results, we tested three different profiles as a function of distance to the satellite centre $(r')$: exponential, Plummer, and Hernquist, with initial scale lengths taken to be 10\% of the initial satellite size (initial truncation radius $r_{\rm tr,\,init}$).
We used the galaxy-halo abundance matching models of \citet{Moster2013} to estimate the maximum dark matter mass that each satellite galaxy contains prior to its infall.
Guided by cosmological galaxy simulations and observations of satellite beyond $\Rvir$ from their hosts, we sampled the initial total mean mass densities from a log-normal distributions centred at $\bar{\rho}_{\rm tr, init}\!=\!2\!\times\!10^{5}\sm\kpc^{-3}$ with a half width of 0.1dex, and tested different initial conditions (IC) as well.
To model the dark-matter profiles of each satellite, we adopted the NFW model, with concentrations from the c-M halo relation \citep{Correa2015c}.
\cMBII{The positions and velocities of the satellites were sampled from the NFW equilibrium distribution of the host cluster (test particles) using the Jeans equations.}
For simplicity, we used an isotropic spherical distribution for the satellites and the cluster potential, and
we neglected the increase in the mass of the cluster by satellite accretion
since the total satellite mass corresponds to $\sim\!3\%$ of the cluster mass.

Lastly, we performed the orbital calculation with an updated version of the \textsc{delorean} code \citep{Blana2020}, integrating the orbits for 10\Gyr to phase-mix the initial distribution, to ensure that all satellites passed through their pericentres.
From this we calculated the tidal field and the tidal mass stripping for each satellite from Eq.~\ref{eq:tdentr} in Sect.~\ref{sec:theo} and determined their half-mass radii and mean mass densities for their total and stellar mass components \citep[see also][]{Ogiya2022}.
From this we obtained the mean dynamical\footnote{Following the dynamics tradition, dynamical is adopted as the total mass measured within a radius.} mass density of each satellite within its truncation radius $r_{\rm tr}$ ($\bar{\rho}_{\rm tr}$), 
and within $2h$ (\textbf{$\bar{\rho}_{2h}$}),
as well as the mean stellar mass density of each satellite within $r_{\rm tr}$ ($\bar{\rho}_{tr}^{\star}$) and
within $2h\star$ ($\bar{\rho}_{2h\star}^{\star}$).
Moreover, we considered two cases: when 100\% of the satellites pass through their orbital pericentres ($r_{\rm per}$) obtaining their maximum densities according to Eq.~\ref{eq:dentrmax}, and when 30\% of the satellites are on their first infall, and, therefore, their truncated mass densities do not yet reached the maximum values.
Then, we explored the resulting distribution of the satellite mass densities with distance using Eq.~\ref{eq:kern}, while comparing with the host's tidal field. 
We also used the moving average kernel of Eq.~\ref{eq:kern} applied to the pericentre distances to compare the pericentre distribution of the satellite population as a function of the cluster-centric distance (\ie $\langle r_{\rm peri} \rangle \propto r$).
Using the toy model, we show in Sect.~\ref{sec:res:toy} how the mean mass densities of the satellite population correlate with the cluster-centric distances and the host tidal field.

\subsection{Simulated systems} 
\label{sec:met:sims}

\cMBII{We used data from cosmological galaxy simulations to probe the host's tidal effects on satellites, which have the advantage over our simple model of having self-gravitating satellite systems that react to the environmental tides self-consistently.
For this, we used clusters from \textsc{Illustris} TNG50 \citep{Nelson2019b, Pillepich2019}, and the TNG100 and TNG300 simulations \citep{Marinacci2018, Naiman2018, Nelson2018, Pillepich2018, Springel2018}.
Physical and numerical parameters are reported in several publications and on the official webpage\footnote{https://www.tng-project.org/}, but we briefly detail the main properties in the following.
The periodic box comoving sizes of TNG50, TNG100 and TNG300 are $51.7\cMpc$, $106.5\cMpc$ and $302.6\cMpc$, respectively.
TNG50 has stellar and dark matter particle masses of $8\times10^4\sm$ and $4.5\times10^5\sm$ with a spatial particle resolution of $290\pc$ and a moving adaptive mesh for the gas, with star-forming cells with a median size of 138\pc, and a recorded minimum value of 8\pc. 
TNG100 has stellar and dark matter particle masses of $1.4\times10^6\sm$ and $7.5\times10^6\sm$, with a resolution of $740\pc$ and a moving adaptive mesh for the gas, with star-forming cells with a median size of 355\pc, and a recorded minimum value of 14\pc.
TNG300 has stellar and dark matter particle masses of $11\times10^6\sm$ and $59\times10^6\sm$, with a resolution of $1.48\kpc$ and a moving adaptive mesh for the gas, with star-forming cells with a median size of 715\pc, and a recorded minimum value of 47\pc.}

\cMBII{We analysed different redshifts, but in this work, we present the results for redshifts $\redshift\!=\!0$ and 1.
After requiring subhaloes to be composed of $>100$ stellar particles each \citep[][]{VandenBosch2018,VandenBosch2018b}, and each cluster to have at least 60 resolved subhaloes, 
we obtained from TNG50, TNG100, and TNG300 a total of $235$ galaxy clusters (and groups) at $\redshift=0$ and $60$ at $\redshift=1$, which we used for the subsequent analysis. We tested the effects of resolution and sampling by increasing the limit by one order of magnitude or the effects of massive subhalo outliers from the main distribution by setting a maximum stellar mass of $10^{10}\sm$. 
We found similar results, leaving our main conclusions unchanged.
From these simulations, we selected two clusters for a more detailed analysis as our fiducial models, taking a Fornax-type cluster analogue from the TNG50 simulation with a virial mass and radius of $7.6\times10^{13}\sm$ and $1115\kpc$, respectively, with 492 satellite galaxies, 
and a Virgo-type cluster from TNG100 with a virial mass and radius of $4.9\times10^{14}\sm$ and $2079\kpc$, respectively.
To explore the impact of different subgrid models in the simulations, we also made a comparison with a cluster from the \textsc{eagle} simulations \citep{Schaye2015}.
The selected cluster and satellite galaxies correspond to a Virgo-type cluster with a virial mass and radius of $3\times10^{14}\sm$ and $1422\kpc$, respectively, containing 271 resolved satellites.
The cluster was taken from the L100N1504 run, which consists of a periodic box of $100\cMpc$ in comoving size, initially containing $1504^3$ gas particles with an initial mass of $1.81\times10^6\sm$, and the same number of dark matter particles with a mass of $9.70\times 10^6\sm$.
The halo catalogues provided in the public database \citep{McAlpine2016} were built using a friend-of-friends (FoF) algorithm, which identifies dark-matter overdensities following \citet{Davis1985}, considering a linking length of 0.2 times the average inter-particle spacing. Baryonic particles were assigned to the FoF halo of their closest dark-matter particle. Subhalo catalogues were built using the \textsc{subfind} algorithm \citep[]{Springel2001,Dolag2009}, which identifies local overdensities using a binding energy criterion for particles within a FoF halo.}

The tidal field profiles of the host clusters ($|\tau|$) were calculated from their total mass profiles (gas, stars, and dark matter).
We verify that these clusters are the most massive structures within four times their virial radii. 
Analysis of more and less massive clusters shows the same behaviour as in our fiducial case.

We computed the virial radius $\Rvir$ of a cluster from the virial mass $M_{\rm vir}$ defined in \citet{MoBoWh2010} as an over-dense region with the physical virial radius defined as
\begin{align}
\Rvir &=\left(\frac{M_{\rm vir}(\redshift)}{{\small 4/3}\,\pi\rho_{\rm crit}(\redshift)\,\Omega_{\rm m}(\redshift)\,\Delta_{\rm vir}(\redshift)}\right)^{1/3},
\label{eq:rvir}
\end{align}
where the cosmological parameters $\rho_{\rm crit}$, $\Omega_{\rm m}$, $\Delta_{\rm vir}$ are the critical density, the matter to critical density ratio, and the spherical collapse over-density criterion $\Delta_{\rm vir}\approx\left(8\pi^2+82\epsilon-39\epsilon^2\right)\left(\epsilon+1\right)^{-1}$ with $\epsilon=\Omega_{\rm m}-1$.

\begin{figure*}[h]
\centering
\includegraphics[width=8.9cm]{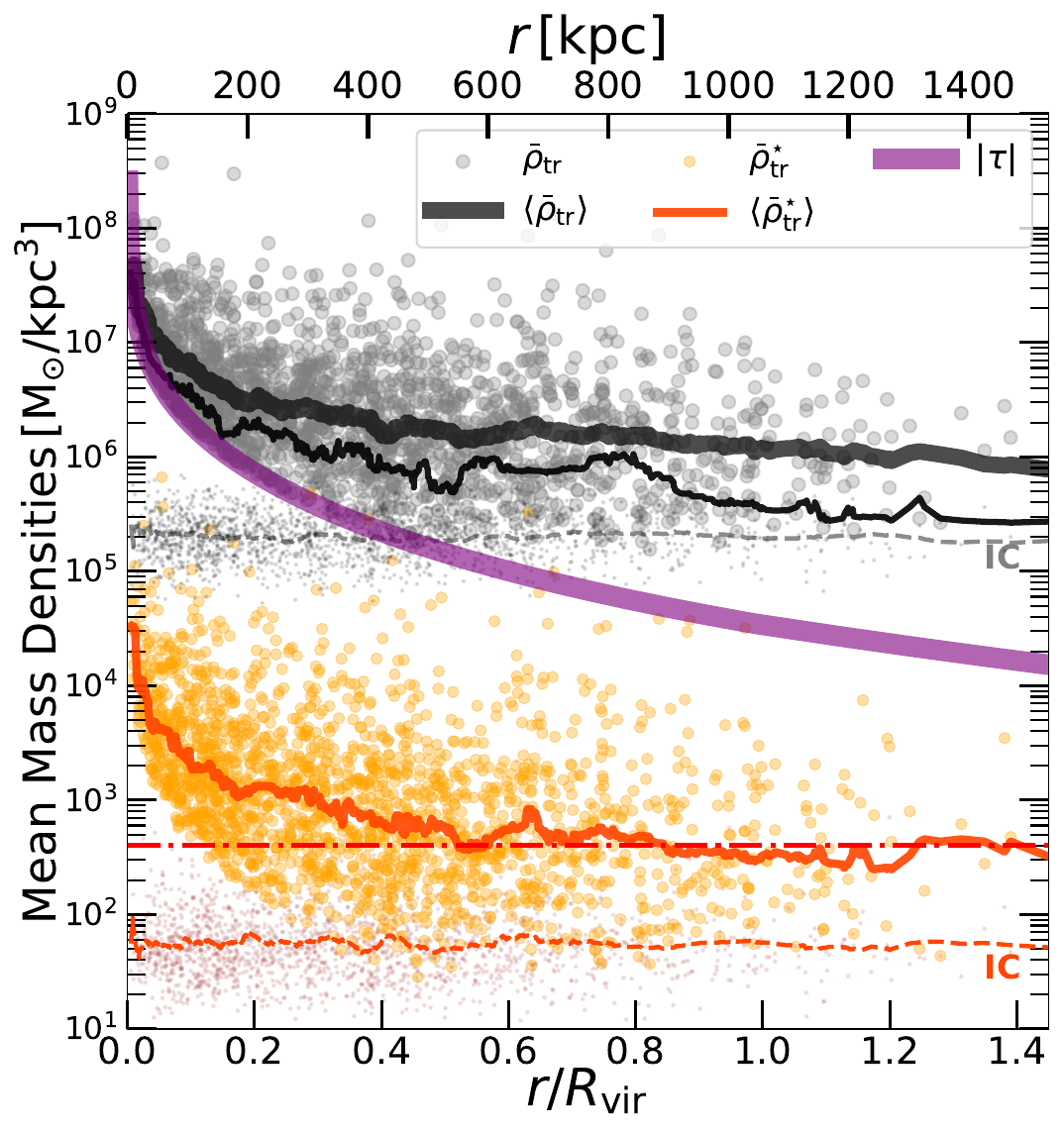}
\hspace{0.2cm}
\includegraphics[width=8.9cm]{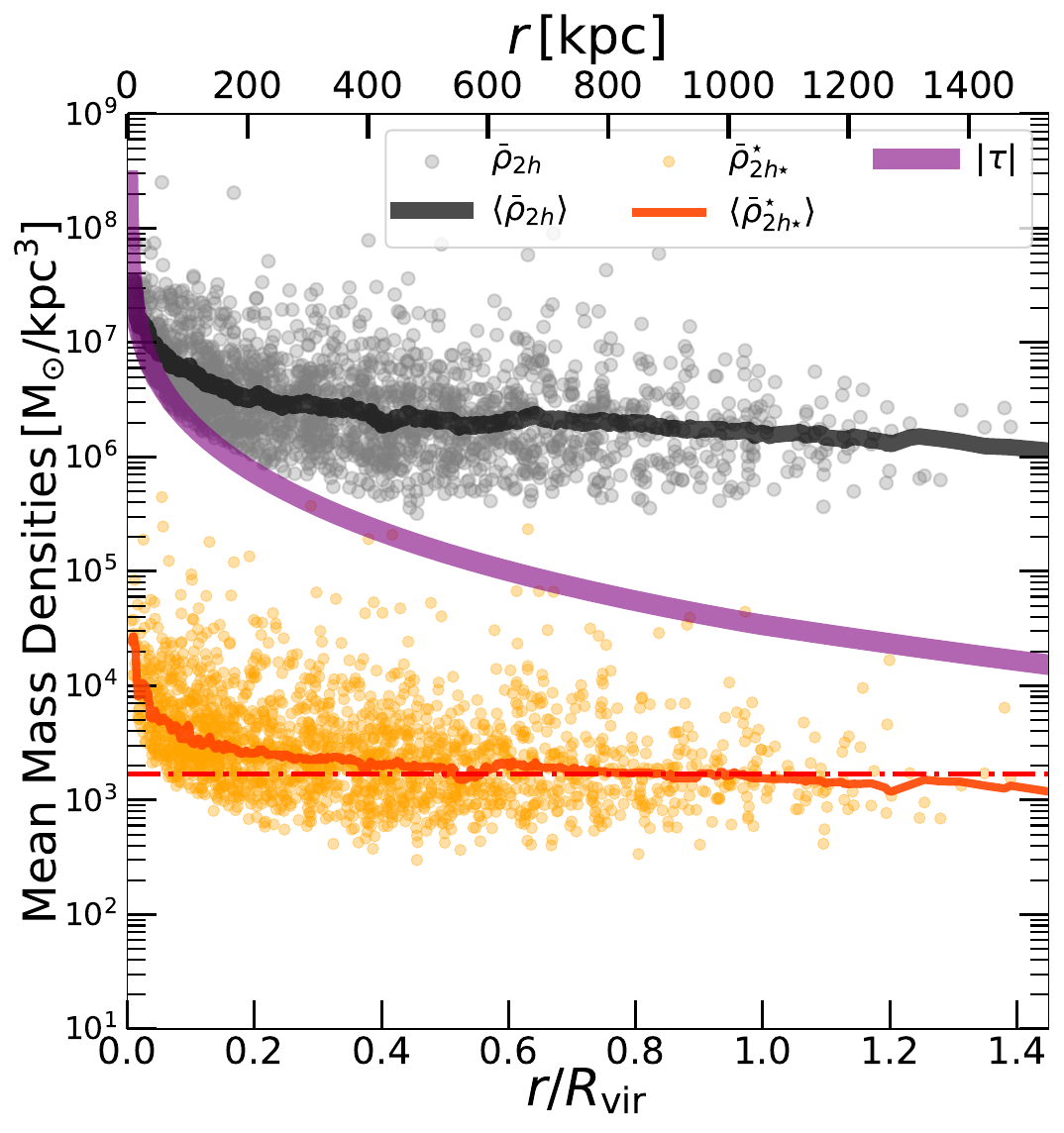}
\vspace{-0.3cm}
\caption{
Results of the toy model showing the imprint of the host tidal field on the mean mass density distribution (circles) of $n_{\rm sat}\!=\!2\times10^3$ satellite galaxies shown as a function their distances to the cluster centre. 
The host has an NFW halo mass and size of a Fornax-type cluster, showing its tidal field ($|\tau|$) in both panels (purple curves).
The satellites are taken from a snapshot after a 10\Gyr orbital integration with \textsc{delorean} \citep{Blana2020}.
{\it Left panel}: 
We show the final total (\ie dynamical) mass mean densities $\bar{\rho}_{\rm tr}$ (large grey circles) and mean stellar mass densities $\bar{\rho}^{\star}_{\rm tr}$ (yellow circles) after 10\Gyr.
The moving average profiles of these densities are calculated with Eq.~\ref{eq:kern} using $q=2\sqrt{n_{\rm sat}}\simeq89$ neighbours and are shown with solid thick black and orange curves. 
The IC of the total and stellar mass densities of the satellites are shown with small points in grey and brown, respectively, showing also their moving average profiles (dashed curves).
In addition, we show the effect of including a population of 30\% of the first in-falling satellites and the resulting moving average (thin black curve).
{\it Right panel}: For the same satellites, we show the total and stellar mass mean densities measured in the central regions of each satellite within twice the half-mass radius of the total mass ($2h$) and the stars ($2h\star$), as well as their moving averages (solid curves). \cMBII{The increase in the moving average profiles towards the cluster centre shows that the shrinking of the truncation radius of each satellite ($r_{\rm tr}$) produces an increasing central mean mass density as a result of the shrinking half-mass radii, $h$ and $h\star$. Moreover, the effect is more noticeable in the inner cluster regions ($r<0.5\Rvir$) where the tidal field is stronger and the satellites have smaller pericentres, making them more susceptible to stronger tidal stripping.} The horizontal dot-dashed lines included in both panels are added for reference to illustrate the deviation of the mean stellar mass densities from constant values.
}
\label{fig:fig_toymod}
\end{figure*}

\subsection{Observed systems} 
\label{sec:met:obs}
We included four observed systems to explore the empirical effect of tidal fields on the mean mass densities of satellites: the galaxy clusters Fornax and Virgo, and the Local Group satellites of the MW and Andromeda (M31) galaxies. 
We selected these systems for tracing as large a host mass range as possible ($10^{12}$-$10^{14.6}\sm$), to ensure deep photometric observations \cMBII{with a large spatial coverage that conveys homogenous and high sample completeness}, and because they are publicly available. We modelled the tidal field of observed clusters using mass profiles with parameters based on the literature. In the following, we briefly describe the main characteristics of each dataset:
\begin{itemize}
\def\labelitemi{\textbullet}
\item\textit{Fornax cluster:} For the Fornax cluster mass profile we adopted parameters determined
by \citet{Schuberth2010} and derived from dynamical models fitted to star cluster kinematics.
We selected their NFW model $a5$ as our fiducial model,
which has a virial radius of $\Rvir\!\simeq\!R_{200}\!=\!1075\kpc$ and mass of $\Mvir\!=\!7.3\times10^{13}\sm$ that agree well with previous estimates \citep{Drinkwater2001}.
Our fiducial model also includes the stellar mass component of NGC\,1399 modelled according to \citet{Schuberth2010}.
For the analysis of the Fornax cluster satellites, we used the {\it Next Generation Fornax Survey} (NGFS) observations obtained with the wide-field Dark Energy Camera mounted on the 4-m Blanco telescope at the CTIO in Chile \citep{Munoz2015, Eigenthaler2018, Ordenes-Briceno2018a,Ordenes-Briceno2018b}, 
having the advantage of being among the widest-deepest surveys on a single galaxy cluster, covering $\sim\!\!50\deg^2$ with a wide photometric filter range between NUV and NIR.
We used the combined catalogues of \citet{Eigenthaler2018} and \citet{Ordenes-Briceno2018a} with 630 satellite galaxies, \cMBII{which contain photometric information and stellar masses based on colour information.}
Taking the NGC\,1399 galaxy as the centre of the cluster, which is at $20.4\Mpc$ \citep{Blakeslee2009}, the sample extends to $R\!=\!1017\kpc\,(2.8\degree)$ in projected radius, which includes faint satellites down to $M_{i'}\!=\!-8.8\mags$ (${\rm m}_{i'}\!=\!22.73\mags$), with objects as extended as $R_{\rm e}\!=\!2.7\kpc\,(27.9\as)$ in the effective radius of S\'ersic.
To probe the relation of Eq.~\ref{eq:tdentrdm*}, we calculated the stellar mean densities of the Fornax satellites within twice their deprojected effective radii $r_{\rm e}$ \citep[adopting the conversion $r_{\rm e}\!=\!4/3 R_{\rm e}$; see][]{Wolf2010}, obtaining the mean stellar mass densities \cMB{$(\bar{\rho}^{\star}_{2h\star})$ for each satellite, where, for simplicity, we approximated $h\star\!\approx\!r_{e}$.}
We tested luminosity-weighted densities, adopting a constant stellar mass-to-light ratio $\Upsilon_{i'}$ in the $i'$-band, finding similar results.\\

\item \textit{Virgo cluster:} To probe the tidal relation in a more massive environment, we selected the Virgo cluster.
This cluster has two main substructures, the more massive Virgo A subgroup with a virial mass and radius of $M_{200}^{\rm VA}\!\!=\!\!4.2\times10^{14}\sm$ and $\Rvir^{\rm VA}\!=\!1550\kpc$ respectively, 
and the Virgo B subgroup with a smaller virial mass and radius of $M_{200}^{\rm VB}\!\!=\!\!10^{14}\sm$ and $\Rvir^{\rm VB}\!=\!960\kpc$ \citep{McLaughlin1999, Ferrarese2012}.
For the analysis of the Virgo cluster satellites, we used the publicly available extended catalogue of \citet{Kim2015} with $\sim$1500 satellites that extend out to $R\simeq3.7\Rvir$ $(5730\kpc)$ in projected radius, containing galaxies brighter than ${\rm m}_{r}\!\simeq\!18.76$\,mag, or in absolute magnitude of approximately $-12.32$ mag, adopting a distance of $16.5\Mpc$ \citep{Mei2007}.
We focused on the Virgo A subgroup and took the central galaxy M87 as the origin, while excluding the subgroup B centred on M49 by masking out satellites within $R=\Rvir^{\rm VB}/2\simeq480\kpc$ in projection. We note that by including all satellites only mildly changes the Virgo A satellite distribution, as Virgo B is separated by $4.4\deg\,(1266\kpc)$ in projection.
We determined the mean stellar mass densities $(\bar{\rho}^{\star}_{2h\star})$ for each satellite using its de-projected stellar half-mass radius \cMB{$h\star$}, using the $r$-band half-light radii and a constant $\Upsilon_r$. We point out that changing $\Upsilon_r$ would only shift vertically the mass-density-distance relation ($\bar{\rho}-r$) in the density axis, while differences of $\Upsilon_r$ within the satellite population introduce insignificant variations, three orders of magnitude smaller than the spread of the distribution in stellar densities, as shown in Sect.~\ref{sec:res:obs}.\\

\item\textit{Local Group systems:} 
To explore the tidal field relation in lower-mass environments, we probed the M31 and MW galaxy systems. 
The mass models of these hosts include their dark haloes, bulges, stellar and gaseous disc components with parameters adapted from \citet{Tamm2012} and \citet{Blana2017,Blana2018} for M31 ($M_{200}\!=\!1.04\times10^{12}\sm$, $R_{200}\!=\!207\kpc$), and from \citet{Bland-Hawthorn2016,Shen2022} for the MW ($M_{200}=1.08\times10^{12}\sm$, $R_{200}=216\kpc$).
We used the catalogue of \citet{Putman2021} with 55 satellites for 
the MW and 41 for M31 and calculated their mean stellar mass densities ($\bar{\rho}^{\star}_{2 h\star}$) within twice their half-mass radii 2$h\star$ , and their central mean dynamical mass densities ($\bar{\rho}_{h\star}$) within $h\star$.

\end{itemize}

\section{Results} 
\label{sec:res}

\subsection{Satellite mass densities: Toy model predictions} 
\label{sec:res:toy}

We start with the results of the toy model
in Fig.~\ref{fig:fig_toymod}, presenting a snapshot after the 10\Gyr integration. 
There we show the total and stellar mass mean densities of each satellite measured at their truncation radii ($r_{\rm tr}$) and twice their half-mass radii ($2h$, $2h\star$) and plotted as functions of their distances to the centre of the cluster. We include the moving average of the mean mass densities of the satellite distribution applying the function in Eq.~\ref{eq:kern}.
In addition, we show the tidal field of the cluster, which has the largest values in the cluster centre.

\cMBII{As Fig.~\ref{fig:fig_toymod} reveals, we find that the mean mass densities of the satellite population have systematically higher values towards the centre of the cluster, resulting in a spatial mean mass density segregation.}
In contrast, we also show the initial satellite mass densities, before being truncated by the tidal field, which show the \cMBII{moving averages} centred around the IC.
This distribution is produced in this model by the synergy between two effects: (i) the profile of the tidal field of the cluster ($|\tau|$) that imprints its shape according to Eqs.~\ref{eq:tdenj} and \ref{eq:dentrmax}, and (ii) the distribution of the pericentre distances that depend on the satellite orbital distribution. 
\cMBII{We analysed the distribution of the satellite pericentres applying the moving average function of Eq.~\ref{eq:kern}, finding that these correlate with the distance $\langle r_{\rm peri}\rangle \propto \langle r\rangle$ with a scatter that depends on the orbital distribution.
A small fraction of satellites ($1.3\%$) with small pericentres ($r\!<\!100\kpc$) are at large distances ($r\!>\!1\Mpc$) in the snapshot taken to produce Fig.~\ref{fig:fig_toymod}.}
These satellites have excentric radial orbits, which explains the presence of some satellites with high mean mass densities in the outskirts of the cluster, since they previously entered the cluster central regions where the tidal forces are stronger.
These could correspond to the backsplash satellite population detected in cosmological galaxy simulations and observations \citep{Gill2005,TeyssierM2012,Garrison-Kimmel2014a,Blana2020,Haggar2020,Diemer2021}.  In general, this small satellite sub-population has insignificant effects on the conclusions.

In addition, we explore how the mass density distribution is affected when we include 30\% of first-infall satellites, initially with low mass densities.
In this case, as shown in Fig.~\ref{fig:fig_toymod}, we find that the moving average slightly shifts the overall distribution towards the initial condition (IC) values. Similarly, the minimum distance of the first-infall satellites ($r_{\rm min}$) corresponds to their current position ($r$), yielding a tighter correlation between $r_{\rm min}$ and $r$ of the entire satellite population (since these distances are the same for the 30\%).

We show the effect of the tidal stripping on the mean mass densities measured in the satellite's central regions, measured at twice the total half-mass radius ($2h$) and at twice the stellar half-mass radius ($2h\star$).
As Fig.~\ref{fig:fig_toymod} (right panel) shows, the central mean mass densities increase towards the cluster centre.
This is a consequence of the shrinking of $h$ and $h\star$ due to the mass stripping of the outer layers of the total and stellar masses that define them (Eqs.~\ref{eq:rh} and \ref{eq:rh*}).
In addition, as expected from Eqs.~\ref{eq:tdenh} and \ref{eq:denh*}, the mean mass density of each satellite measured at its half-mass radius (\eg $h\star$) is higher than at larger radii (\eg $r_{\rm tr}$).

Furthermore, we see in Fig.~\ref{fig:fig_toymod} that the moving average of the total and stellar mass mean density starts to increase within $r< 0.5\,\Rvir$, being slightly steeper for the stellar mass densities. 
This can be seen in the figure for mass densities measured at the truncation radius (left panel), or at their half-mass radii ($h$ or $h\star$, right panel).
Besides the Hernquist profile used for the satellite mass models shown in Fig.~\ref{fig:fig_toymod}, we also tested exponential and Plummer profiles, resulting in a similar shape of the moving average of the satellite mass densities.
In the following Sect.~\ref{sec:res:sim} we show the moving average of stellar mass densities for models that include more realistic physical processes.

\begin{figure}[h]
\centering
\includegraphics[width=9.3cm,trim={0.2cm 0.25cm 0.25cm 0.1cm},clip]{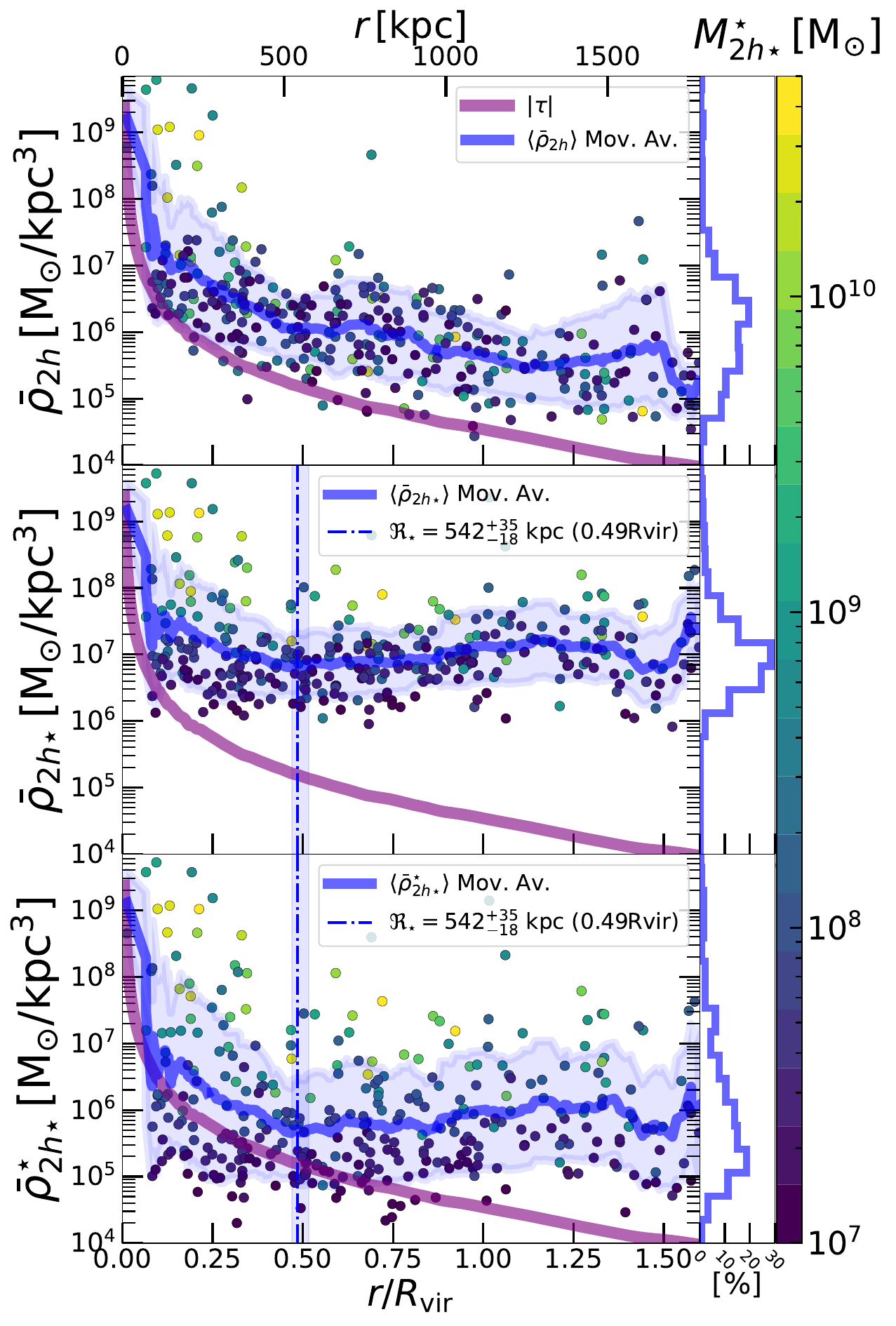}
%\vspace{-0.2cm}
\caption{
 Mean mass densities of 492 satellite galaxies (circles) as a function of distance to their hosting TNG50 Fornax-type cluster,
 with colours indicating their stellar masses (colour bar). 
The cluster has a virial mass and radius of 
$M_{\rm vir}\!=\!7.6\times10^{13}\sm$ and $\Rvir\!=\!1075\kpc$, respectively, and corresponds to a snapshot at redshift $\redshift\!=\!0$.
Its tidal field radial profile ($|\tau|$) is shown in all panels (purple curve).
Shown as a function of cluster-centric distance from the top to bottom panels are: 
the dynamical mass densities of each satellite $\bar{\rho}_{2h}$ (top), measured within $2h$;
 the central mean dynamical mass densities $\bar{\rho}_{2h\star}$ within $2h\star$ (middle),
 and the central stellar mass densities $\bar{\rho}^{\star}_{2h\star}$ within $2h\star$ (bottom).
The blue curves in each plot show the moving average (Mov. Av.) of different mass densities for all the satellites using Eq.~\ref{eq:kern}.
The measurement of the transition radius ($\Rr$ metric [ii]) is shown with a vertical dot-dashed line (middle and bottom panels). 
Histograms are shown in the right column sub-panels to show the shape of the mean mass density distributions.
}
\label{fig:TNG50cluster}
\end{figure}

\subsection{Satellite mass densities: Galaxy simulations} 
\label{sec:res:sim}
The toy model offers a simplified modelling of the tidal evolution of satellite galaxies, 
predicting the resulting radial distribution of the mean mass densities of the satellite population, 
and the strength of the tidal field of the host cluster that drives the evolution.
However, the satellites are represented by a rigid model, which has limitations in including additional physical processes, such as tidal shocks, galaxy harassment, star formation, and gas and ram-pressure stripping. 
Therefore, we performed an analysis of satellite galaxies in cosmological galaxy simulations that include these processes.

We start by presenting the analysis of a Fornax-type analogue cluster in detail, selecting a cluster with a mass of $M\!=\!7.6\times10^{13}\sm$ from the Illustris TNG50 simulation.
We see in Fig.~\ref{fig:TNG50cluster} (top panel) that the dynamical mean mass density profile of the satellites $(\bar{\rho}_{2h})$ follows the tidal relation of Eq.~\ref{eq:tdenh},
with satellites being tidally shaped and having mass densities that are systematically higher in the central region of the cluster.
As the tidal field strength of the host cluster increases towards the cluster centre, it clearly demonstrates its effect on shaping the mass density profile of the satellite population ($\langle \bar{\rho}_{2h}\rangle$).
Moreover, we observe a similar behaviour to the satellite mean mass density profile predicted in the toy model (Fig.~\ref{fig:fig_toymod}), \cMBII{where most satellites have $\bar{\rho}_{2h}$ values  larger than the tidal field at any distance $r$, because they have already passed their pericentres where $|\tau|$ was larger.}
The presence of some satellites having mean mass densities lower than the tidal field strength in Fig.~\ref{fig:TNG50cluster} (top panel) reveals that secondary effects, such as the satellite's internal processes (star-formation, morphological substructure, etc.), orbital characteristics, or the presence of substructures within the galaxy cluster, may play some role for individual satellites.  
Furthermore, a careful inspection of the density profile in Fig.~\ref{fig:TNG50cluster} (top panel) reveals that the gradient increases within $r<{0.5}\,\Rvir$ towards the cluster centre.
It worthy to point out that at these short distances, the tidal field strength of the cluster surpasses the lowest mean mass densities of satellites coming from the outskirts of the cluster \ie where $|\tau|>\bar{\rho}_{2h}\sim10^{5}\sm\kpc^{-3}$.

Next, we examine the central mean dynamical-mass densities of the satellites ($\bar{\rho}_{2h\star}$; see Fig.~\ref{fig:TNG50cluster}, middle panel). 
\cMBII{We find that in the cluster outskirts ($r>0.5\Rvir$) the shape of the moving average of the satellite mass densities $\langle\bar{\rho}_{2h\star}\rangle$ is consistent with a small positive slope and a roughly constant value beyond $r\!\!\approx\!\Rvir$.}
Inside $r\!\!\approx\!\!{0.5}\,\Rvir$, the central regions within the satellites become increasingly more tidally affected, having their outer layers stripped, which shifts the population's moving average to higher densities that increase towards the cluster centre. The mass density profile $\langle\bar{\rho}_{2h\star}\rangle$ has a negative gradient, similar to the behaviour of the mean (total) dynamical-mass density profile $\langle\bar{\rho}_{2h}\rangle$ (top panel).

\begin{figure}[t]
\begin{center}
\hspace*{-0.3cm}
\includegraphics[width=9.5cm]{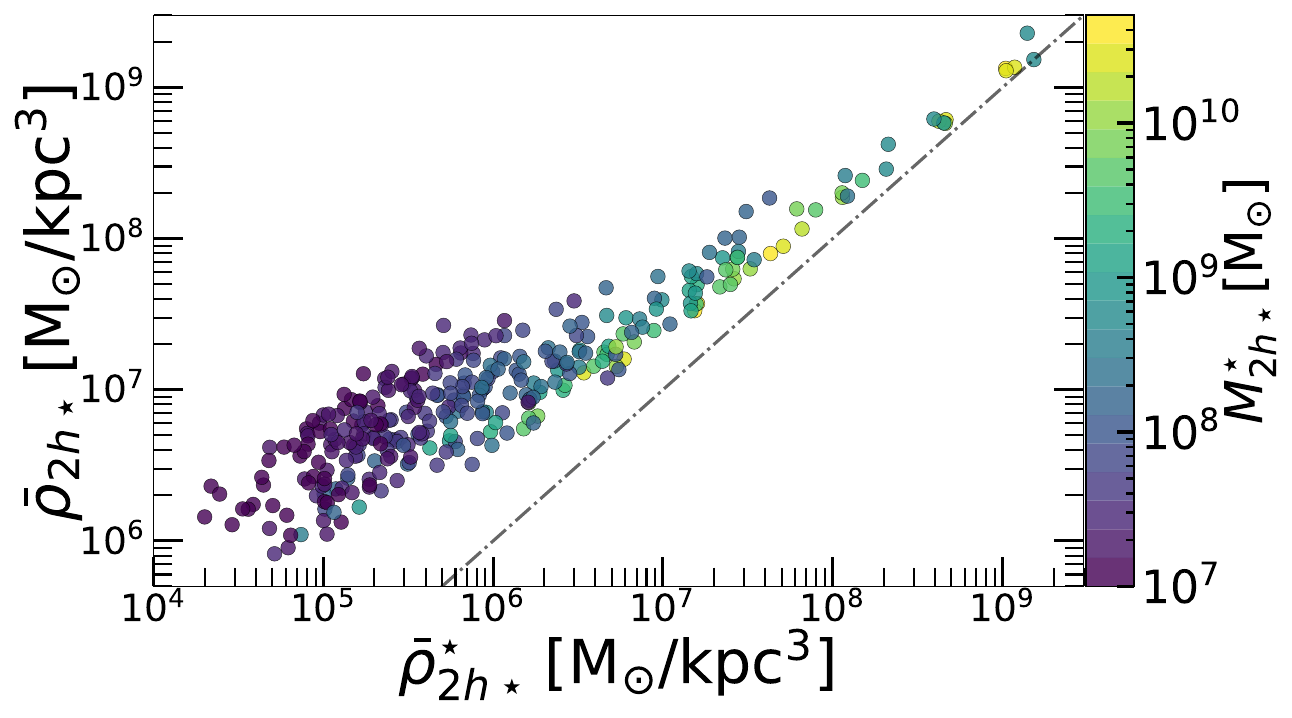}
\vspace{-0.5cm}
\caption{Comparison between the mean central dynamical and stellar mass densities for satellites in the TNG50 Fornax-type cluster. The symbol colour encodes stellar mass, and the identity function is shown as the dot-dashed line.
}
\label{fig:fig_tng50_den_den}
\end{center}
\end{figure}

Following the analysis of our toy model, we also examine the central mean stellar mass densities of satellites ($\bar{\rho}^{\star}_{2h\star}$) as a function of cluster-centric distance\footnote{We note that the central stellar mass density profile measured within one stellar half-mass radius ($h\star$) exhibits the same characteristics.}, shown in Fig.~\ref{fig:TNG50cluster} (bottom panel).
\cMBII{We find a distribution with more scatter than the central dynamical-mass densities, which is due to the variations of the baryon-to-dark matter fractions between satellites.}
The moving average of the stellar mass density has a roughly constant value of $\langle\bar{\rho}^{\star}_{2h\star}\rangle\approx10^6\sm\kpc^3$ at large distances $r\gtrsim1\Rvir$. 
It decreases slightly within $r\lesssim1\,\Rvir$, reaching a value of $\langle\bar{\rho}^{\star}_{2h\star}\rangle\approx 5\times10^5\sm\kpc^3$. 
Further in at $r\lesssim{0.5}\,\Rvir$, the average stellar mass density profile starts to increase significantly, following the cluster's tidal field strength.
The correlation between the central dynamical and stellar mass density profiles can be appreciated by comparing the middle and bottom panels in Fig.~\ref{fig:TNG50cluster}, \cMBII{and in Fig.~\ref{fig:fig_tng50_den_den} where we directly compare both density variables, finding that the scatter is less than 1\,dex.} 
Overall we find that the moving average of the total, central, and stellar mass density profiles ($\langle\bar{\rho}_{2h}\rangle$, $\langle\bar{\rho}_{2h\star}\rangle$, $\langle\bar{\rho}_{2h\star}^{\star}\rangle$) behave in a similar way within $r<{0.5}\,\Rvir$, where the tidal field of the cluster dominates.

Comparing with the toy model predictions (Fig.~\ref{fig:fig_toymod}) we see that the mean stellar mass densities of the toy model are shifted to values lower compared to those in the TNG simulations. However, the shape of both profiles are very similar, especially showing the same transition at $r\simeq{0.5}\,\Rvir$.

\begin{figure*}[t]
\centering
\includegraphics[width=9.1cm]{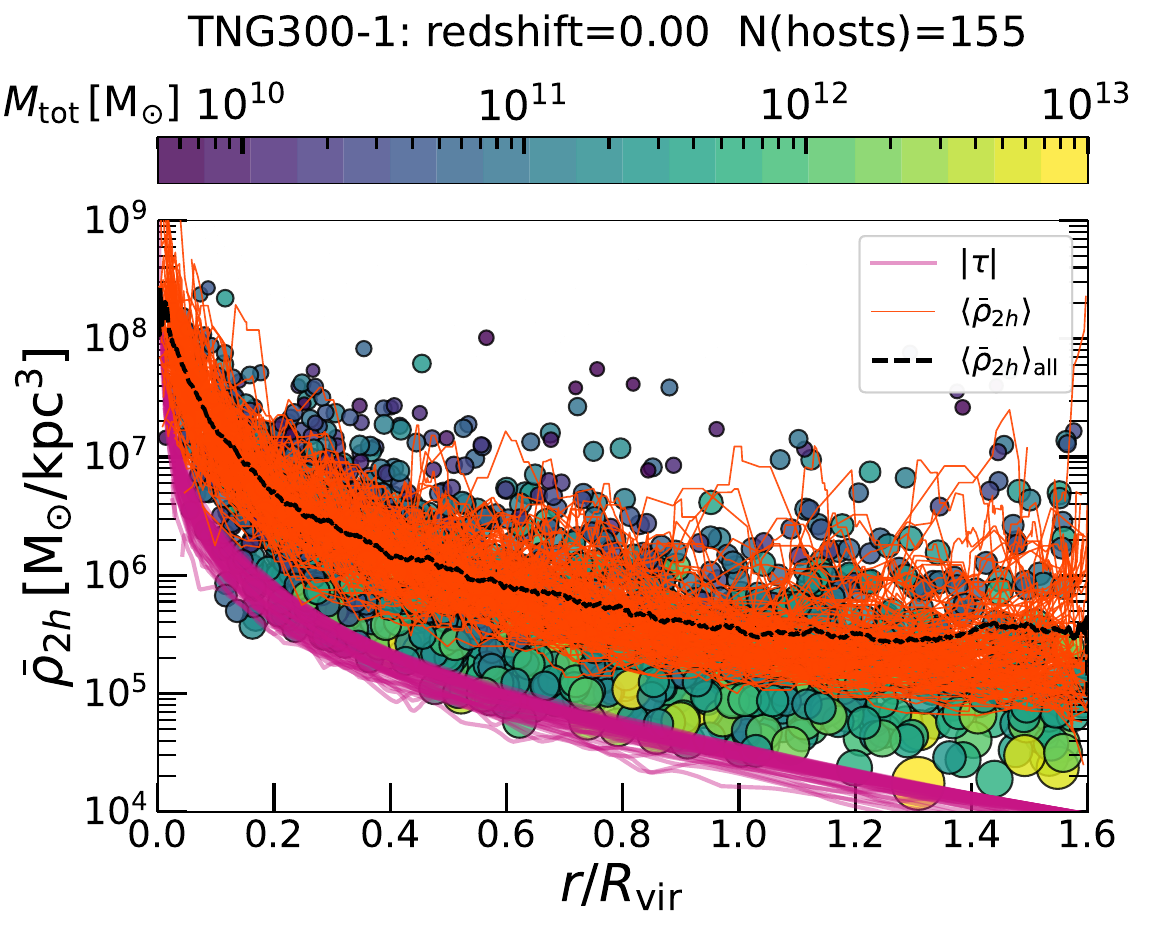}\hspace{0.0cm}
\includegraphics[width=9.1cm]{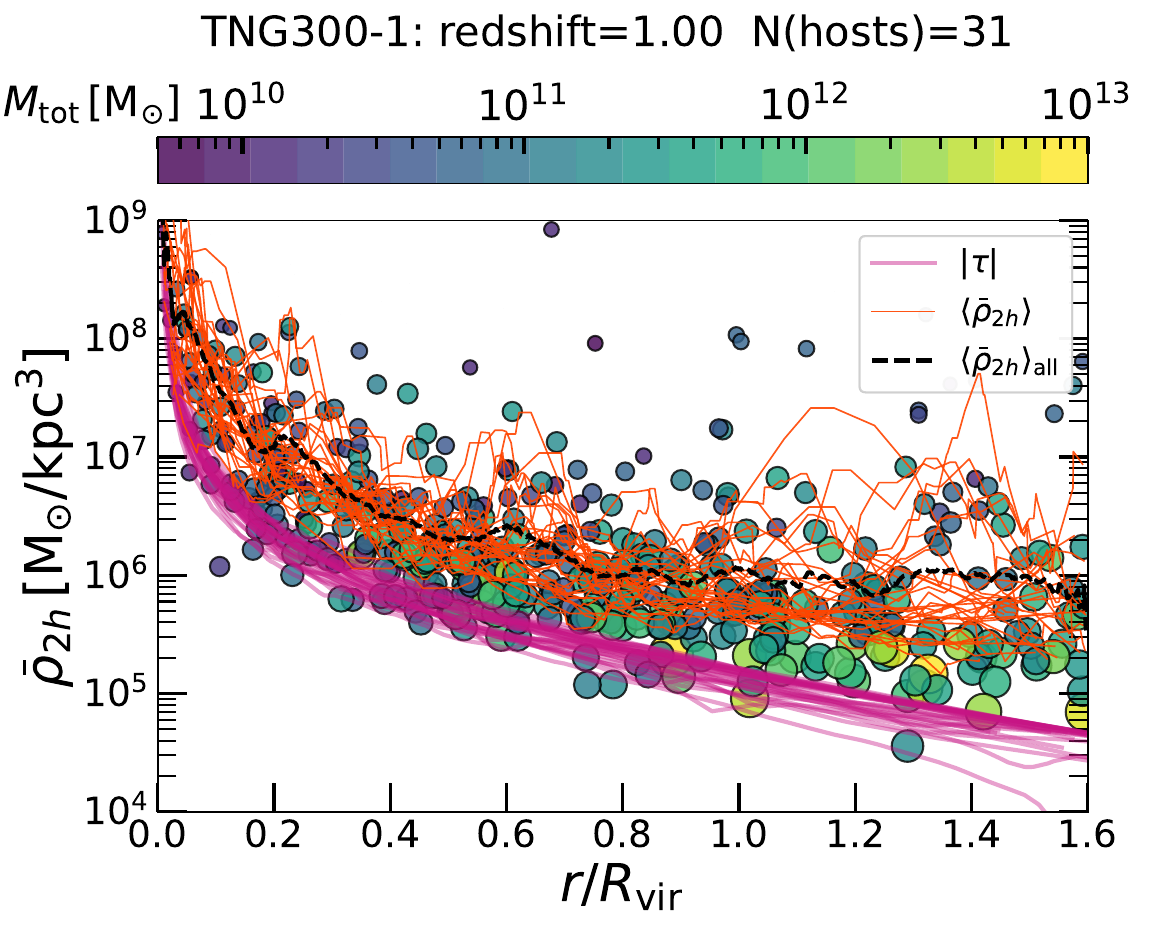}

\hspace{-0.25cm}\includegraphics[width=9.2cm]{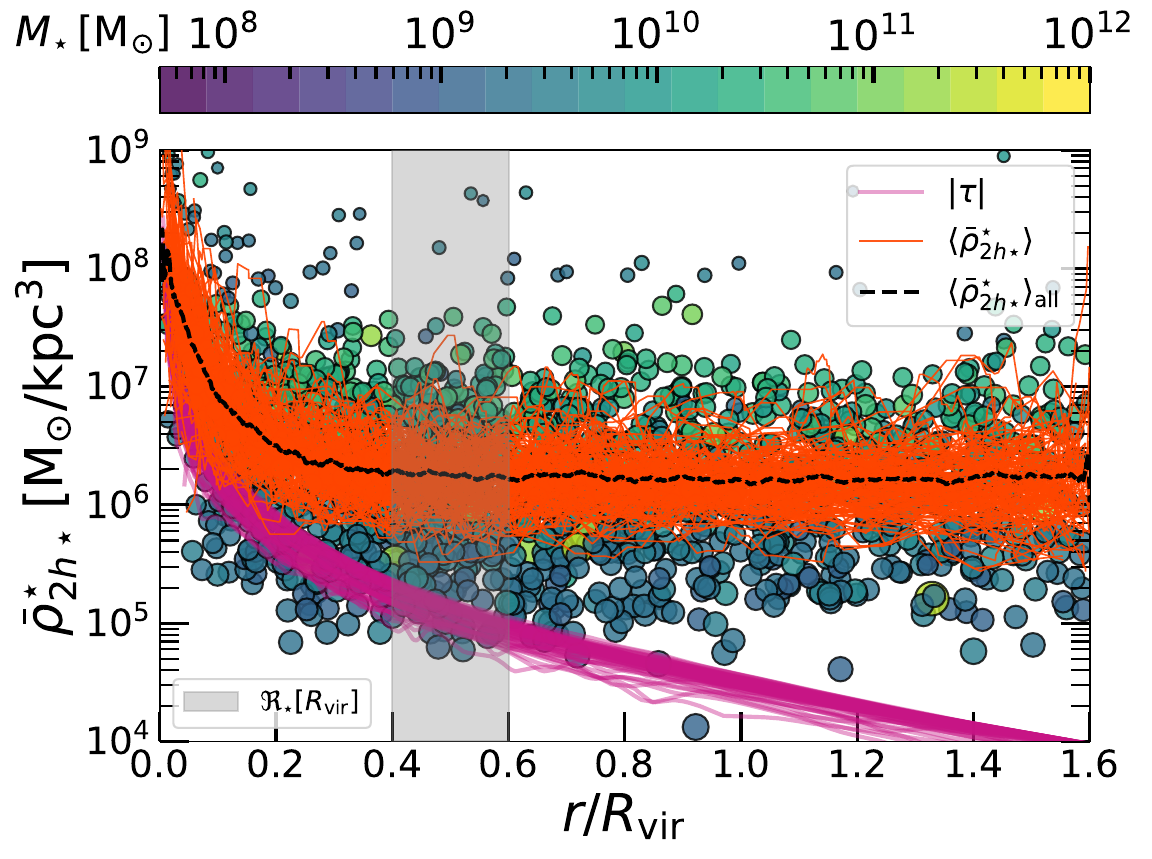}\hspace{-0.15cm}
\includegraphics[width=9.2cm]{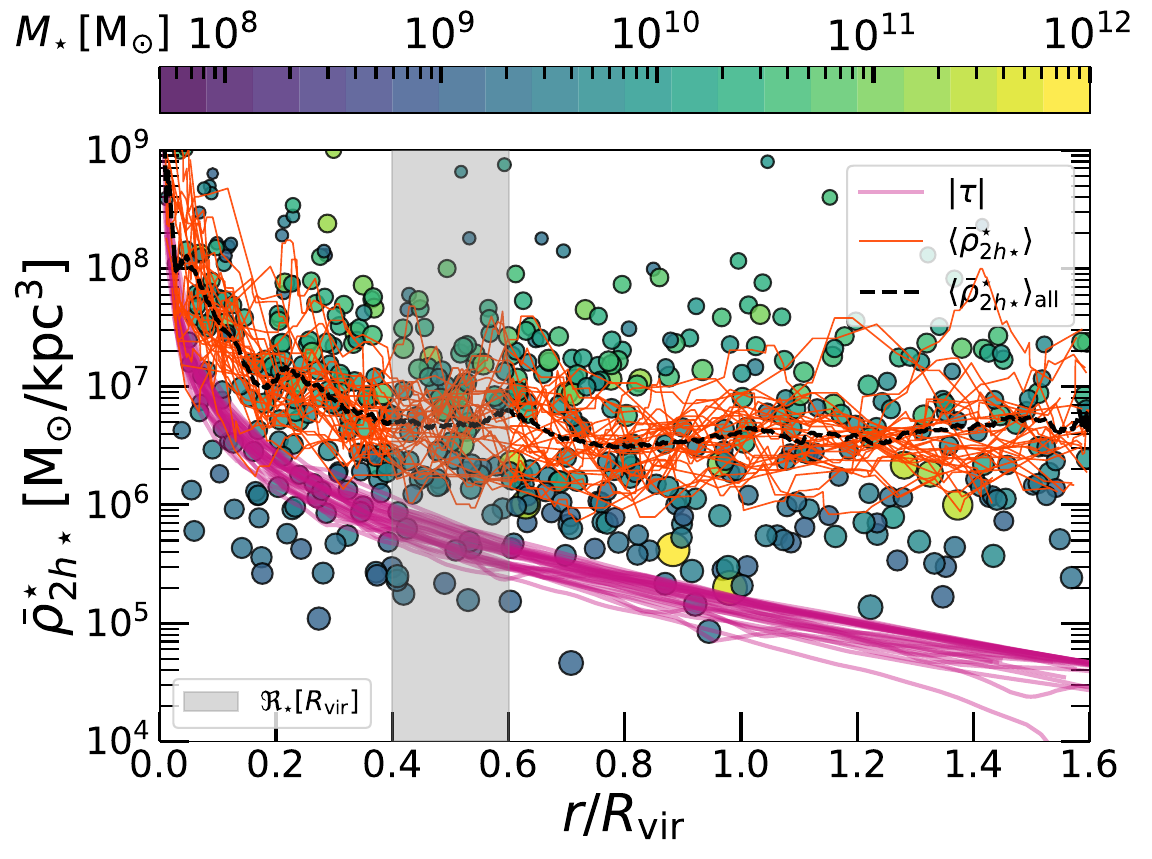}
%\vspace{-0.2cm}
\caption{
\cMB{Mass densities of satellite galaxies vs their cluster centric distances for 155 host galaxy clusters from a snapshot at redshift $\redshift\!=\!0$ (left column) and for 31 clusters at $\redshift\!=\!1$ (right column), taken from TNG300. 
The tidal field profiles of all the clusters are shown in each panel ($|\tau|$ in purple curves).
Top panels show the total mass mean densities of the subhaloes (circles, colour-coded by total subhalo masses),
and the bottom panels show the central mean stellar mass densities (circles, colour-coded by stellar masses).
To avoid overcrowding, we plotted a sub-sample with 10\% (left column) and 20\% (right column) of the satellites (circles).
Circle sizes are proportional to $h$ (top panels) or $h\star$ (bottom panels).
In each panel the moving average satellite mean mass density profile of each cluster is shown in with an orange curve.
In addition, it is shown the global moving average in dashed black curve that is calculated with the moving averages of all clusters. All are moving averages are calculated with Eq.~\ref{eq:kern}.
The transition radii ($\Rr$), measured in each cluster, are located within the vertical grey region (bottom panels), which are determined from the moving average of the mean stellar mass densities $\langle \bar{\rho}^{\star}_{2h\star} \rangle$, marking where the $\langle \bar{\rho}^{\star}_{2h\star} \rangle$ start to increase further in. 
We present a similar analysis of TNG100 and TNG50 in Figs.~\ref{fig:TNG100} and \ref{fig:TNG50}, respectively.}}
\label{fig:TNG300}
\end{figure*}

The TNG simulations and our toy model indicate that mass density profiles undergo a transition at $r\approx{0.5}\,\Rvir$, within which the cluster's tidal field strength appears to dominate satellite densities, while further out we observe roughly constant, perhaps slightly increasing satellite densities as a function of cluster-centric distance. 
In the following, we seek to quantify this \cMBII{distance at $r\!\!\approx0.5\,\Rvir$} as the `transition radius' parameter \Rr. 
\cMBII{We defined two metrics to measure \Rr: using the minimum of the moving average of the mean stellar mass profile $\langle\bar{\rho}^{\star}_{2h\star}\rangle$ within $r<\Rvir$ (\Rr[i]), and using the minimum of the derivative of the moving average profile (\Rr[ii]). 
Later, in Sect.~\ref{sec:dis:Rr}, we provide a complete detailed description of two metrics that we define in Eq.~\ref{eq:rtran1} and \ref{eq:rtran2} to measure this feature, as well as estimates of the uncertainties in different simulated and observed systems (Table~\ref{tab:results}).}
Here we report that the transition radius measured using $\langle\bar{\rho}^{\star}_{2h\star}\rangle$ for this TNG cluster is $\Rr\!=\!0.49\,\Rvir\,(542\kpc)$, which is marked with the vertical line in Fig.~\ref{fig:TNG50cluster} (bottom panel).
In the same figure, we also plot \Rr in the middle panel to show that it matches the profile shape of central dynamical mass densities ($\langle\bar{\rho}_{2h\star}\rangle$).

Furthermore, we also analyse in detail two additional clusters with masses similar to the Virgo cluster ($M_{\rm vir}\!=\!4.2\times10^{14}\sm$).
One cluster is selected from Illustris TNG100, while the other was selected from the \textsc{Eagle} simulations \citep{Schaye2015} to explore how different star-formation prescriptions and sub-grid models impact our results. 
\cMBII{In general, we find that the satellites in these two cluster models exhibit similar behaviour in their mean mass density profiles ($\bar{\rho}_{2h}$, $\bar{\rho}_{2h\star}^{\star}$), having progressively declining values within $r\lesssim{0.5}\,\Rvir$ and a transition radius \Rr in the mean stellar mass density profiles.} 

\begin{figure*}[t]
\centering
\includegraphics[width=8.8cm]{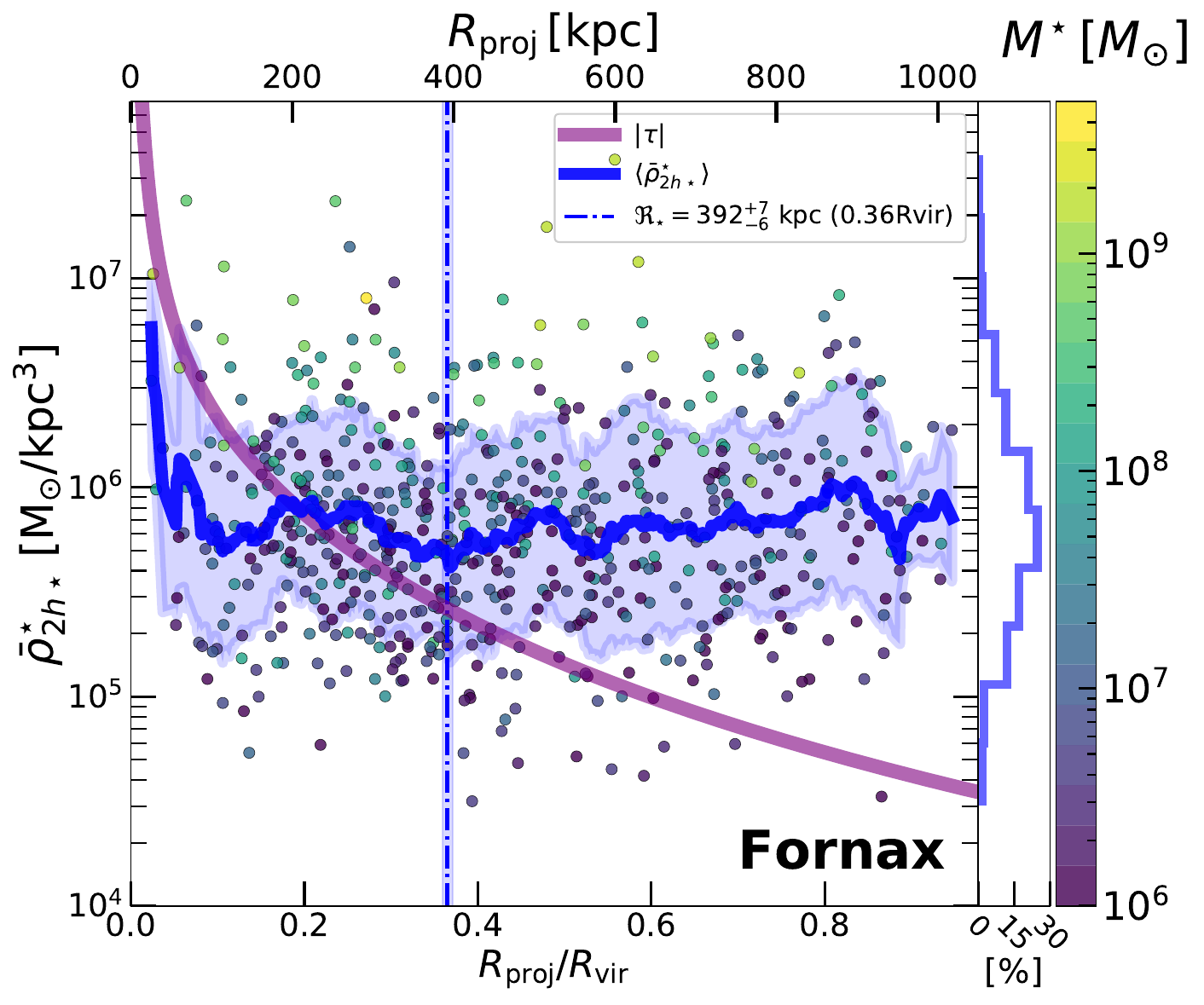}
\includegraphics[width=8.8cm]{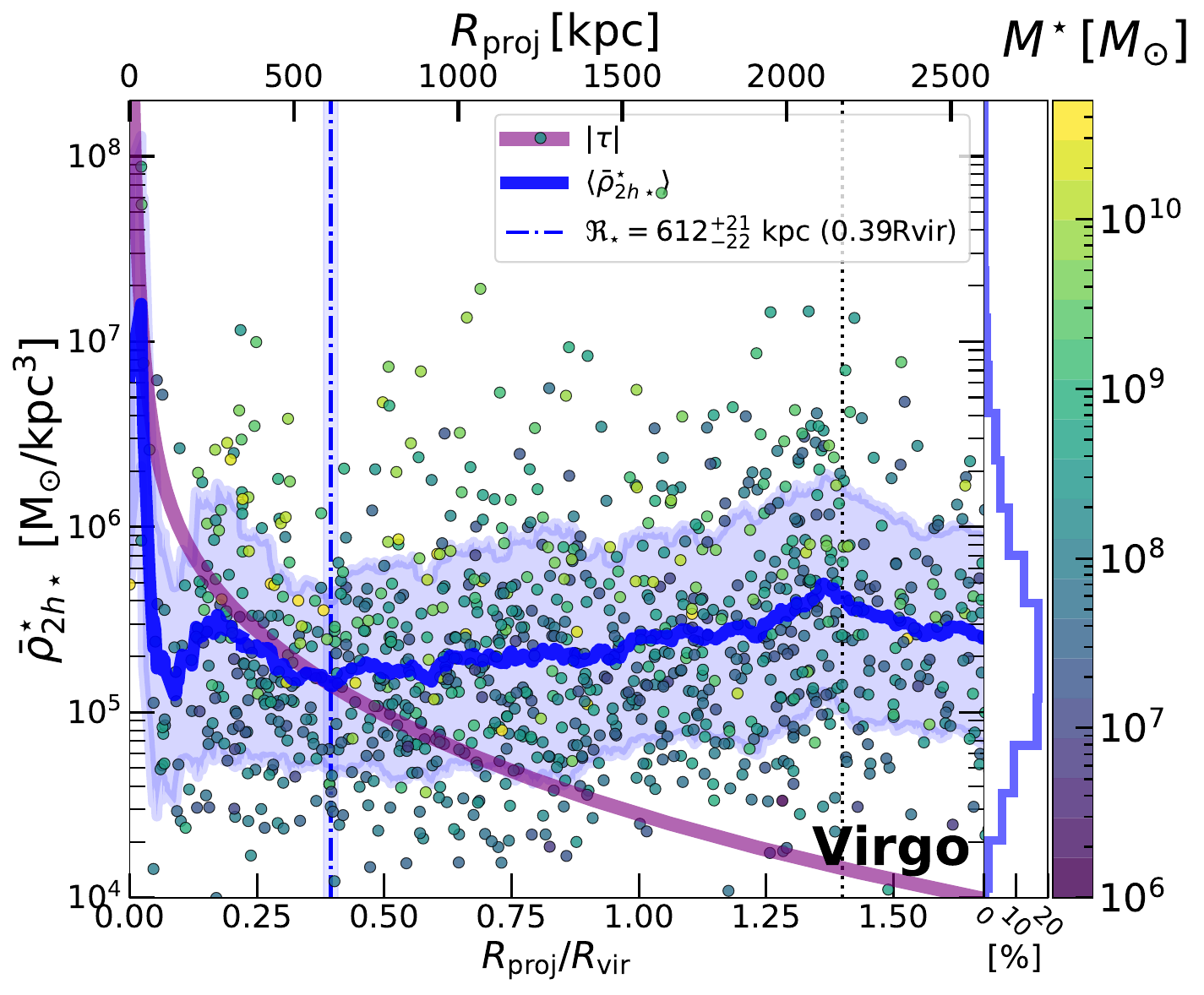}
\includegraphics[width=8.8cm]{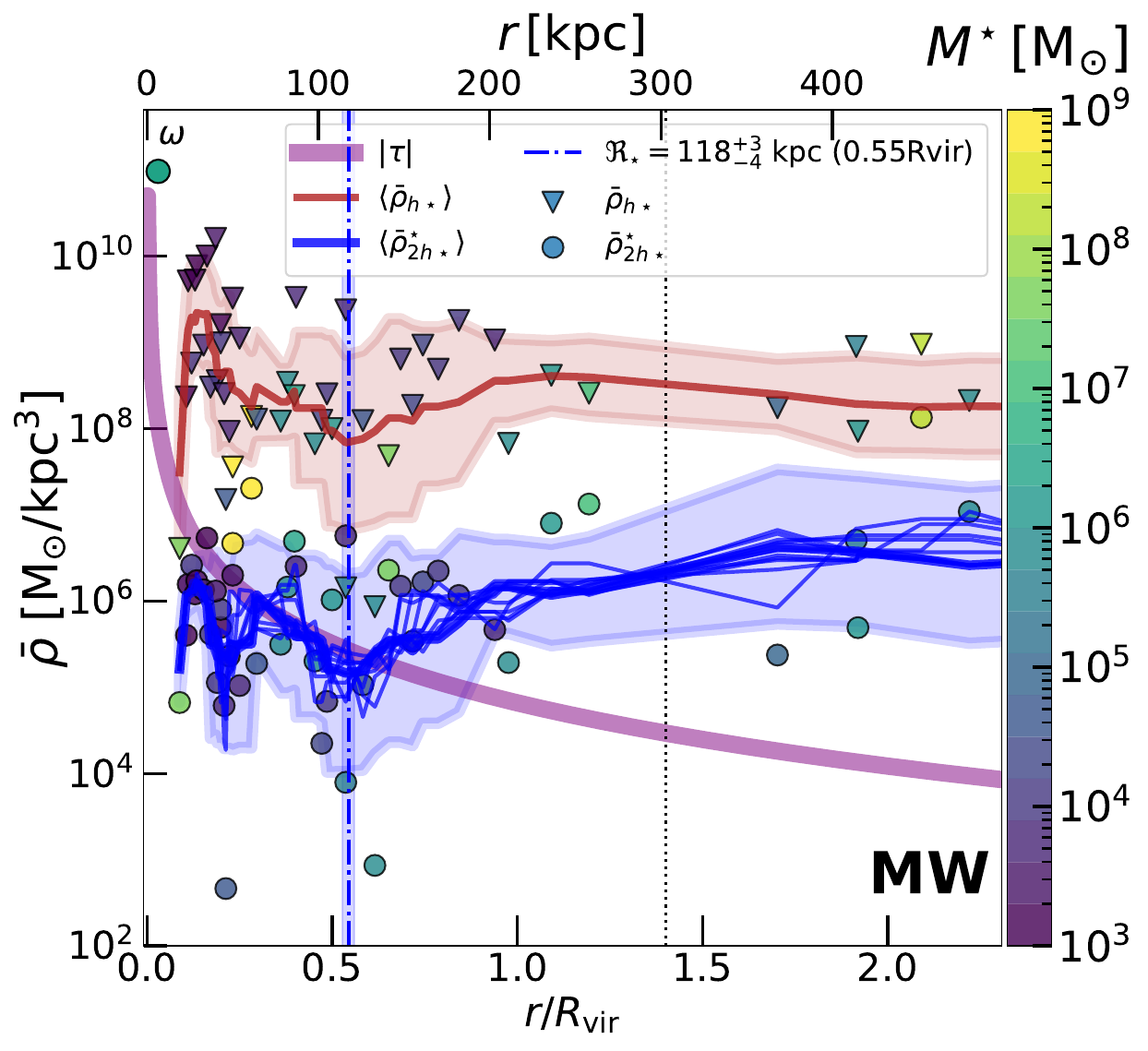}  
\includegraphics[width=8.8cm]{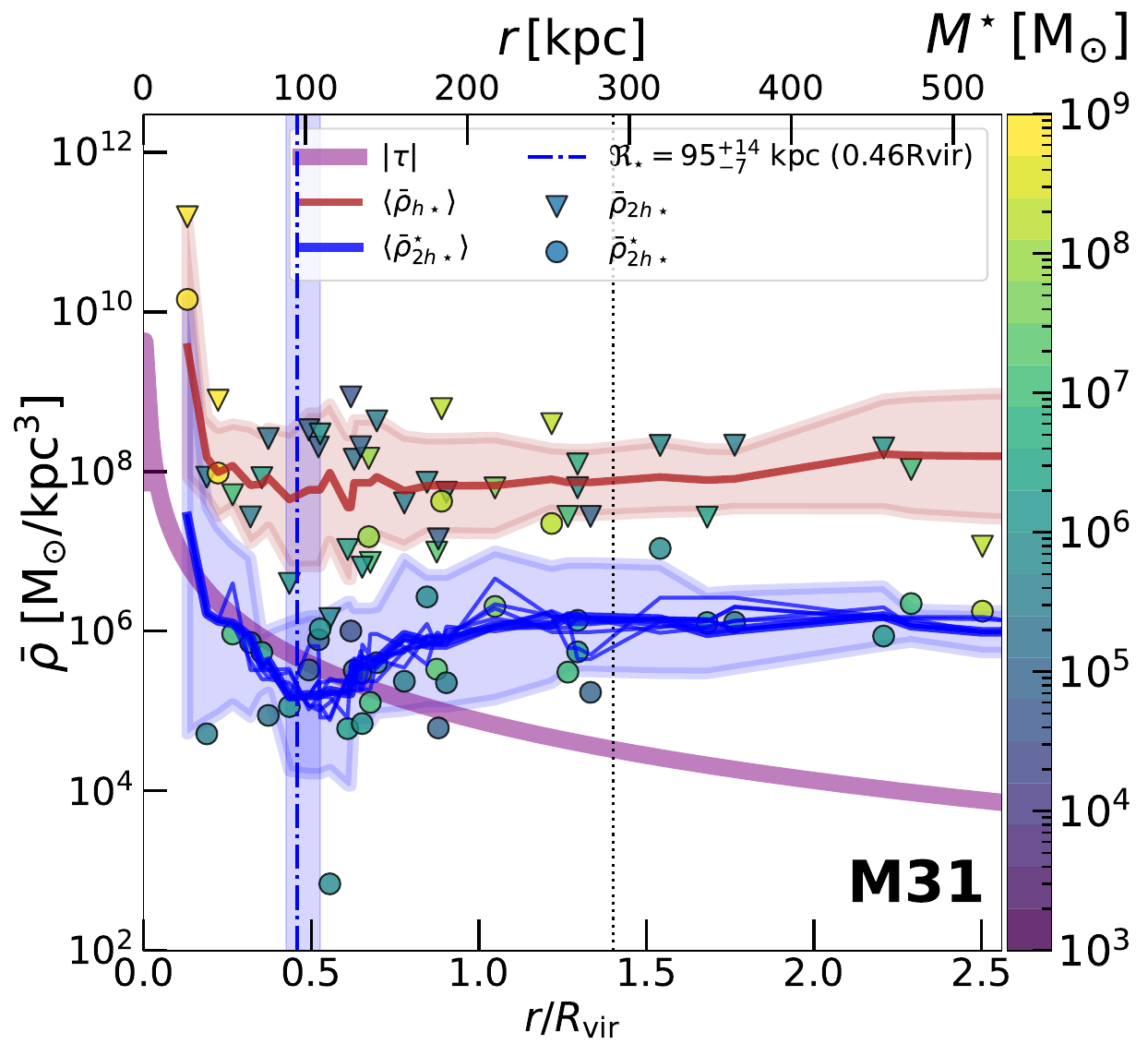}    
%\vspace{-0.3cm}
\caption{
Satellite galaxy mean stellar mass densities ($\bar{\rho}^{\star}_{2h\star}$ as circles) of four observed systems: 
the Fornax galaxy cluster (top left panel), the Virgo cluster (top right panel), 
the MW (bottom left panel), and M31 (bottom right panel), 
with the satellite stellar masses indicated in colour.
In all the panels, we show the tidal field for the host systems ($|\tau|$ in purple curves). 
In each panel we include the moving average profiles of the mean stellar mass densities using Eq.~\ref{eq:kern} with $q\!=\!\sqrt{n_{\rm sat}}$ (blue curves, with the blue areas casting one standard deviation).
The measured transition radii $\Rr$ are shown for each system (dash-dotted vertical lines), and the expected location of the splashback radii $\sim 1.4\Rvir$ (vertical dotted lines).
\cMBII{For MW and M31, due to the low number of satellites, we add thinner blue curves that show moving averages $\langle\bar{\rho}^{\star}_{2h\star}\rangle$ where each is calculated with number of neighbours between $q\!=\!0.5 \sqrt{n_{\rm sat}}$ and $2\sqrt{n_{\rm sat}}$. This indicates that, independent of $q$, all the curves show a low value of $\langle\bar{\rho}^{\star}_{2h\star}\rangle$ at \Rr.}
In addition, for MW and M31 we show the central mean dynamical mass densities of each satellite measured within $h\star$ (triangles) 
and the respective moving average profiles (brown curves, bottom panels). 
\cMBII{This can be compared with other studies in the literature on the $\bar{\rho}-r$ relation for the MW satellites mentioned in Sect.~\ref{sec:theo}, which show that using $r$ instead of the pericentre also reveals an anti-correlation with distance ($\bar{\rho}\propto1/r$). Moreover, as predicted by the simulations, the central dynamical mass densities show a change in the moving average profile where \Rr is measured (see Fig.~\ref{fig:TNG50cluster}, middle panel).}
Histograms included in the top panels show the overall shape and peak of the distribution of $\bar{\rho}^{\star}_{2h\star}$ for the satellites of each cluster.
}
\label{fig:obs}
\end{figure*}

Galaxy clusters and groups are in constant dynamical evolution presenting significant substructures due to massive subgroups of galaxies being accreted, and with occasional major mergers at different redshifts \citep{McGee2009,Pallero2019,Kuchner2022}. 
Therefore, we also analyse a total of 235 galaxy clusters \cMBII{and groups} with a range of masses 
between $9\times10^{12}\sm$ and $3\times10^{15}\sm$,
and at different redshifts from the Illustris TNG simulations, 
\cMBII{to explore possible perturbations in the mass density profiles of satellite populations due to infalling substructures in the cluster/group environments.}
In Fig.~\ref{fig:TNG300} we show the dynamical mass densities and the central stellar mass densities of satellites \cMBII{of 155 clusters in the TNG300 simulation at redshifts $\redshift\!=\!0$ and 31 clusters at $\redshift=1$}, which have the largest number of satellites.

In order to study how the satellite mean mass densities behave in lower mass host environments, we present in Figs.~\ref{fig:TNG100} and \ref{fig:TNG50} similar results analysing the TNG100 and TNG50 simulations.
In general, we find consistent results \cMBII{in the three sets of simulations (TNG50, TNG100, TNG300), where the satellite populations exhibit the same cluster-centric behaviour in their mass density profiles as discussed earlier for the Fornax-type cluster.
We measured the transition radius in each of these clusters and groups in TNG300, TNG100, and TNG50, finding typical values in the range of $\Rr=0.4\Rvir$ and $0.6\Rvir$. 
Moreover, clusters can present substructures due to different processes, such as major or minor mergers of groups (visible as wiggles in the orange curves in Figs.~\ref{fig:TNG300}, \ref{fig:TNG100} and \ref{fig:TNG50}). 
The analysis at redshifts $\redshift\!=\!1$ and $0$ shows a consistent mass density profile behaviour with cluster-centric distance. 
However, we defer the characterisation of the temporal evolution of the satellite mass density profiles to future works, as this requires the careful assembly of observed satellite samples at corresponding redshifts, which are currently unavailable.}

Until now, we have only examined the total and stellar mass densities of satellite galaxies. 
Although the total (dynamical) mass densities include the gas component, we have not mentioned how the gas properties of satellites may change with distance to the host cluster.
\cMBII{As rapid gas-mass loss could be driving internal changes in the stellar and dark matter distribution, we proceed to measure the gas mass content of satellites and the resulting mass densities.}
We find that within $r\lesssim{0.5}\,\Rvir$ only a small percentage of satellites ($\lesssim\!2\%$) have gas mass densities comparable to the values of stellar mass densities, having gas-to-baryon mass fractions of up to 10\%.
Later in Sect.~\ref{sec:dis:trace} we discuss the gas mass properties in more detail, but in general we find that the gas-rich satellites are not the dominant population in the central regions of clusters. 

\subsection{Satellite mass densities: Observations} 
\label{sec:res:obs}

\cMBII{In this section we explore the ($\bar{\rho}-r$) relation for a selection of satellite galaxies in observed systems.}
For this we show the mean stellar mass densities of satellites in the Fornax and Virgo galaxy clusters, which have the advantage of having a large observed satellite samples. We also include data from Local Group satellites that have accurate measurements of their internal kinematics and photometry, as well as 3D distances to their MW and M31 host, which allow us to determine their central dynamical and stellar mass densities.

\begin{figure}[tp]
\centering
\includegraphics[width=0.47\textwidth]{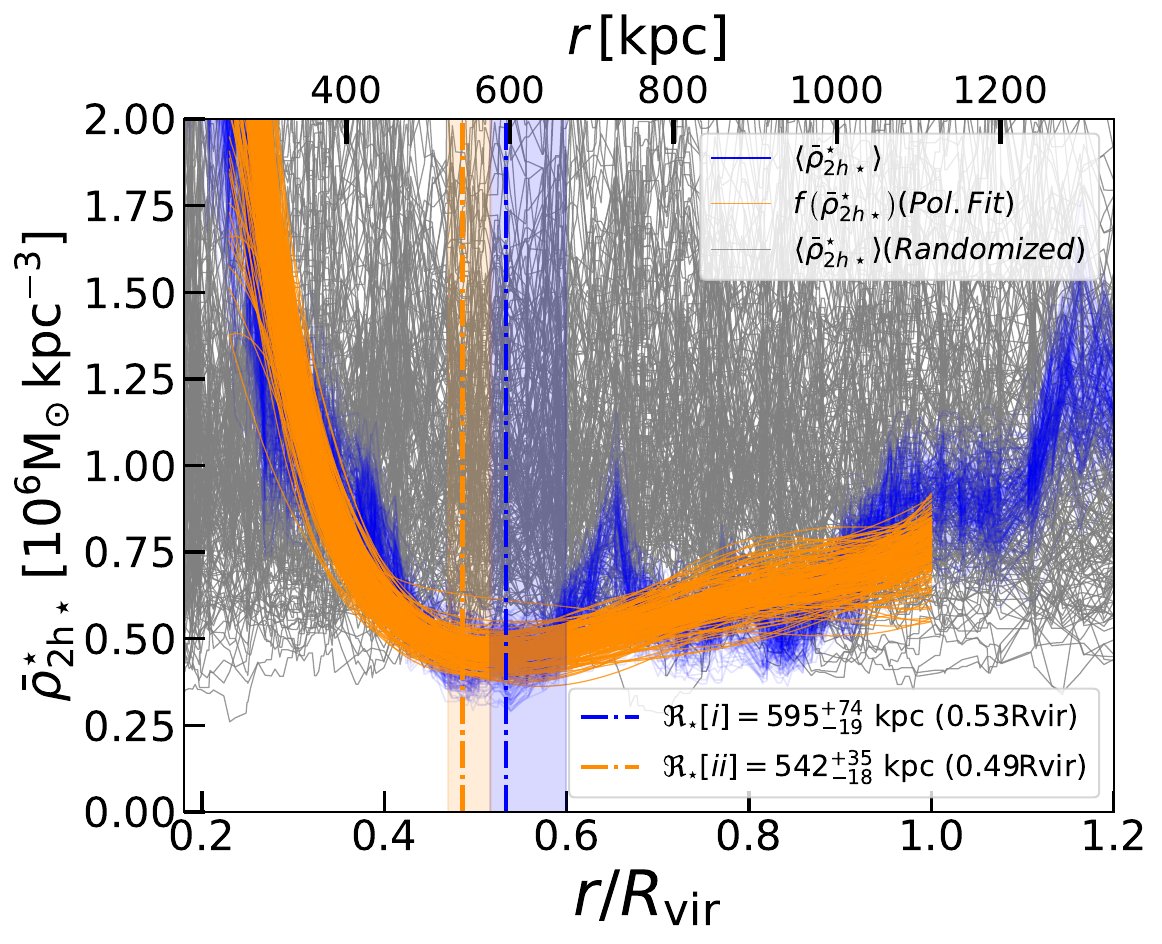}

\includegraphics[width=0.47\textwidth]{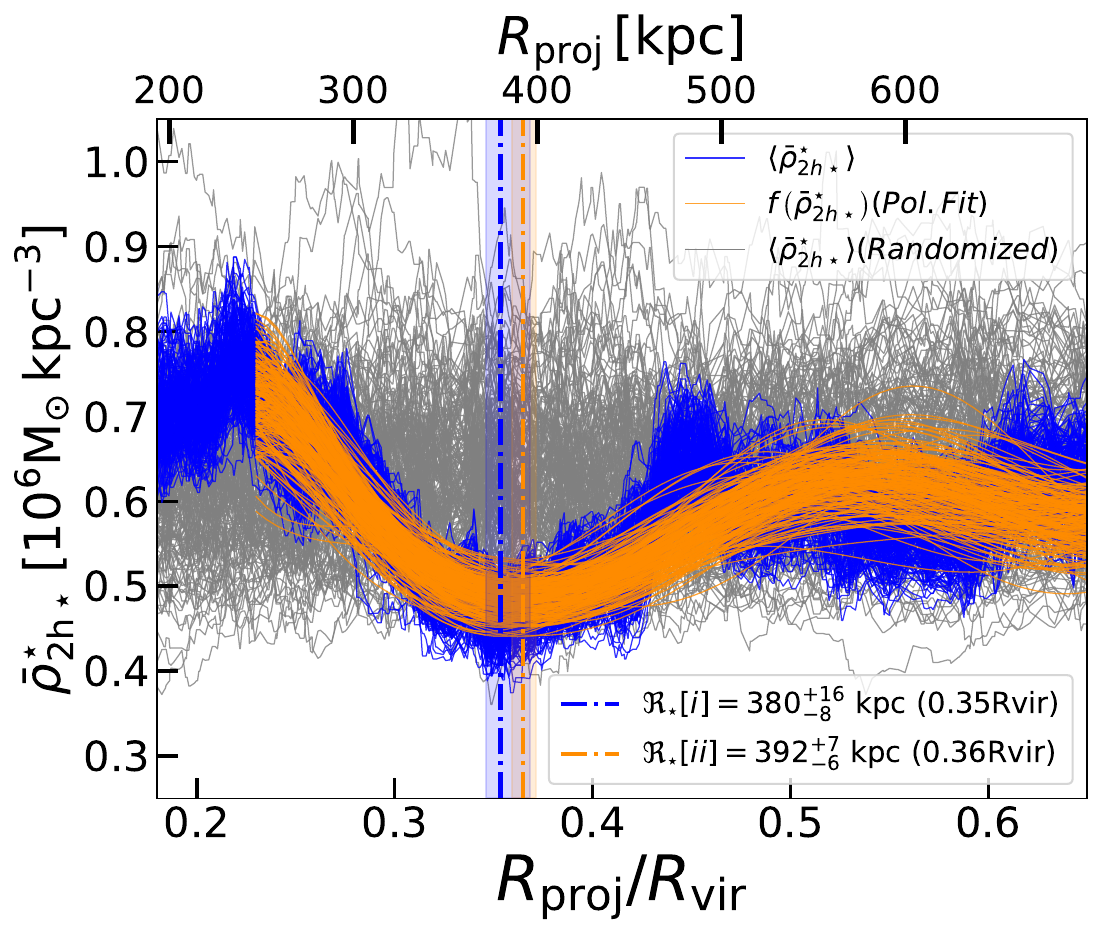}
%\vspace{-0.3cm}
\caption{
Measurement examples of the transition radius $\Rr$:  
the Fornax-type cluster (top panel) from TNG50, and the Fornax cluster (bottom panel).
From a total of $10^6$ sub-sample realisations and the $10^6$ moving average curves of the satellite mean stellar mass densities, 
we show here 200 (blue curves). Each curve is calculated with Eq.~\ref{eq:kern} with a sub-sample of $n_{\rm sub,\,sat}\!=\!n_{\rm sat}\!-\!\sqrt{n_{\rm sat}}$ satellites from a total of $n_{\rm sat}$.
The neighbour number is a random number between $q\!=\!2\sqrt{n_{\rm sub}}$ and $4\sqrt{n_{\rm sub}}$. 
\cMBII{The first method of measurement ($\Rr[i]$) uses Eq.~\ref{eq:rtran1} and is shown with vertical dot-dashed blue lines and their 1-sigma confidence region. The second method ($\Rr[ii]$) uses Eq.~\ref{eq:rtran2} and is shown with vertical dot dash orange lines and their 1-sigma confidence region.
Bootstrapping reveals a small scatter in the moving average profiles, due to the satellite-to-satellite sub-sample variations and the different number of neighbours used.}
Lastly, we also show the moving average after first randomising (shuffling) the mean mass densities of the satellites, which erases the signature in the satellite population (see Null-hypothesis test in Sect.~\ref{sec:dis:null}).
}
\label{fig:measureRrFornax}
\end{figure}

We present the radial mass density distribution profiles for these four observed satellite systems in Fig.~\ref{fig:obs}.
Overall, they share a distribution of the stellar mass densities similar to simulations, 
where the moving average $\langle\bar{\rho}^{\star}_{2h\star} \rangle$ 
\cMB{decreases slightly from the outskirts of the hosts ($r\!\gtrsim\!\Rvir$) inwards until $r\approx 0.4-0.5\,\Rvir$, where a minimum is reached and $\langle\bar{\rho}^{\star}_{2h\star}\rangle$ starts to increase further inwards. 
Later in Sect.~\ref{sec:dis:Rr} we show in detail how this transition radius ($\Rr$) in the moving average profile of the satellite mean stellar mass densities is revealed on a linear scale (see Fig.~\ref{fig:measureRrFornax}).}
Furthermore, we find that the location of $\Rr$ in the observations (Fornax, Virgo, and M31) is 10\% smaller in $\Rvir$ units than in the simulations. 
\cMBII{Such differences could be attributed to total mass overestimation in observations, which could result in larger values of $\Rvir$ (Eq.~\ref{eq:rvir}), and/or due to projection effects of the satellite spatial distribution.}
In fact, in all simulations the parameter $\Rr$, measured in projection, has values systematically smaller in $\Rvir$ units than measured in three dimensions, as expected for most radial mass distributions (see Sect.~\ref{sec:dis:proj}).

\cMBII{Towards the cluster centres, the moving average profiles of Fornax and Virgo in Fig.~\ref{fig:obs} reveal a dip in the mass densities at $R\sim0.1\Rvir$, and increasing values towards the centre within.}
\cMBII{This lower mass density dip may be attributed to projection effects, where distant interlopers with lower mass densities positioned along the line-of-sight in radial filaments decrease the projected moving average of the satellite mean mass densities in the central cluster regions (see Sect.~\ref{sec:dis:proj}).}
\cMBII{Another cause could be the presence of central massive cD galaxies in these two clusters (M87 in Virgo and 1399 in Fornax).
The tidal shocks and dynamical friction that these central galaxies produce could be strong enough to quickly cannibalise neighbouring satellites, reducing their masses and densities more efficiently.}
Future, more detailed comparisons with simulations could reveal whether these deviations are produced by stochastic substructure fluctuations and line-of-sight effects, or whether important physical processes will require further improvement in simulations \citep[\eg see][]{VandenBosch2018}.

\cMBII{Here we mostly focused on the satellite stellar mass densities, which can be readily determined in large surveys using stellar population synthesis models. 
The dynamical mass densities, on the other hand, require relatively high-resolution spectroscopy to measure the velocity dispersion of each satellite galaxy, which is observationally challenging to perform for the hundreds or thousands of satellites in galaxy clusters and groups.
Nevertheless, we present here two important examples, the MW and M31 systems, the proximity of which allows for more complete kinematic estimates, despite their lower sample statistics containing a few dozen satellites each.
An inspection of the bottom panels in Fig.~\ref{fig:obs} shows that the central mean dynamical-mass densities $\bar{\rho}_{h\star}$ of satellites in the MW and M31 exhibit systematically higher average densities towards to the centre of the hosting galaxies. This demonstrates that the ($\bar{\rho}-r$) relation found in simulations is also traceable in the Local Group, as other studies have shown using satellite orbital pericentres \citep{Kaplinghat2019,Hayashi2020,Pace2022,Cardona-Barrero2023}.}
Moreover, while the dynamical-mass densities have a radial trend similar to their stellar counterparts, the former are shifted to larger values because of their dominant dark matter components, in addition to the measurement being performed within $1h\star$ instead $2h\star$, decreasing the volume by a factor of eight. 
The $\langle\bar{\rho}_{h\star}\rangle$ profiles show a similar behaviour as the simulations (Fig.~\ref{fig:TNG50cluster}), with a weak decrease in the average satellite density from the outskirts towards the transition radius $\Rr$. 
Within this region, the mean satellite densities ($\bar{\rho}_{h\star}$ and $\bar{\rho}_{2h\star}^{\star}$) increase significantly towards the host centre. However, the low number of satellites (55 for
the MW and 41 for M31) introduces fluctuations due to stochastic satellite-to-satellite density variations.
\cMBII{As example, in the MW the moving average of the satellite stellar and dynamical mass densities decreases in the very central region ($r<0.2\,\Rvir$). 
This is a consequence of including satellites that are strongly tidally disrupted, such as the dSph Sagittarius and Boötes III, which have large stellar half-mass radii at $h\star=2.6\kpc$ and $h\star=1.3\kpc$, respectively, as well as low luminosities producing low stellar mass densities. 
However, their central dynamical mass densities are still higher than the local MW tidal field, following the ($\bar{\rho}-r$) relation (see purple curves in Fig.~\ref{fig:obs}).}

\section{Discussion}
\label{sec:dis}
\cMB{Here we discuss in more detail the measurement procedure of the transition radius \Rr 
and possible systematic effects that could influence its estimation. 
We also discuss physical processes that could impact the stellar mass densities of satellite galaxies, and how these results relate to other works in the literature.}

\subsection{Measuring the transition radius $\Rr$}
\label{sec:dis:Rr}

\begin{figure*}
\begin{center}
\includegraphics[width=0.485\textwidth]{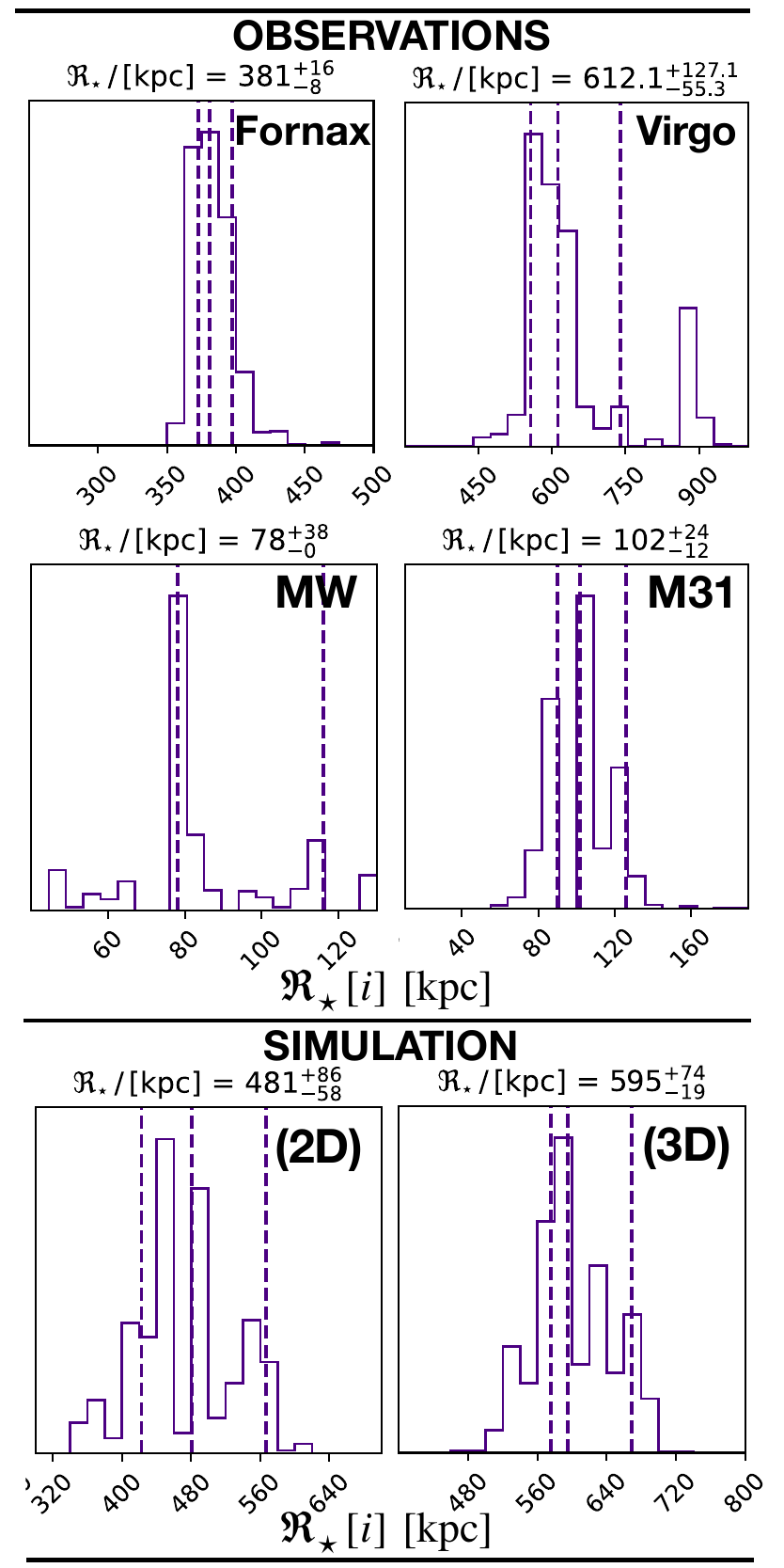}
\includegraphics[width=0.485\textwidth]{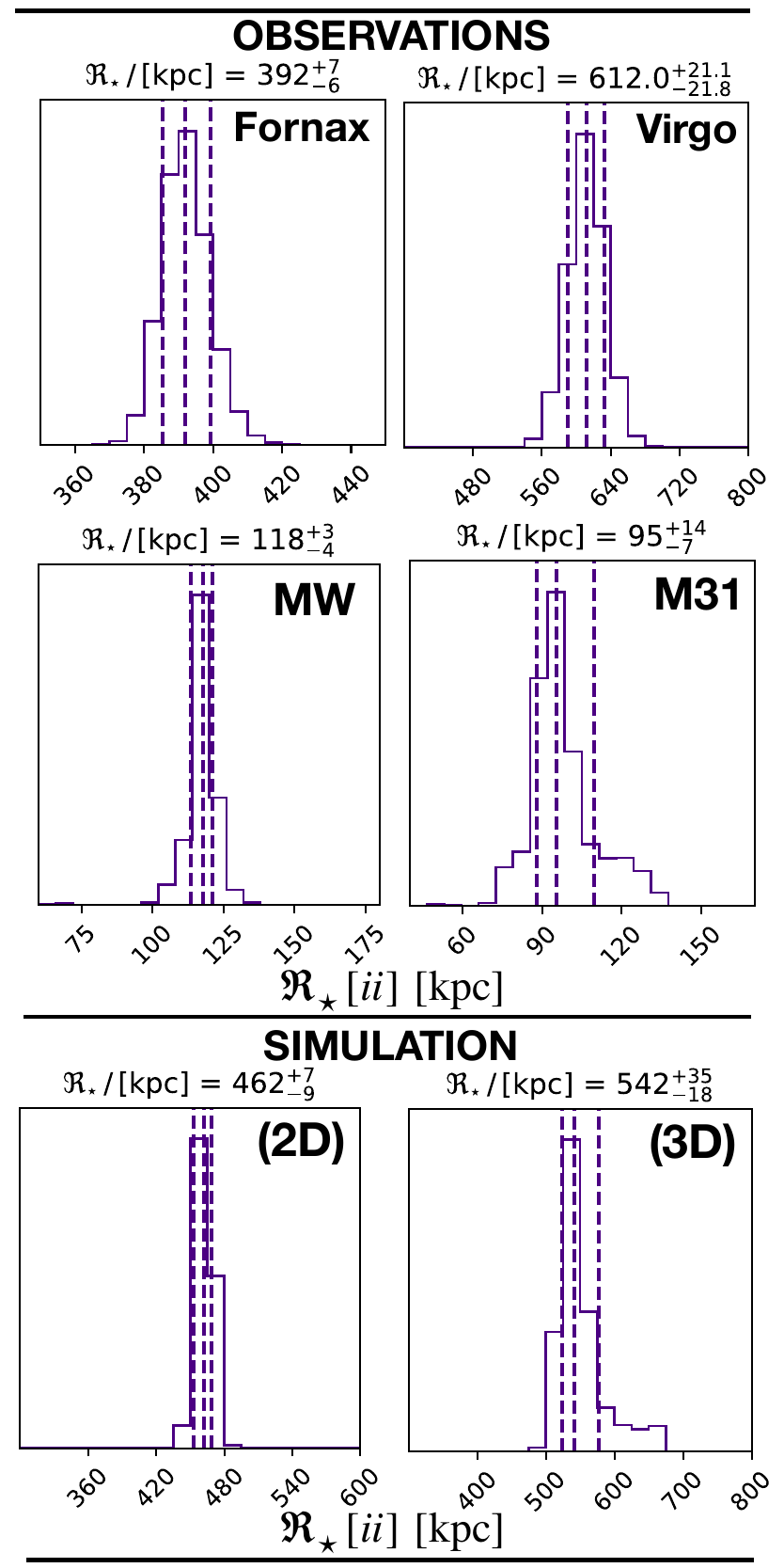}
%\vspace{-0.2cm}
\caption{Distribution of measurements of $\Rr$, marginalising over the distribution of $\langle\bar{\rho}^{\star}_{2h\star}\rangle_{\Rr}$ (minimum in the moving average of the mean stellar mass densities profiles) using methods [i] (left panel) with Eq.~\ref{eq:rtran1} and [ii] (right panel) with Eq.~\ref{eq:rtran2}. 
The top four rows of panels show the results for the observed systems (Fornax, Virgo, MW, and M31), and the bottom row panels present the simulated TNG50 Fornax-type cluster, 
with \Rr measured in projection (2D) and in 3D.}
\label{fig:measureRr}
\end{center}
\end{figure*}

\cMB{We show, using the toy model, simulations, and observations, that the stellar and dynamical mean mass densities of satellite galaxies tend to be higher in the central regions of their hosting environments.
Furthermore, within a transition radius $\Rr$ the moving average profile of the satellite central stellar mass densities $\langle \bar{\rho}^{\star}\rangle$ increases towards the cluster/group centre.}

\cMBII{In Sect.~\ref{sec:res:sim} we qualitatively describe this transition radius $\Rr$ using the profiles of $\langle \bar{\rho}^{\star}\rangle$, and plot the corresponding \Rr values in the figures where we presented the data of simulations and observations (see Figs.~\ref{fig:TNG50cluster}, \ref{fig:TNG300}, and \ref{fig:obs}).}
Here we describe in detail how we measure $\Rr$ and the possible systematic uncertainties of its measurement.
We define two metrics, showing two examples in Fig.~\ref{fig:measureRrFornax} with simulations and observations, where we measure \Rr according to the following equations:
\begin{align}
 \Rr [i] &:= R\left(\rhoavf_{\Re\!\star}\right)\label{eq:rtran1}\\
 \Rr [ii] &:= R\left(\partial_{r} \rhoavf\!=\!0\right) \label{eq:rtran2}
\end{align}
The first method ($[i]$) measures a minimum in the moving average of the satellite mean stellar mass density profile in the region $0.2<R/\Rvir<1$, that is, where $\rhoavf_{\Re\!\star}:=\rhoavf({\rm min})$.
For this, we show in Fig.~\ref{fig:measureRrFornax} data from simulations (Fornax-type cluster, top panel),
and from observations (Fornax cluster, bottom panel) showing the application of this metric in a linear scale where the minimum is better revealed.

The second method ($[ii]$) uses a smoothing function (polynomial of order 7) that is fitted directly to the data to calculate the profile of the derivative. 
The order of the polynomial is obtained from optimising \cMBII{it to well approximate the moving average profile near its minimum.}
\cMB{From this $\Rr$ is found where the derivative in the radial range of $0.2<R/\Rvir<1$ becomes zero.
Systems that have substructures can produce additional zero roots, as seen in the Fornax cluster (Fig.\ref{fig:measureRrFornax}, bottom panel).
Therefore, we select the first zero that appears in the range $0.2<R/\Rvir<1$ starting from the lower bound, 
\cMBII{which is where the polynomial function fits the profile well near the minimum.}
Even though both methods are arbitrary and empirical they agree well, as shown in Fig.~\ref{fig:measureRrFornax}, with both capturing the trends of the moving averages.
Such methods with derivatives of mass density profiles are also commonly applied to measure changes in density profile slopes to detect and quantify substructures in dark and stellar haloes \citep{ONeil2021,Aung2021}.}

\begin{table}[t]
\begin{center}
\caption{Measurements of $\Rr$ {in nearby environments and simulations}.}
\vspace{-0.5cm}
 \begin{threeparttable}
\setlength{\tabcolsep}{1pt}
\renewcommand{\arraystretch}{1.2}
\begin{tabular}{|l | r | r| }
 \hline
 System & $[i]\,\Re_{\star}\,[{\rm kpc}]\,(R_{\rm vir})$ & $[ii]\,\Re_{\star}\,[{\rm kpc}](R_{\rm vir})$ \\ [0.5ex] 
 \hline\hline
 Fornax Cluster & $380^{+16}_{-8}\,(0.35)$ & $392^{+7}_{-6}\,(0.36)$\\ 
 \hline
 Virgo Cluster (A) & $612^{+127}_{-55}\,(0.39)$ & $612^{+21}_{-22}\,(0.39)$  \\
 \hline
 Milky Way & $78^{+38}_{-1}\,(0.36)$ & $118^{+3}_{-4}\,(0.54)$ \\
 \hline
 Andromeda (M31) & $102^{+24}_{-12}\,(0.49)$ & $95^{+14}_{-7}\,(0.46)$\\
 \hline\hline
  Fornax-type (TNG50) (2D) & $481^{+86}_{-58}\,(0.45)$ & $462^{+7}_{-9}\,(0.42)$\\
   Fornax-type (TNG50) (3D) & $595^{+74}_{-19}\,(0.54)$ & $542^{+35}_{-18}\,(0.49)$\\
    \hline
   Virgo-type (TNG100) (2D) & $846^{+66}_{-20}\,(0.41)$ & $889^{+19}_{-19}\,(0.43)$\\
   Virgo-type (TNG100) (3D) & $828^{+50}_{-68}\,(0.40)$ & $914^{+11}_{-11}\,(0.44)$\\\hline   
   Virgo-type (\textsc{eagle}) (2D) & $747^{+96}_{-74}\,(0.51)$ & $707^{+20}_{-17}\,(0.50)$\\
   Virgo-type (\textsc{eagle}) (3D) & $823^{+59}_{-59}\,(0.58)$ & $745^{+16}_{-14}\,(0.52)$\\\hline   
\end{tabular}
\renewcommand{\arraystretch}{0.75}
\label{tab:results}
\begin{tablenotes}
  \item Note 1: The adopted virial radii for each system are: Fornax $1075\kpc$ \citep{Schuberth2010}, Virgo A 1550\kpc \citep{McLaughlin1999} , M31 207\kpc \citep{Tamm2012}, MW 216\kpc \citep{Bland-Hawthorn2016,Shen2022}, Fornax-type TNG50 cluster 1115\kpc, Virgo-type cluster TNG100 2079\kpc, and Virgo-type cluster \textsc{eagle} 1442\kpc.
\end{tablenotes}
\renewcommand{\arraystretch}{1}
\vspace{-0.5cm}
\end{threeparttable}
\end{center}
\end{table}

Given that the first method can be affected by the arbitrary number of neighbours $(q)$ used to calculate the moving average, 
and that both methods can be affected by the sample number of satellites, substructures, and outliers, we adopted a Monte Carlo technique to measure \Rr and its uncertainties.
For this we performed $10^6$ sub-sample realisations to measure \Rr each time applying a bootstrap (Jackknife) method taking a sub-sample with $n_{\rm sub\,sat}=n_{\rm sat}-\sqrt{n_{\rm sat}}$ satellites for each system with $n_{\rm sat}$ satellites, while taking a range of neighbours between $q=2\sqrt{n_{\rm sub,\,sat}}$ and $4\sqrt{n_{\rm sub,\,sat}}$. 
This generates $10^6$ moving average profiles $\langle \bar{\rho}^{\star}_{2h\star}\rangle$ that can slightly vary due to the
data sub-sample. We show in Fig.~\ref{fig:measureRrFornax} the moving averages for 200 realisations as an example.

\cMB{We tested the propagation of photometric errors (20\% for the luminosity and 10\% for the half-light radius) into the sub-sampling by adding noise to mean stellar mass density of each satellite (observations and simulation) from a normal distribution with a standard deviation of 30\% ($\sigma_{\bar{\rho}\star}$).
This results in a slight increase in the spread of $\Delta\langle \bar{\rho}^{\star}_{2h\star}\rangle\approx10^5\sm\kpc^{-3}$, as shown in Fig.~\ref{fig:measureRrFornax}. 
Furthermore, we also tested a modified version of Eq.~\ref{eq:kern} to calculate the moving averages of the mean stellar mass densities by weighting them with the inverse of the photometric errors ($w\!=\!1/(\sigma_{\bar{\rho}\star})^2$), finding similar values of \Rr. 
Therefore, in our fiducial procedure, we decide to omit the contribution of photometric errors.}
The resulting distributions of the measurements of \Rr for the different observed and simulated systems and their confidence ranges are presented in Fig.~\ref{fig:measureRr}
and summarised in Table~\ref{tab:results}.
The distributions of $\Rr$ show that the metric that uses the derivative ($\Rr [ii]$) is more stable than that that uses the minimum ($\Rr [i]$).
This is because $\Rr [i]$ directly measures the minimum in the moving average, which can be affected by outliers during the bootstrap re-sampling, while $\Rr [ii]$ is obtained from fitting a function of the re-sampled distributions, reducing stochasticity. 
\cMBII{Furthermore, we also find that $\Rr [i]$ can be more stable when the spatial completeness of the satellite information is limited, or significant substructures perturb the moving average of the satellite mass densities.}

We find that the transition radius \Rr has consistent values in virial radius units ($\Rr\!\simeq\!0.5\,\Rvir$) across the studied environments in simulations and observations. 
Furthermore, we also measure \Rr in the simulations in projection (2D),
where for each of the $10^6$ realisations we randomly choose a projection (Table \ref{tab:results}).
The comparison with \Rr measured from the 3D distribution reveals that the projected value is systematically smaller by $\sim\!15$\%, as generally expected for a projected mass distribution \citep[see][]{Wolf2010}.

\subsection{Null-hypothesis test}
\label{sec:dis:null}

\begin{figure}[t]
\includegraphics[width=8.5cm]{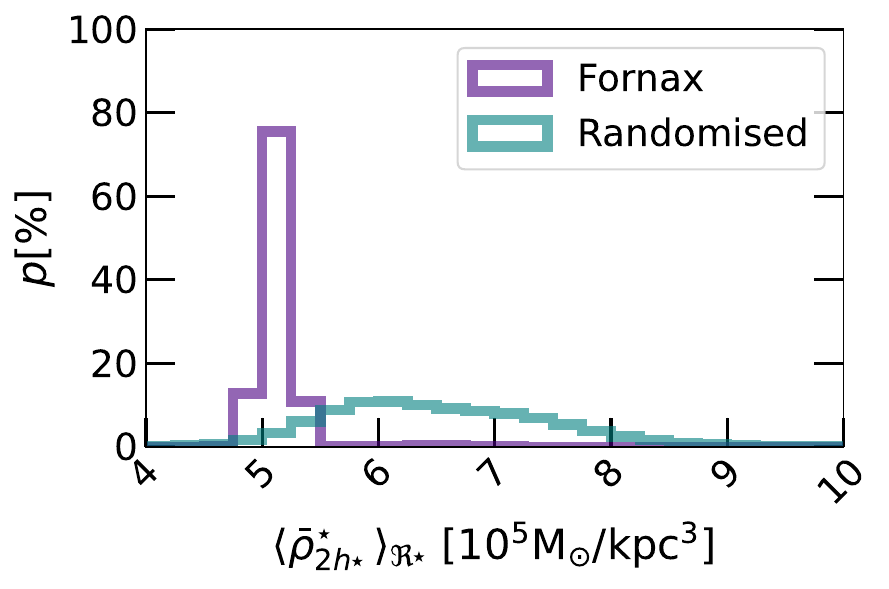}
%\vspace*{-0.3cm} 
\caption{Null-hypothesis test: Example with the Fornax galaxy cluster.
\cMB{For each of the $10^6$ measurements of $\Rr$ ([ii]), the corresponding distribution of the minimum values of the moving average curves of the mean stellar mass densities is shown (purple histogram).}
The distribution of the measurements of $\Rr$ for the randomised (shuffled) satellite densities (turquoise histogram) is produced by randomly shuffling the density values between the satellites while keeping their spatial positions within Fornax.
\cMB{A more visual comparison is shown in Fig.~\ref{fig:measureRrFornax}.}
Using these distributions we calculate that the probability of the randomised distributions resulting in densities as low as or lower than the most probable value measured in Fornax (\ie $\sim\!5.1\times10^5\sm\kpc^{-3}$) is of $p_{\rm NH}\!=\!4.0\%$.
We use the same procedure to calculate these probabilities, $p_{\rm NH}$, for Virgo, MW, M31 and certain simulation examples.}
\label{fig:fig_null}
\end{figure}

\cMBII{The signature of \Rr is detected at $\Rr\!\simeq\!0.5\,\Rvir$, where we find a minimum and a change of the moving average of the mean mass density profiles of the satellite population as a function of cluster-centric distance. Within this distance, the average density profile of satellites increases towards the centre.
Therefore, a coherent global structure where this mass density minimum occurs is encoded by the mean mass densities of satellites. 
If this coherent structure were produced randomly, statistical realisations would be expected to produce such a minimum, or one of even larger magnitude, very frequently.}
Therefore, we performed a null hypothesis test to estimate the significance of the \Rr measurement
in the four analysed observed systems (MW, M31, Fornax, Virgo) and in two clusters from the TNG simulations (Fornax-type and Virgo-type
clusters).
For this, we randomly shuffled the satellite stellar-mass densities, erasing the profile encoded in the satellite mass density distribution $\langle \bar{\rho}^{\star}_{2h\star} \rangle$.
In Fig.~\ref{fig:measureRrFornax} we present two examples, each with 200 moving averages of total $10^6$ realisations of the satellite populations, showing data from simulations (Fornax-type), observations (Fornax) and randomised data.
Then we measure \Rr and the moving average stellar density $\rhoavf$ at \Rr with methods [i] (Eq.~\ref{eq:rtran1}) and [ii] (Eq.~\ref{eq:rtran2})
in the original data and the randomised data.
With this we calculate the probability that a minimum of this significance is randomly generated, shown in Fig.~\ref{fig:fig_null} for the Fornax cluster sample.
In this exercise, we find that the \Rr in Fornax has a $p_{\rm NH}\!=\!4.0\%$ probability of being randomly generated with a minimum density as low as the observed one.
Following the same procedure for the other observed systems, we find probabilities of \Rr being randomly generated of 0.1\% in Virgo, 17\% in the MW, and 26\% in M31, with the latter estimates being \cMBII{larger than in Virgo and Fornax} due to the limited sample statistics.
The compound probability that \Rr is generated simultaneously at random in four observed systems is $1.8\cdot10^{-6}$, an unlikely scenario corresponding to $\sim 5\sigma$. The compound probability would become even smaller if we included the probabilities from all clusters and groups from the cosmological galaxy simulation data.

\subsection{Satellite sample completeness, mass range, gas, and other effects on the tracers of $\Rr$}
\label{sec:dis:trace}
We test the impact of satellite completeness on the mass density profiles, and on the measurement of \Rr. 
For the observations, we tested the satellite samples of the Fornax cluster, which has satellites as faint as 22.7 mag ($i$-band), and the Virgo cluster, with a satellite sample as faint as 18.7 mag ($r$-band).
We moved our detection threshold to 2 mag higher luminosities and found similar values of $\Rr[ii]=379^{+14}_{-9}\kpc$ for Fornax, which is only 3\% smaller than for the full dataset.
Similarly, for simulations we tested increasing the threshold of the minimum total mass (number of particles) of the subhaloes of the Fornax-type cluster by one order of magnitude larger than our fiducial configuration, finding a similar value of \Rr within the scatter.

\begin{figure}[b]
\begin{center}
\includegraphics[width=9.0cm]{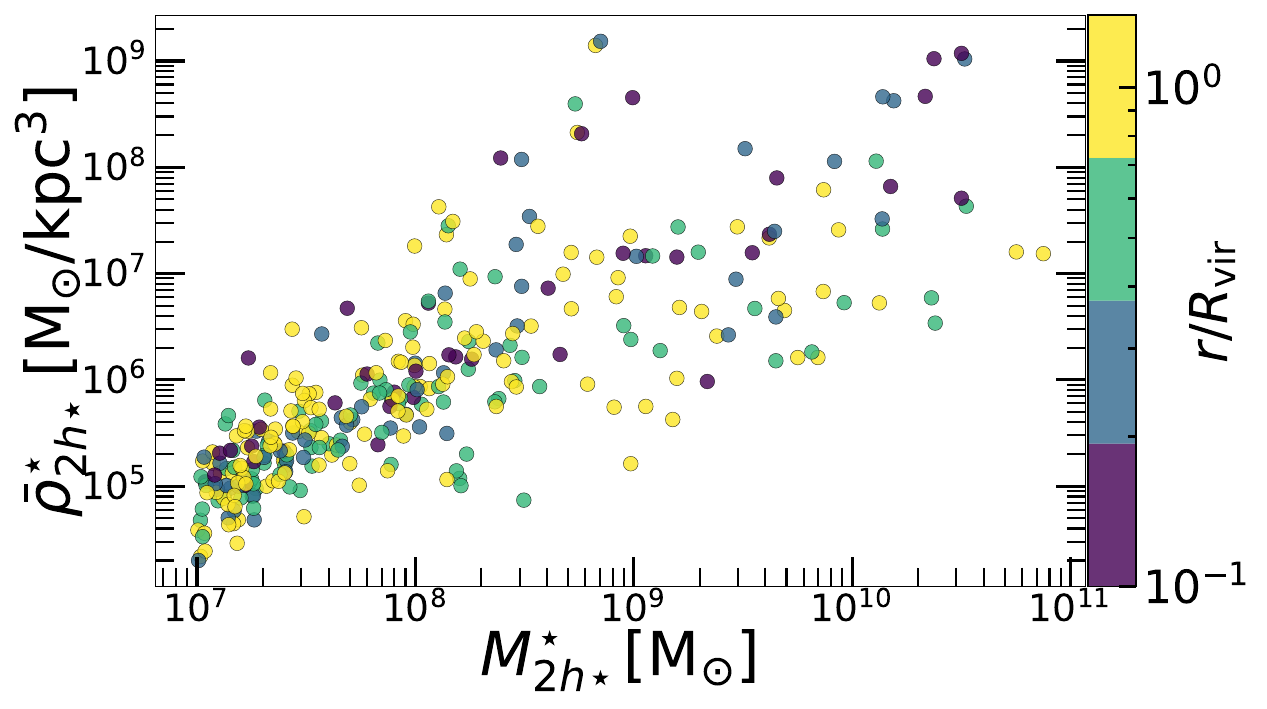}
%\vspace{-0.7cm}
\caption{Satellite galaxies in the TNG50 Fornax-type cluster, showing their central stellar mass densities as a function of their stellar masses. 
The colour-coding here shows their cluster-centric distances, which indicate that massive satellites can be found both in the outskirts and in the internal regions of the cluster.
}
\label{fig:fig_tng50_den_mass}
\end{center}
\end{figure}

\begin{figure}[t!]
\begin{center}
\includegraphics[width=8.7cm]{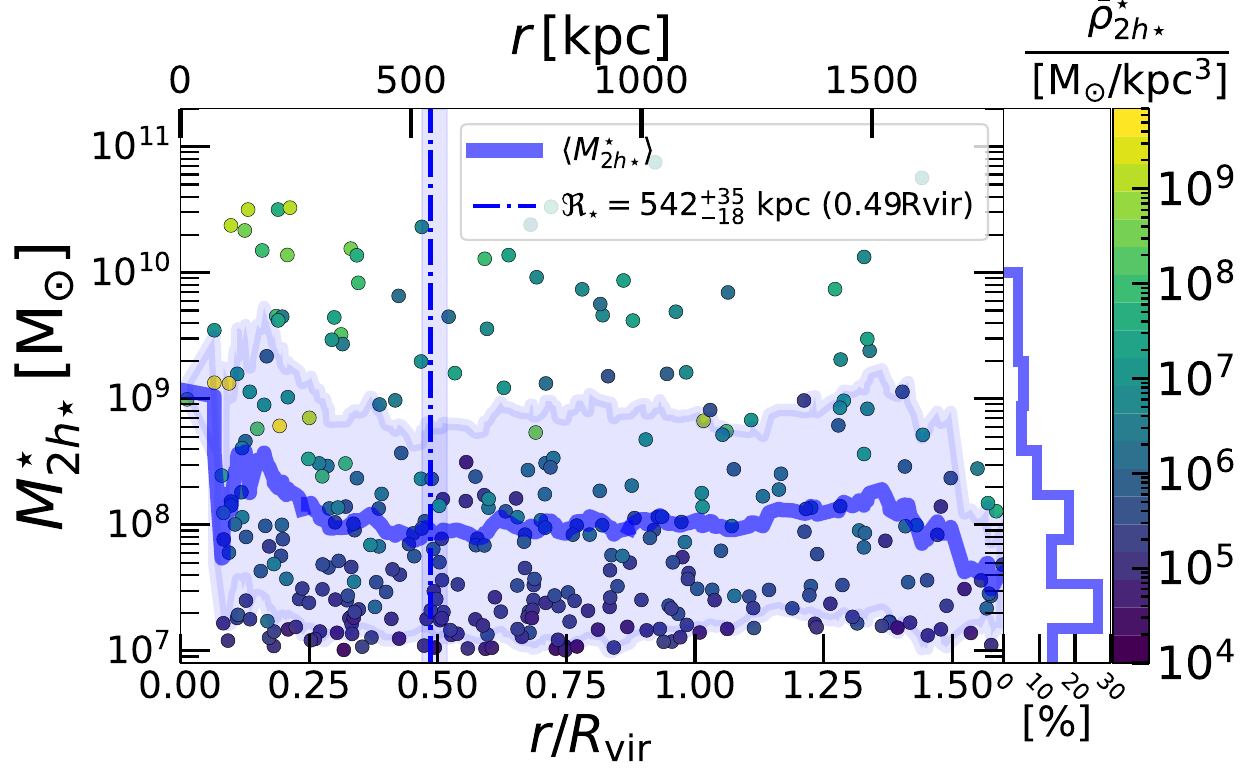}

\includegraphics[width=8.7cm]{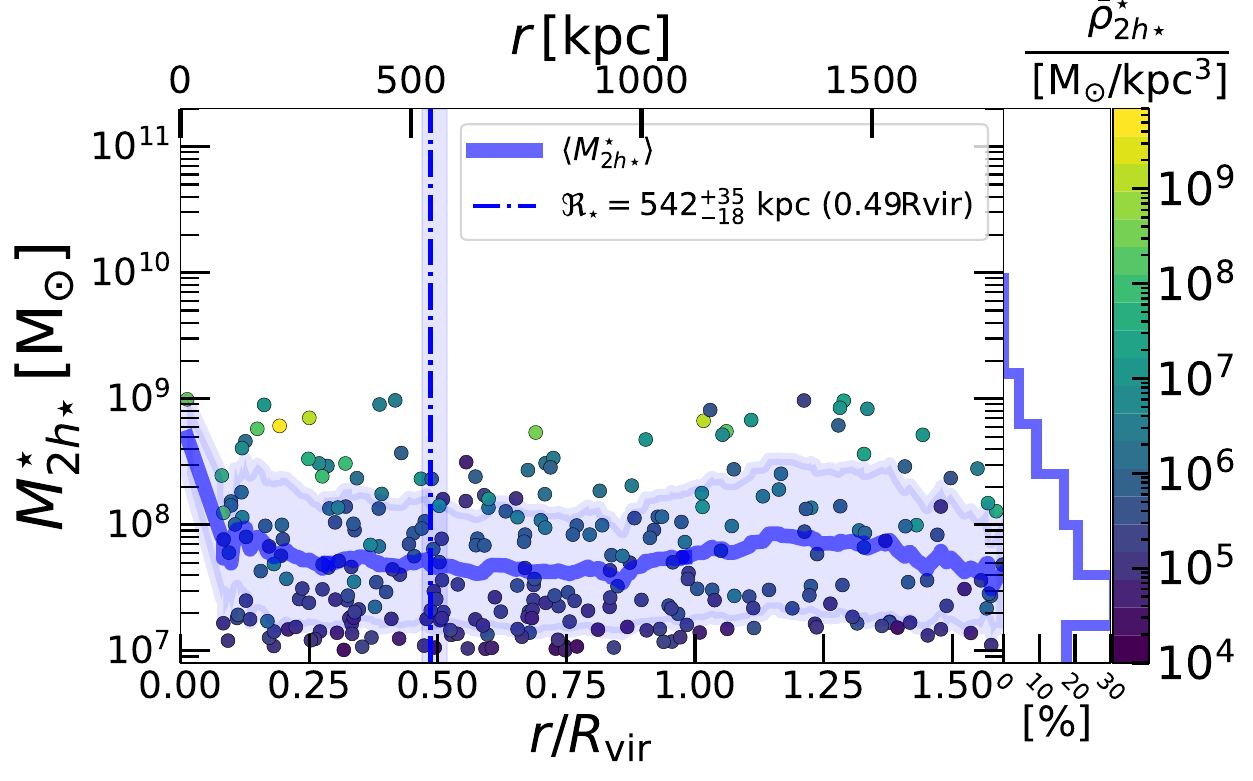}

\includegraphics[width=9.1cm]{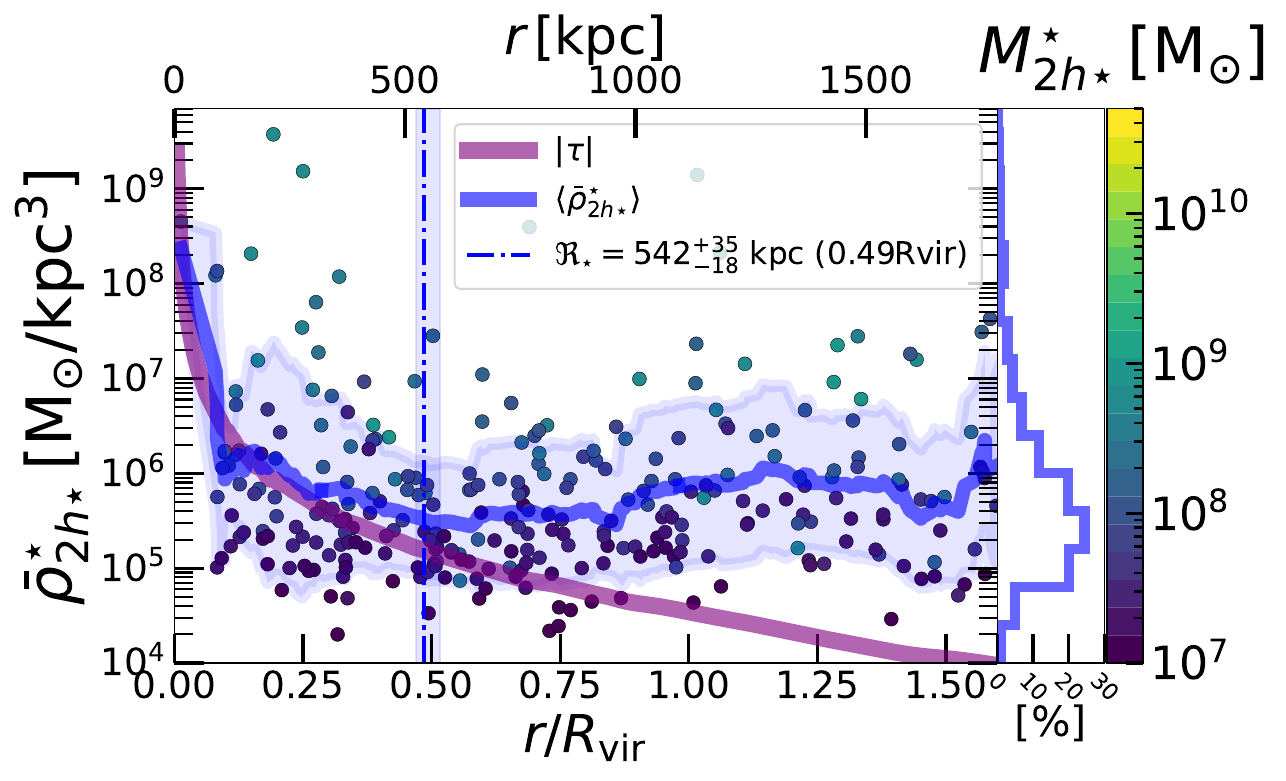}
%\vspace{-0.7cm} 
\caption{Satellite galaxies in the TNG50 Fornax-type cluster. 
Top and Middle panels: Satellite stellar masses and moving average (calculated with Eq.~\ref{eq:kern}) shown as a function of their cluster-centric distances and colour-coded by their stellar mass densities. 
In the middle panel, we masked out satellites with $M^{\star}_{2h\star}>10^9\sm$.
Bottom panel: Central mean stellar mass densities of the satellites and the moving average as a function of their distances shown for satellites with stellar masses within $M^{\star}_{2h\star}\leq10^9\sm$.
The transition radius \Rr is also plotted (vertical blue line), indicating that within \Rr the mean stellar mass densities still increase despite the removal of massive satellite.
The cluster tidal field is also shown ($\tau$).  
Histograms are plotted in the right sub-panels to show their distributions.}
\label{fig:fig_tng50_mass_dist}
\end{center}
\end{figure}

We also explored how different satellite density tracers may change the measurement of \Rr. 
For example, in the case of the observations, we tested the Fornax cluster by using the satellite luminosity mean densities derived from their $i$-band observations, 
finding a value of \Rr similar to the value estimated with the mean stellar mass densities.
In the case of the simulations, we find that the value of \Rr varies only within the scatter when using different mass density tracers, such as the central dynamical mass mean density ($\bar{\rho}_{2h\star}$) shown in Fig.~\ref{fig:TNG50cluster} (middle panel), or
the central mean stellar mass density within one stellar half-mass radius $\bar{\rho}^{\star}_{h\star}$.
This co-variation of the central mean stellar density measured at different radii ($h\star$ or $2h\star$) and with different tracers (light, stellar, and dynamical mass) is a consequence of the half-mass radius variation (Eq.~\ref{eq:rh} and \ref{eq:rh*}) and the mass stripping driven by the cluster tidal field.
More importantly, these tests show that, regardless of how the tracer is chosen, as long as it is used consistently for the entire satellite sample, it can trace the mass density profile and the location of the transition radius \Rr.

We also tested whether the satellite mass densities profiles may be driven by mass segregation, with the most massive galaxies located in the centre of clusters.
For this we analysed in detail the Fornax-type cluster from the TNG50 simulation, 
which shows a correlation between stellar mass densities and stellar masses (Fig.~\ref{fig:fig_tng50_den_mass}).
The relation shows a large scatter without any systematic bias of finding a preferred satellite mass at a given cluster-centric radius. We calculate the moving average of the stellar masses as a function of distance using Eq.~\ref{eq:kern}, where Fig.~\ref{fig:fig_tng50_mass_dist} (top panel) shows that within the transition radius \Rr the stellar masses start to increase on average.
To test whether the stellar masses could be driving the mass density profile, we masked out the massive satellites, leaving only satellites with stellar masses smaller than $M^{\star}_{2h\star}\!=\!10^{9}\sm$. We find that the increase in the stellar mass moving average in the centre is negligible when the masking is considered (Fig.~\ref{fig:fig_tng50_mass_dist}, middle panel). 
Furthermore, the stellar mass densities of the satellite sample with the masked massive satellites still shows a significant increment of the moving average (Fig.~\ref{fig:fig_tng50_mass_dist}, bottom panel), increasing from a minimum of $\langle\bar{\rho}^{\star}_{2h\star}\rangle\!\simeq\!3\times10^5\sm\kpc^{-3}$ near \Rr to $\sim\!20\times10^5\sm\kpc^{-3}$ within $r<0.2\,\Rvir$.
This shows that low-mass satellites still have systematically higher mass densities in the central region of the cluster.
Moreover, as the dynamical friction strength depends directly on the satellite's mass and is inversely proportional to the satellite orbital velocity ($f_{\rm DF}\propto m_{\rm sat}/v_{\rm sat}^2$) it can drive massive satellites to the central region of the cluster faster than less massive satellites. 
However, this only changes the orbital distribution of the satellites, while the mean mass density of each satellite continues to be shaped by the tidal field of the host during its orbit, stripping the outer material of the satellite according to Eq.~\ref{eq:tdentrdm*}.\\
A similar argument can be stated for galaxy harassment, where the cumulative effects of satellite-satellite collisions within a satellite population must obey the tidal shaping of the overall cluster potential, as shown by simulations.

We also estimated whether the gas component in satellite galaxies is significantly affecting the stellar and dark matter distribution.
Satellite gas depletion through ram-pressure stripping, tidal stripping, and stellar feedback, can perturb the satellite's potential, while star formation converts gas into stars, changing the stellar-mass fraction.
Therefore, we inspect the radial distribution of the satellite gas masses and densities in the studied environments.
For this, we again used data from cosmological galaxy simulations that provide access to a complete satellite sample with all their gas phases.
Overall, our analysis of the gas properties in satellite galaxies shows that within the transition radius ($r\lesssim\Rr$), most satellites have very low gas-to-baryon mass fractions, with only 1.5\% of the satellites having gas-to-baryon fractions above 10\%, and increasing to 13\% within the splashback radius ($r\lesssim R_{\rm sp}$).
\cMBII{Therefore, we find that the impact of the gas mass densities on the stellar and total mass densities of the analysed environments is minimal in the central region ($r\lesssim0.5\Rvir$), but it matters for larger distances ($r\gtrsim\Rvir$).
Moreover, data from the observed environments show that most satellites located in central regions lack significant amounts of gas, as shown in the MW and M31 \citep{Putman2021}, the Fornax cluster \citep{Loni2021,Kleiner2023}, and in the Virgo cluster \citep[\eg][]{Chung2009a,Yoon2017}.
Further analysis goes beyond the scope of this work, but there is a vast literature related to the detailed quenching of satellite galaxies at different mass regimes \citep[e.g.][]{Gunn1972,Mori2000,Faerman2013,Wetzel2013,Emerick2016,Donnari2018,Lotz2019,Roberts2019,Tonnesen2019,Tonnesen2021,Pallero2022,Herzog2022,Rodriguez2022,Blana2024}.}

\subsection{Projection effects and sky observables}
\label{sec:dis:proj}

\begin{figure}[b]
\includegraphics[width=9.2cm,trim={0.2cm 0.25cm 0.25cm 0.1cm},clip]{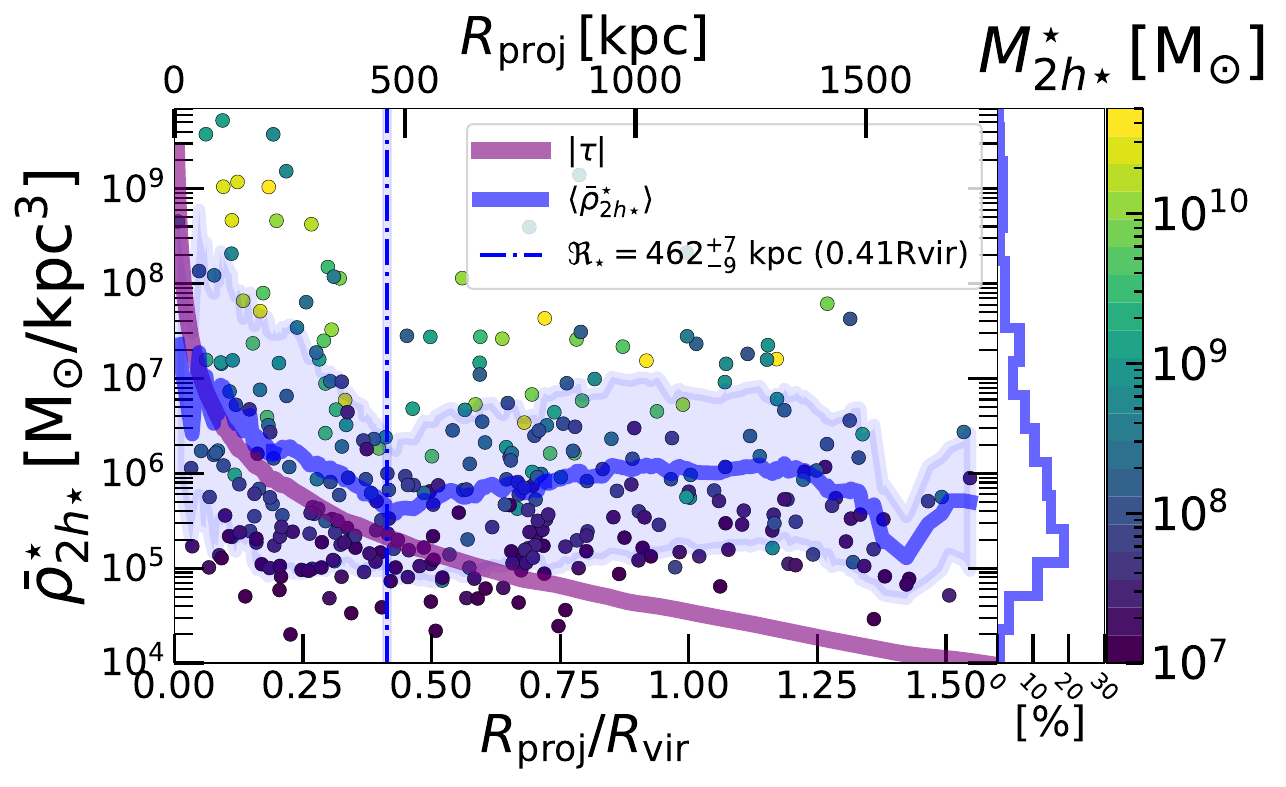}
%\vspace{-0.6cm}
\caption{
 Mean stellar mass densities ($\bar{\rho}^{\star}_{2h\star}$) of 492 satellite galaxies (circles) as a function of their projected distances to their hosting TNG50 Fornax-type cluster,
 with colours indicating their stellar masses. The deprojected profile is shown in  Fig.~\ref{fig:TNG50cluster} (bottom panel). 
 The host tidal field profile ($|\tau|$) is shown with the purple curve.
The blue curve shows the moving average profile using Eq.~\ref{eq:kern}.
The measurement of the transition radius ($\Rr$) is performed using the projected profile $R_{\rm proj}$ (vertical dot-dash line).
The histogram in the right sub-panel shows the shape of the satellite mass density distribution.
}
\label{fig:TNG50clusterproj}
\end{figure}

We explored with simulations how the satellite stellar mass density distribution is affected by projection effects to understand its impact on the moving average satellite density profiles and the transition radius \Rr.
For this, we analysed the Fornax-type cluster from TNG50, where we randomly project the cluster to measured the projected moving average of the satellite stellar mass density profile $\langle\bar{\rho}^{\star}_{2h\star}\rangle$.
As shown in Fig.~\ref{fig:TNG50clusterproj}, the projected profile of $\langle\bar{\rho}^{\star}_{2h\star}\rangle$ shows the same behaviour as in the case measured with the 3D distances $r$ (Fig.~\ref{fig:TNG50cluster}, bottom panel) reaching increasingly higher values in the central region of the cluster.
However, the transition radius \Rr is slightly shifted towards smaller cluster-centric distances at $\Rr\!=\!0.42\,\Rvir\,(462\kpc$; see Table~\ref{tab:results}).
This smaller projected \Rr is likely produced by the projected distribution of the satellites, similar to the projected 3D mass profiles \citep{Wolf2010}.
Additionally, in the very centre of the cluster ($r<0.2\Rvir$) the mean mass densities appear lower because of the distant satellites that appear closer in projection (interlopers).
\cMBII{This systematic could partially explain the variations in the shape of the observed satellite mass density profiles in Fig.~\ref{fig:obs} (see Sect.~\ref{sec:res:obs}).}
In addition, the random orientation of the cluster along the observer's line of sight can also affect the profile of $\langle\bar{\rho}^{\star}_{2h\star}\rangle$, especially in clusters that have massive filaments and merging substructures.

\subsection{Our results and the literature}
\label{sec:dis:lit}

As covered in Sect.~\ref{sec:theo:tdenrel}, the mean mass densities of satellite galaxies are expected to be a function of the distance to the centre of the host system because of its varying tidal field.
Studies focused on MW satellites have identified this correlation in the central dynamical mass densities of satellites and their orbital pericentres \citep[$\bar{\rho}\!-\!r_{\rm peri}$; see][]{Kaplinghat2019,Hayashi2020,Pace2022,Cardona-Barrero2023}.
Here, we expand on this further, finding that:
i) the mean stellar mass densities (and light) of the satellite population also trace this relation at their current positions, giving rise to the stellar-mass density-distance relation ($\bar{\rho}^{\star}\!-\!r$), and relate this to the host's tidal field; 
ii) the behaviour of the $\bar{\rho}^{\star}\!-\!r$ relation changes at the transition radius at $\Rr\!\simeq\!0.5\,\Rvir$ where the host tidal field is weak;
and iii) these correlations can be identified in systems with a wide range of host masses ($10^{12}$-$10^{15.5}\sm$) considering simulations and observations.

Furthermore, to investigate the pristine scaling relations of satellite galaxies, one must account for the effects of the host tidal field and the observed $\bar{\rho}\!-\!r$ relation, which reflects how satellites have been transformed since their infall into denser environments.
In particular, tidal evolution impacts the mass-size scaling relations such as the fundamental plane in early-type galaxies \citep{Djorgovski1987}, the (stellar) mass–size relation in early-type galaxies \citep{Kauffmann2003a,Kauffmann2003b} and in late-type galaxies \citep{Kormendy2013a,Saglia2016}, which can also depend on more detailed morphologies of the galaxies (spheroid versus disc-like) \citep{Dressler1980,Kuchner2017}.
Moreover, observational evidence suggests that the environment affects the galaxy mass-size relation \citep{Kuchner2017,Demers2019,Matharu2019,Venhola2019,Joshi2020}.
For example, \citet{Chamba2024} report a similar finding to our analysis of the Fornax cluster satellites. They find that satellites within the cluster are systematically smaller than in the field,
which they attribute to rapid gas removal in the cluster environment, truncating satellites \citep[see also][]{Asencio2022}.
Instead here we find that, with our toy model and simulations, tidal truncation by the host cluster can explain the relation between the satellite spatial distribution and their stellar and total mean mass densities. 
The effect on the stellar mass densities strongly depends on cluster-centric distance, starting within the transition radius at $\Rr\!\simeq\!0.5\,\Rvir$, and becoming rapidly stronger towards the centre. 
However, projection effects can increase the scatter in cluster centres at $R_{\rm proj}<0.1\Rvir$.

\cMBII{Moreover, the $\bar{\rho}\!-\!r$ relation is intimately connected to the morphology-density relation \citep{Dressler1980,Fasano2015}; both depend on the cluster-centric distances to the most crowded regions.
A growing body of work shows that a satellite's orbital history leaves a clear imprint on its evolution, with the satellite's present orbital phase being encoded in its phase-space coordinates and stellar population parameters \citep[\eg][]{Wetzel2011, Yoon2017, Rhee2017,Smith2022,Montero-Dorta2024}.
In particular, the timing of first infall and pericentric passage strongly influences the star formation, mass, and morphological properties of satellites across different regions of the cluster \citep{Rhee2017,Lotz2019,Pallero2019,Smith2022},
a conclusion reinforced by the strong radial colour gradient of satellites found within 300\kpc of the Fornax cluster centre \citep{Eigenthaler2018,Spavone2020}. 
Our results indicate that the interplay between the host's tidal-field topology and the satellites' orbital histories shapes the observed radial pattern in their mean mass densities.}

\section{Summary and conclusions}
\label{sec:con} 

In this study, we investigated how the mean mass densities of satellite galaxies vary with their cluster-centric distances ($\bar{\rho}-r$) and with the strength of the host's tidal field. To this end, we combined three complementary approaches: an idealised tidal-stripping toy model (Sect.~\ref{sec:res:toy}), satellite populations drawn from cosmological hydrodynamical simulations (Sect.~\ref{sec:res:sim}), and deep panchromatic observational data (Sect.~\ref{sec:res:obs}).
Our main findings are the following:

\begin{enumerate}[label=\Roman*]
\setlength\itemsep{0.4em}
\item The tidal-stripping toy model and the IllustrisTNG simulations consistently show that satellites become denser towards the cluster centre. We find that the total mean mass densities scale as $\langle\bar{\rho}_{2h}\rangle\propto1/r$ and $\langle\bar{\rho}_{2h}\rangle\propto |\tau|$, where $|\tau|$ is the tidal field strength (Figs.~\ref{fig:fig_toymod} and \ref{fig:TNG50cluster}).
Moreover, the same relation holds for the satellites' central mean stellar mass densities, $(\bar{\rho}^{\star }_{2h\star})$ out to a well-defined transition radius at $\Rr\!\!\approx\!0.5\Rvir$.
\cMBII{Beyond \Rr the host tidal field weakens below the densities of the least-bound satellites, where stripping is negligible. As a result, $\bar{\rho}^{\star}_{2h\star}$ flattens or rises slightly towards the virial and splash-back radii ($\sim\!1.4\Rvir$).}

\item We examined 296 clusters in the \textsc{illustris} TNG50, TNG100, and TNG300 cosmological galaxy simulation volumes at redshifts $\redshift\!=\!0$ and 1 (Figs.~\ref{fig:TNG300}, \ref{fig:TNG100}, \ref{fig:TNG50}). The $\bar{\rho}^{\star}_{2h\star}-r$ transition appears in every run, with $\Rr\sim0.5\,\Rvir$ in 3D, and somewhat smaller values once projection effects are included. Detailed comparisons with Fornax-type and Virgo-type clusters in both TNG and \textsc{eagle} simulations confirm these trends.

\item \cMBII{Four observed satellite galaxy populations -- in the Virgo and Fornax galaxy clusters, as well as in the MW and the M31 galaxies -- show the same behaviour in their $\bar{\rho}^{\star}_{2h\star}-r$ scaling relations (Fig.~\ref{fig:obs}). The measured \Rr/\Rvir match the simulations, although inside $r\lesssim0.2\Rvir$ the observed densities are lower. This offset may reflect projection bias, but could also signal a difference in the actual stripping and/or disruption efficiency, which occurs in cluster cores, warranting further study.}

\item \cMBII{Our null-hypothesis statistical tests of the observations (Sect.~\ref{sec:dis:null}, Figs.~\ref{fig:measureRrFornax}, \ref{fig:fig_null}) show that the density transition radius \Rr is highly significant. The joint probability of obtaining this feature in all the observed environments purely by chance is approximately $\sim\!2\!\cdot\!10^{-6}$ ($\simeq5\sigma$).}

\item Where dynamical masses are available for the Local Group satellites and the simulations, the central dynamical-mass density correlates with both their central stellar-mass density and tidal field strength (Fig.~\ref{fig:TNG50cluster} middle panel, Fig.~\ref{fig:obs} bottom panels). The resulting \Rr values agree with those derived from the stellar mass components, reinforcing the connection between the satellite structure and the host's tidal field strengths.
\end{enumerate}

\cMBII{We find that the distribution of the satellites' mean mass densities is governed by three interlocking factors acting simultaneously:
(1) the tidal field of the host cluster,  
(2) the satellites' orbital histories and their cluster-centric distances, and  
(3) the satellites' initial mass densities, with this last point revealing that neither mass nor size on their own dictate how tidal forces reshape satellites.
}

\cMBII{These findings tighten the link between the satellites' internal properties and external tidal environment, which drive their evolution. Extending the study of the ($\bar{\rho}-r$) relation to loose groups and rich clusters will reveal how the tidal evolution of satellites is interconnected with large-scale processes, such as the baryon-to-dark matter fraction, star formation, and ram pressure, among others. 
This in turn will refine classic scaling laws, including the morphology-density and the galaxy mass-size relations.
}

Despite these insights, our analysis still carries limitations and opens several promising avenues for further investigation.
\cMBII{Preliminary snapshots from the IllustrisTNG simulations at $z\!=\!1$ suggest that the radial density profile and the transition radius are in place early. Ongoing work (Bla\~na et al. {\it in prep.}) will pin down when the $\bar{\rho}^{\star}_{2h\star}-r$ scaling relation is established. 
}
\cMB{Our current analysis adopts a spherically symmetric host potential.
Mapping the full tidal-tensor field will allow us to quantify anisotropic compression, stretching, and shear, and to predict where satellite galaxies and star clusters are most susceptible to stripping.
}

\cMBII{Because the transition at \Rr is encoded in the satellites' stellar-mass density alone, it can be measured without kinematic data. This provides a novel and practical way to constrain the tidal field and perhaps even the dark-matter distribution in distant systems where spectroscopic observations are unavailable or limited.
}

\begin{acknowledgements}
The authors deeply thank Mary Putman and her collaborators, as well as Suk Kim and collaborators, for making their respective datasets freely available. 
Mat\'ias~Bla\~na also thanks Andreas~Burkert, Ra\'ul~Angulo, and Claudia~Pulsoni for their helpful comments and discussions, and Carolina~Agurto, Julio~Carballo, Claudio~Bla\~na and Ana~D\'iaz for their support.
The authors also thank the anonymous referee for comments that improved the manuscript.
The authors thank the \textsc{TNG} collaboration for making their data, software and server capability freely available. 
M.~Bla\~na thanks ANID for funding through the Fellowship FONDECYT Postdoctorado 2021 No.~3210592.
Y.~Ordenes-Brice\~no thanks ANID for the FONDECYT Postdoctorado 2021 No.~3210442 and ESO comit\'e mixto 2024.
T.~Puzia acknowledges support through FONDECYT Regular No.~1201016, 
P.B.~Tissera acknowledges funding through FONDECYT Regular No.~1200703, 
and M.~Mora acknowledges funding through FONDO CONICYT-GEMINI Posición postdoctoral.
E.~Johnston acknowledges support from FONDECYT de Iniciaci\'on en Investigaci\'on 2020 Project No.~11200263.
D.~Pallero thanks ANID for funding through the Fellowship FONDECYT Postdoctorado 2024 No.~3230379.
T.~Ziliotto acknowledges funding from the European Union’s Horizon 2020 research and innovation programme under the Marie Skłodowska-Curie Grant Agreement No. 101034319 and from the European Union – NextGenerationEU.
We also acknowledge support through ANID CATA-Basal FB210003.
The authors deeply thank the citizens of Chile for their tax contributions to the national development of science and this project in these difficult post-pandemic times. 
The authors extend their gratitude to all the researchers whose studies have been crucial to this work.
\end{acknowledgements}

\bibliographystyle{aa}
\bibliography{aa49520-24}

\onecolumn

\begin{appendix}
\section{Complementary material}
\label{sec:app:extra}

\begin{figure*}[h]
\centering\hspace{-0.2cm}
\includegraphics[width=9cm]{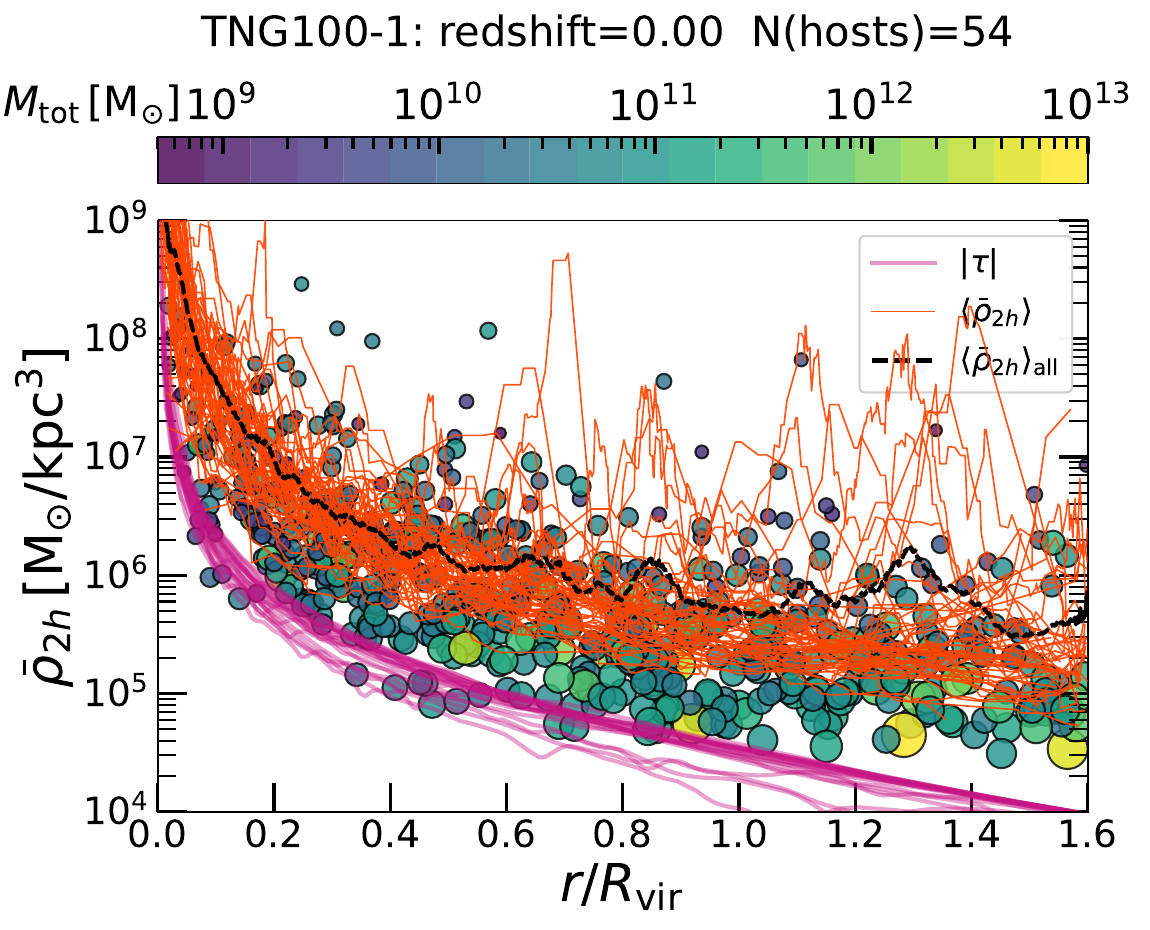}\hspace{0.2cm}
\includegraphics[width=9cm]{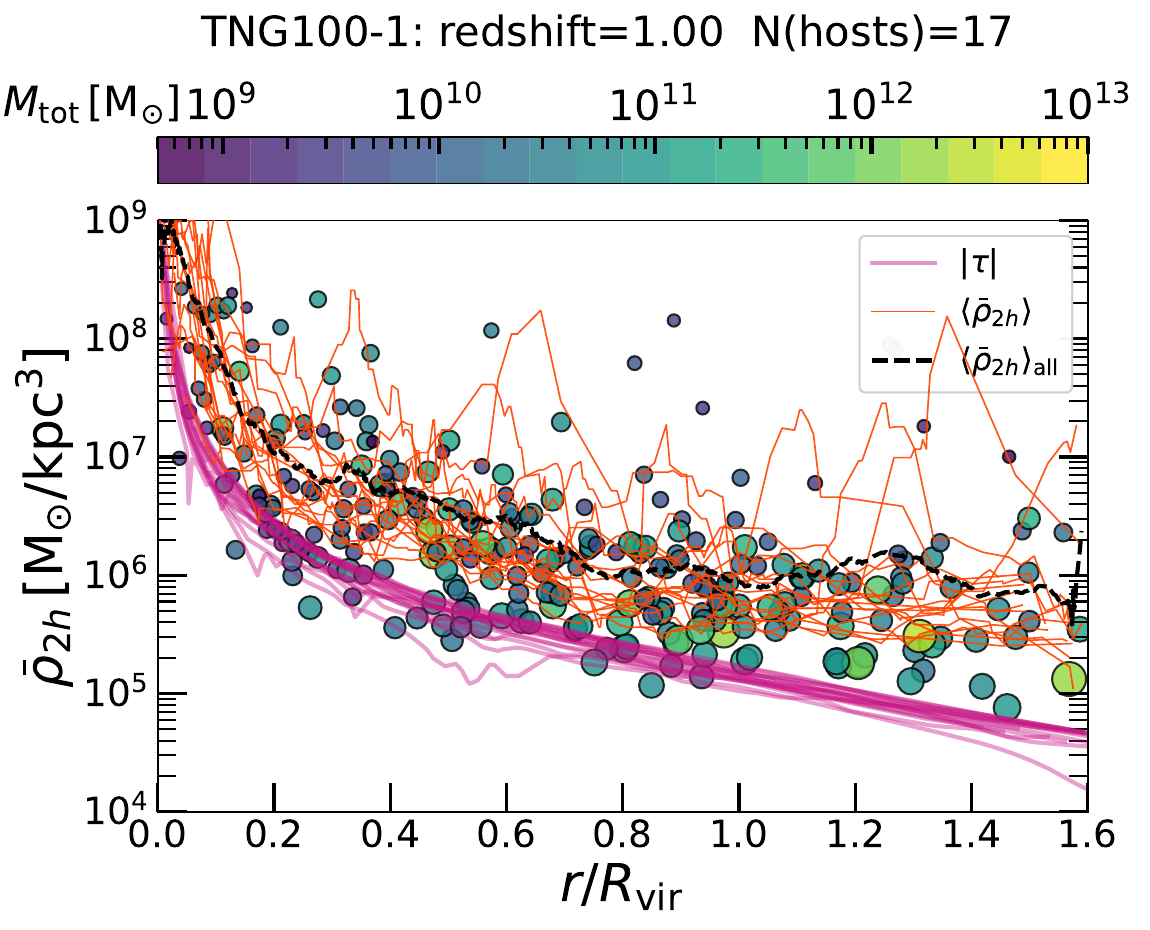}

\hspace{-0.15cm}
\includegraphics[width=9.1cm]{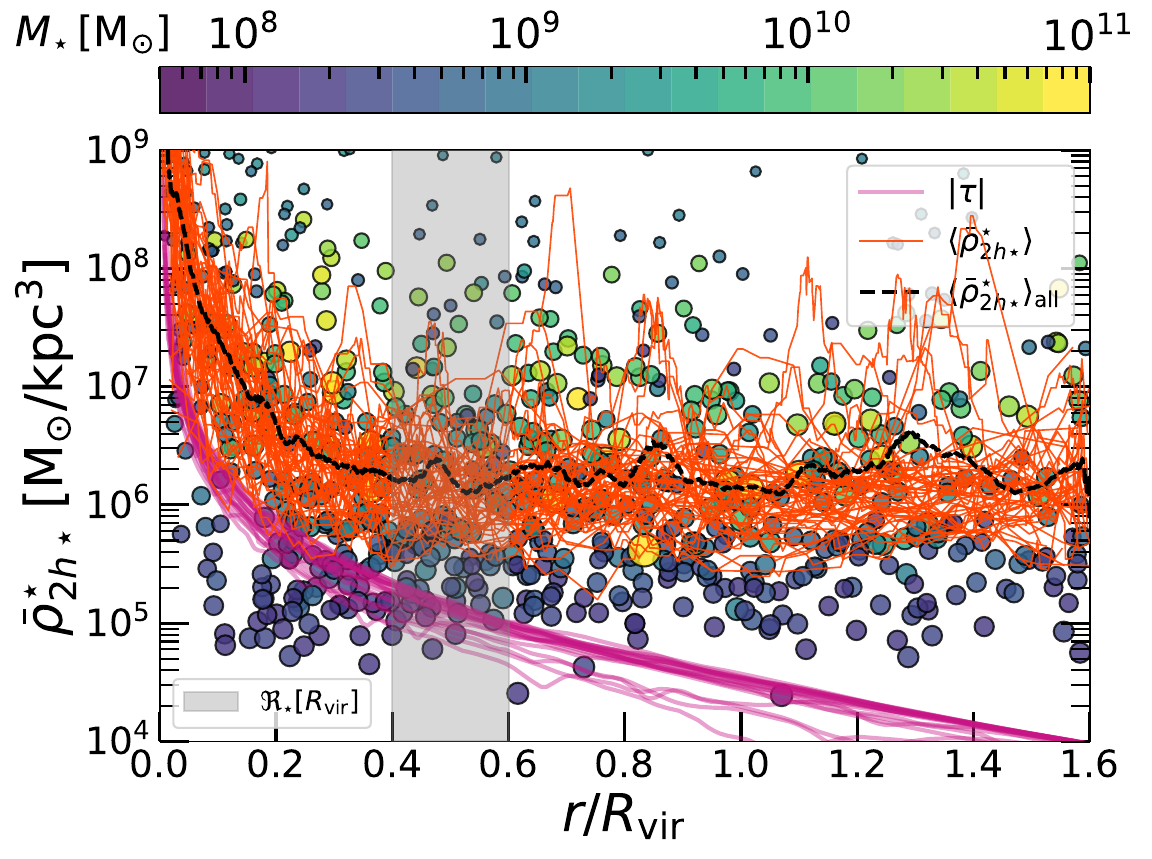}\hspace{0.1cm}
\includegraphics[width=9.1cm]{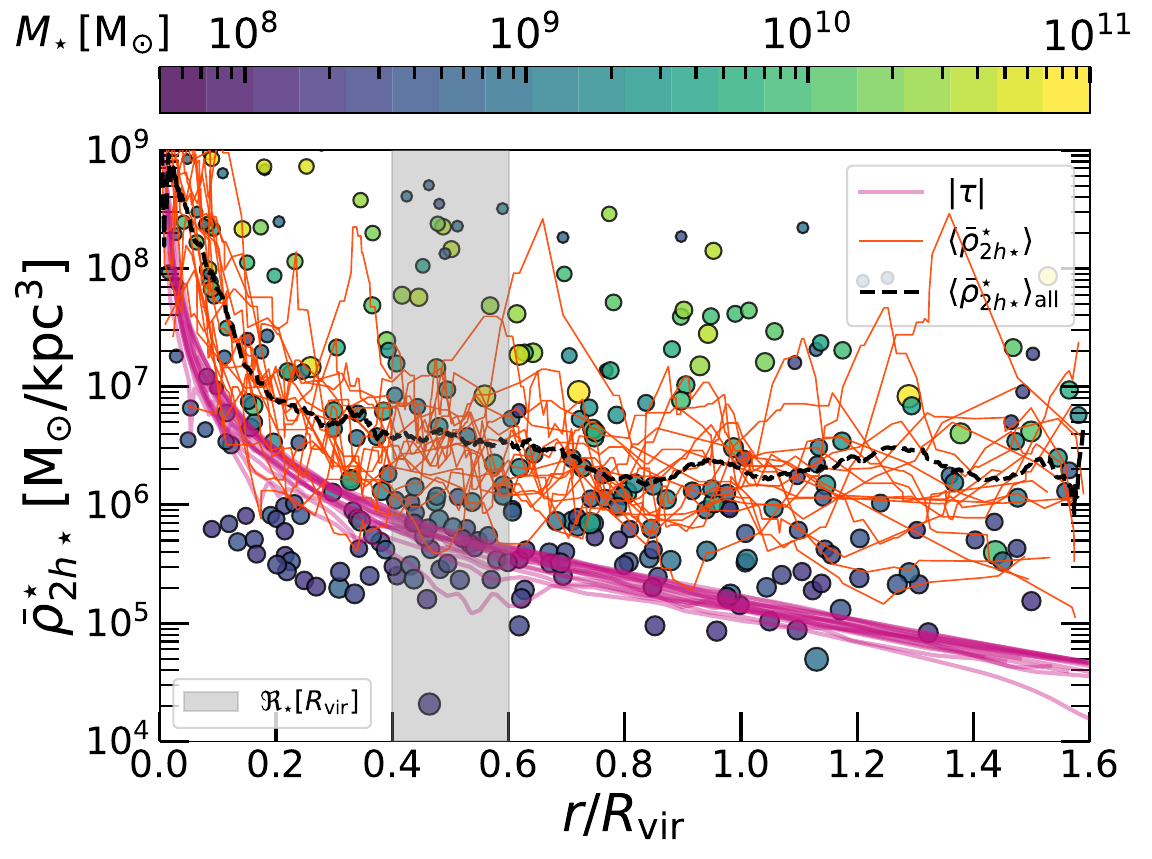}
\caption{Mass densities of satellite galaxies vs their cluster centric distances for 54 host galaxy clusters/groups from a snapshot at redshift $\redshift\!=\!0$ (left column) and for 17 clusters/groups at $\redshift\!=\!1$ (right column), taken from TNG100 (equivalent analysis for TNG300 in Fig.~\ref{fig:TNG300}). 
The tidal field profiles of all the clusters are shown in each panel ($|\tau|$ in purple curves).
Top panels show the total mass mean densities of the subhaloes (circles, colour-coded by total subhalo masses),
and the bottom panels show the central mean stellar mass densities (circles, colour-coded by stellar masses).
To avoid overcrowding, we plotted a sub-sample with 10\% (left column) and 20\% (right column) of the satellites (circles).
Circle sizes are proportional to $h$ (top panels) or $h\star$ (bottom panels).
In each panel the moving average satellite mean mass density profile of each cluster is shown in with an orange curve.
In addition, it is shown the global moving average in dashed black curve that is calculated with the moving averages of all clusters. All are moving averages are calculated with Eq.~\ref{eq:kern}.
The transition radii ($\Rr$), measured in each cluster, are located within the vertical grey region (bottom panels), which are determined from the moving average of the mean stellar mass densities $\langle \bar{\rho}^{\star}_{2h\star} \rangle$, marking where the $\langle \bar{\rho}^{\star}_{2h\star} \rangle$ start to increase further in. 
We present a similar analysis of TNG50 in Figs.~\ref{fig:TNG50}.
}
\label{fig:TNG100}
\end{figure*}

\begin{figure*}
\centering\hspace{-0.4cm}
\includegraphics[width=9.cm]{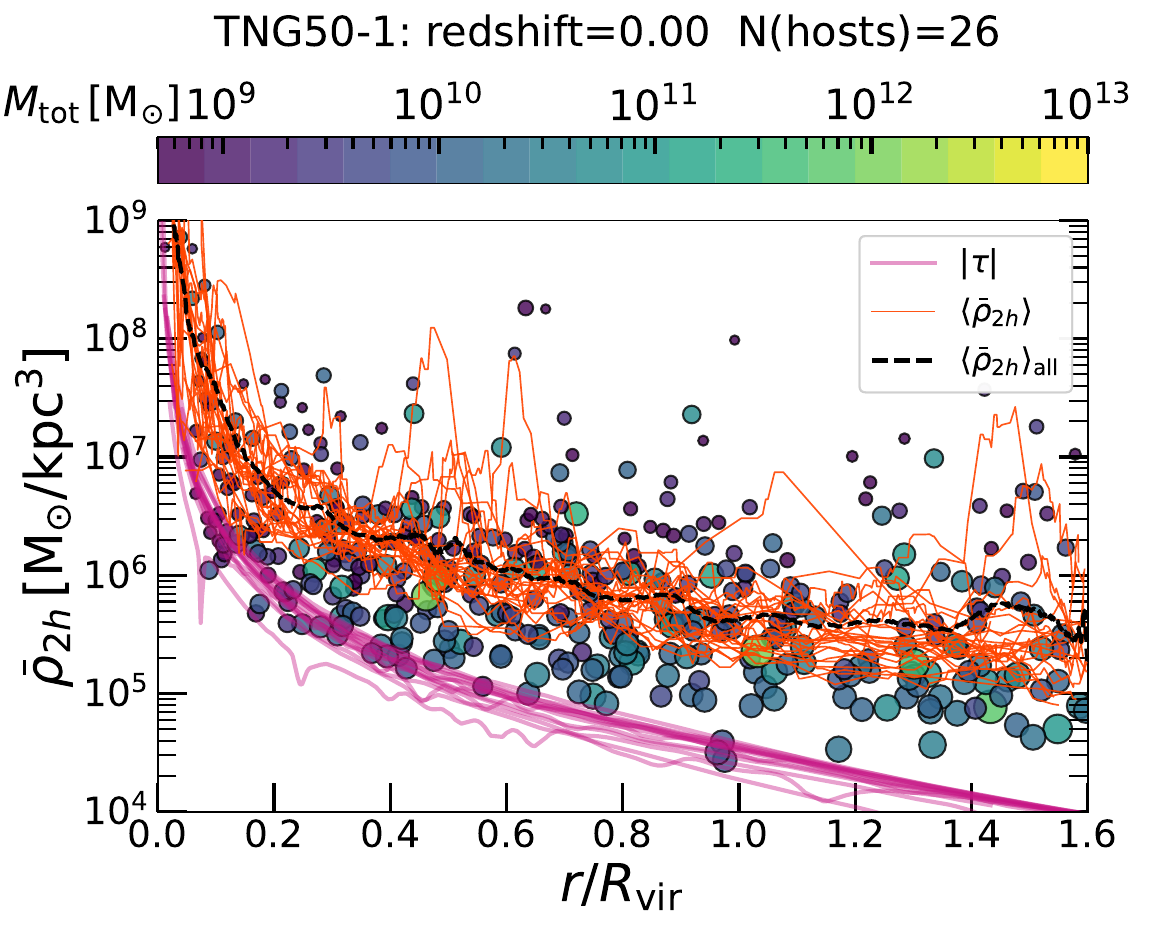}\hspace{0.2cm}
\includegraphics[width=9.cm]{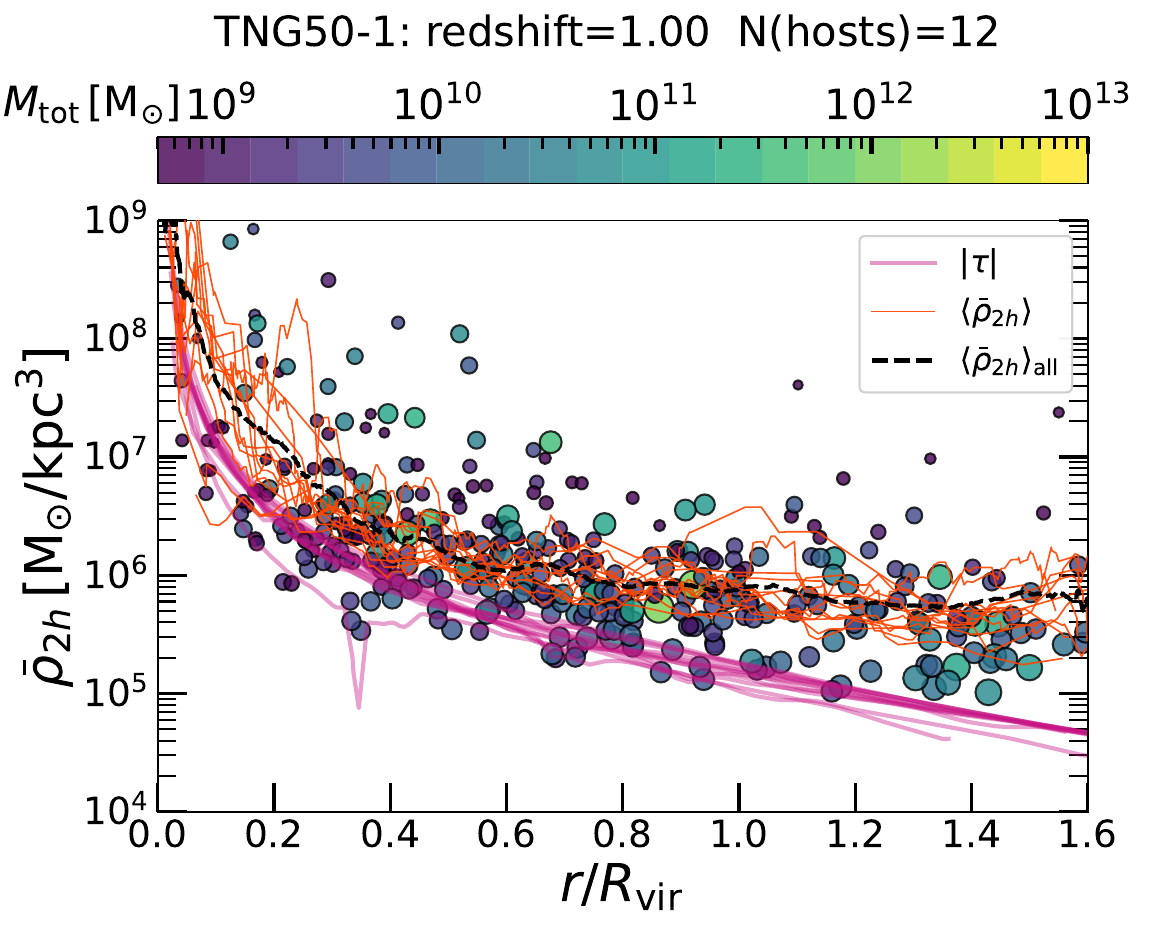}

\hspace{-0.15cm}
\includegraphics[width=9.1cm]{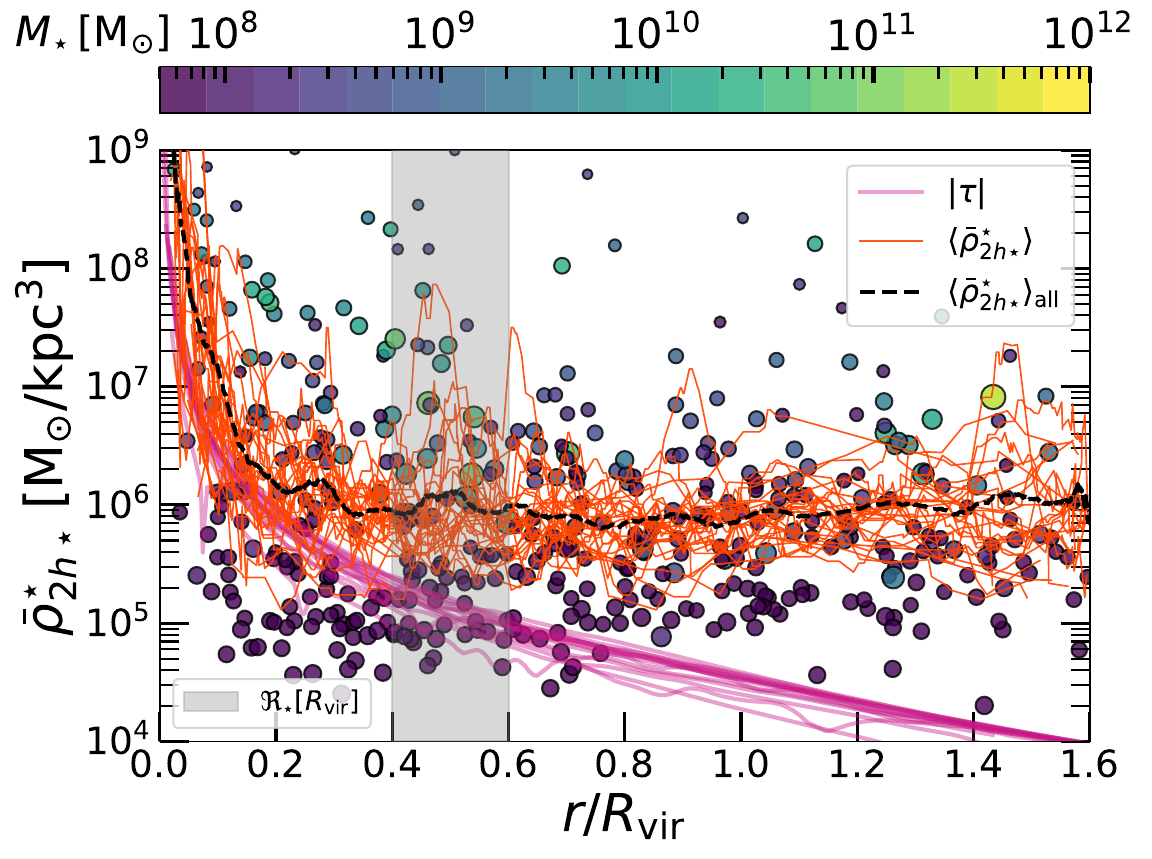}\hspace{0.1cm}
\includegraphics[width=9.1cm]{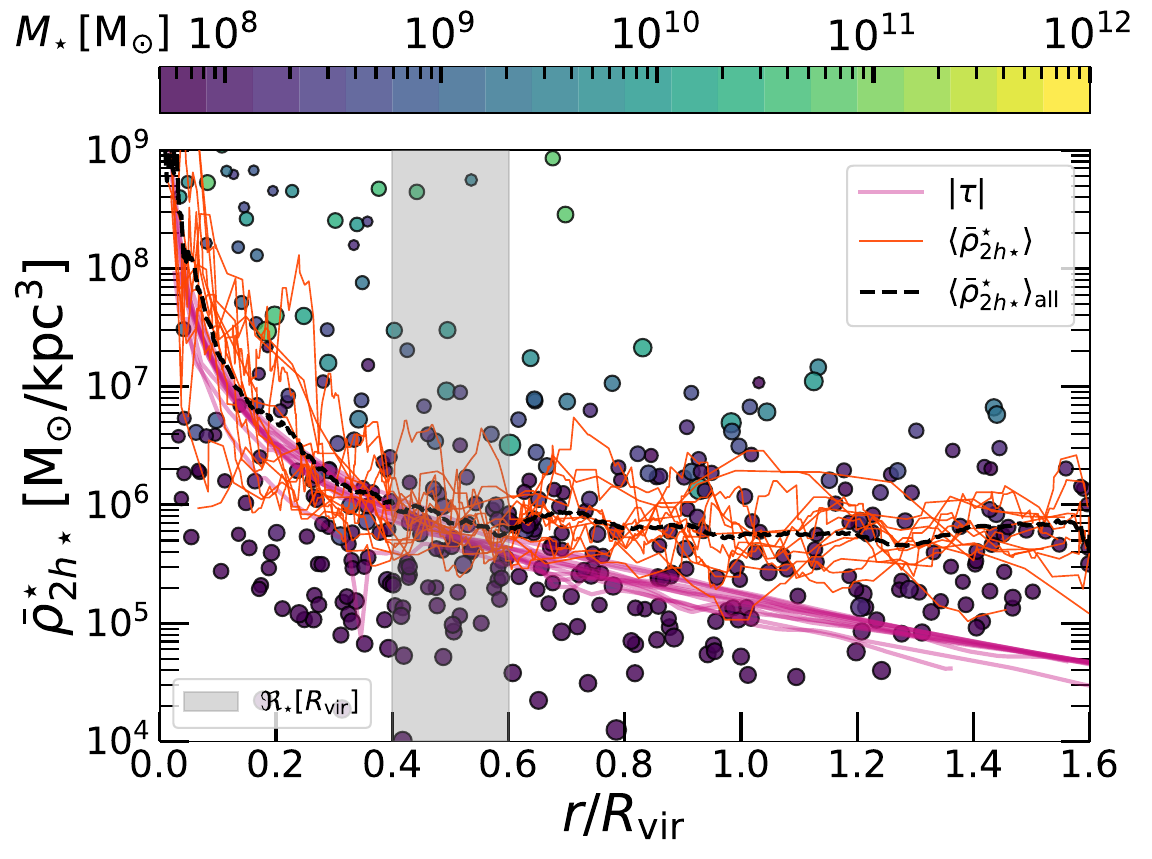}
\caption{
Mass densities of satellite galaxies vs their cluster centric distances for 26 host galaxy clusters/groups from a snapshot at redshift $\redshift\!=\!0$ (left column) and for 12 clusters/groups at $\redshift\!=\!1$ (right column), taken from TNG50 (equivalent analysis for TNG300 in Fig.~\ref{fig:TNG300}). 
The tidal field profiles of all the clusters are shown in each panel ($|\tau|$ in purple curves).
Top panels show the total mass mean densities of the subhaloes (circles, colour-coded by total subhalo masses),
and the bottom panels show the central mean stellar mass densities (circles, colour-coded by stellar masses).
To avoid overcrowding, we plotted a sub-sample with 10\% (left column) and 20\% (right column) of the satellites (circles).
Circle sizes are proportional to $h$ (top panels) or $h\star$ (bottom panels).
In each panel the moving average satellite mean mass density profile of each cluster is shown in with an orange curve.
In addition, it is shown the global moving average in dashed black curve that is calculated with the moving averages of all clusters. All are moving averages are calculated with Eq.~\ref{eq:kern}.
The transition radii ($\Rr$), measured in each cluster, are located within the vertical grey region (bottom panels), which are determined from the moving average of the mean stellar mass densities $\langle \bar{\rho}^{\star}_{2h\star} \rangle$, marking where the $\langle \bar{\rho}^{\star}_{2h\star} \rangle$ start to increase further in. 
We present a similar analysis of TNG100 in Fig.~\ref{fig:TNG100}.
}
\label{fig:TNG50}
\end{figure*}

\cMBII{We present the mean mass density profiles of satellite galaxies as a function of their cluster-centric distances in Fig.~\ref{fig:TNG100}, taken from TNG100 for 54 host galaxy clusters from a snapshot at redshift $\redshift\!=\!0$ (left column) and 17 clusters at $\redshift\!=\!1$ (right column). In Fig.~\ref{fig:TNG50} we show the same relations for 26 host galaxy clusters/groups from a snapshot at redshift $\redshift\!=\!0$ (left column) and 12 clusters at $\redshift\!=\!1$ (right column) taken from TNG50.}

\cMBII{As with the TNG300 cluster (Fig.~\ref{fig:TNG300} in Sect.~\ref{sec:res:sim}), we see that the mean dynamical-mass densities of the satellites ($\bar{\rho}_{2h}$) in TNG100 and TNG50 (top panels in Figs.~\ref{fig:TNG100} and \ref{fig:TNG50}) are shaped by the host tidal field, where the low density subhaloes are clearly limited by the tidal field strength according to Eq.~\ref{eq:tdentr} ($\bar{\rho}_{2h}>|\tau|$).
Moreover, the mean stellar-mass mean densities also show a correlation with the tidal field strength that extends out to the transition radius at $\Rr\simeq 0.5\Rvir$.
Some clusters present high-density variations due to infalling substructure around the main trend, as shown by the density moving averages (wiggles in some orange curves in Figs.~\ref{fig:TNG300}, \ref{fig:TNG50} and \ref{fig:TNG100}).
However, the main trend for the total- and stellar-mass density profiles remains the same across the TNG simulations, as shown by the moving average of the stacked profiles (dashed black curves in all panels in Figs.~\ref{fig:TNG300}, \ref{fig:TNG100} and \ref{fig:TNG50}.}

\end{appendix}
\end{document}